\documentclass[11pt]{article}

\usepackage{jcappub} 
\usepackage[normalem]{ulem}
\usepackage[T1]{fontenc}
\usepackage{slashed,amsmath,amsfonts,amssymb,bbm}
\usepackage{mathrsfs,wasysym}
\usepackage{units}
\usepackage{graphicx}
\usepackage{makecell}
\usepackage{booktabs,multirow}
\usepackage{commath}
\usepackage{xspace}
\usepackage[dvipsnames]{xcolor} 
\usepackage{comment}
\usepackage{bbm}
\usepackage[leftcaption]{sidecap}
\usepackage{wrapfig}
\definecolor{darkblue}{rgb}{0,0,0.5}
\definecolor{darkgreen}{rgb}{0.0,0.5,0.2}
\definecolor{darkred}{rgb}{0.6,0,0}
\usepackage{hyperref}
\usepackage[capitalise]{cleveref}
\usepackage{orcidlink}

\renewcommand{\to}{\rightarrow}

\newcommand{\snudd}{\texttt{SNuDD}\xspace}

\newcommand{\Ux}{$U(1)_X$\xspace}
\newcommand{\Udark}{$U(1)_{D}$\xspace}
\newcommand{\Ume}{$U(1)_{L_\mu-L_e}$\xspace}
\newcommand{\Uet}{$U(1)_{L_e-L_\tau}$\xspace}
\newcommand{\Umt}{$U(1)_{L_\mu-L_\tau}$\xspace}
\newcommand{\Uij}{$U(1)_{L_i-L_j}$\xspace}
\newcommand{\UBL}{$U(1)_{B-L}$\xspace}

\newcommand{\cevns}{CE$\nu$NS\xspace}

\newcommand{\Trh}{$T_{\rm rh}$\xspace}

\usepackage{tikz}
\usepackage{tikz-feynman}

\tikzfeynmanset{
yourboson/.style={
decoration={coil,aspect=0,amplitude=1.5mm,segment length=4mm,pre,pre length=0pt, post,post length=0pt},decorate}
}

\tikzfeynmanset{
yourboson2/.style={
decoration={coil,aspect=0,amplitude=1.5mm,segment length=4.1mm,pre,pre length=0pt, post,post length=0pt},decorate}
}

\subheader{\rm IFT-UAM/CSIC-26-27}

\title
{
New benchmarks for direct detection of freeze-in dark matter in vector portal models
}

\author[a]{David Cerde\~no\,\orcidlink{0000-0002-7649-1956},}
\author[a]{Patrick Foldenauer\,\orcidlink{0000-0003-4334-4228},}
\author[a,b]{Rafael L\'opez No\'e\,\orcidlink{0009-0002-6076-4988},}
\author[c]{\'Oscar Zapata\,\orcidlink{0000-0001-5533-4014}}
\affiliation[a]{Instituto de F\'isica Te\'orica IFT-UAM/CSIC, 
Cantoblanco, E-28049, Madrid, Spain}
\affiliation[b]{Departamento de F\'isica Te\'orica, Universidad Aut\'onoma de Madrid, Cantoblanco,\\ E-28049, Madrid, Spain}
\affiliation[c]{Instituto de F\'isica, Universidad de Antioquia, Calle 70 \# 52-21, Apartado A\'ereo 1226, Medell\'in, Colombia}
\emailAdd{davidg.cerdeno@ift.csic.es}
\emailAdd{patrick.foldenauer@csic.es}
\emailAdd{rafael.lopezn@uam.es}
\emailAdd{oalberto.zapata@udea.edu.co}

\abstract
{
We investigate the freeze-in of MeV-scale fermionic dark matter (DM) that couples to the Standard Model via a new vector mediator to assess the potential that future direct detection experiments have to observe new physics in either the DM or neutrino sectors. We study the minimal kinetic mixing dark photon of a secluded $U(1)_D$ as well as gauge bosons of the anomaly-free $U(1)_{L_i-L_j}$, with $i,j=e,\mu,\tau$, and $U(1)_{B-L}$ gauge extensions, exploring the impact of low reheating temperatures on the DM production rates. For the ultralight dark photon scenario, we show that current experimental constraints from electron recoil data in DAMIC-M and PandaX-4T can be avoided if the DM fermion is only a subcomponent (smaller than 40\%) of the total cold DM and that future detectors can be sensitive to a DM fraction below 1\% for masses above 1~MeV. For a massive dark photon, there are allowed regions of the parameter space with masses in the range 50 MeV $\lesssim m_{\rm DM}\lesssim$ 500 MeV that can be within the reach of direct detection experiments through nuclear recoils if freeze-in occurred at a low reheating temperature. Finally, the case of $U(1)_{L_i-L_j}$ and $U(1)_{B-L}$ is particularly interesting since the discovery of new physics can come from either the DM or the neutrino sector, which features new interactions. We find that freeze-in at low reheating temperatures can reproduce the observed abundance in large parts of the parameter space up to gauge couplings of $g_X\sim10^{-2}$ for MeV DM. Most notably, direct detection experiments will be sensitive to considerable parts of this parameter space in nuclear recoils for 50 MeV $\lesssim m_{\rm DM}\lesssim$ 500 MeV. Additionally, the enhanced signal from solar neutrino coherent scattering is observable in these scenarios, which can serve as a further handle to identify the underlying particle physics model.
}

\makeatletter
\gdef\@fpheader{}
\makeatother

\begin{document}
\notoc
\maketitle

\section{Introduction}
\label{sec:intro}

Direct dark matter (DM) detection experiments have become extremely versatile tools with which to test the nature of this abundant and exotic new form of matter. They attempt to observe the collisions of DM particles with the target material of a detector that is typically placed underground and shielded to reduce the flux of cosmic rays and other unwanted sources of background.

For several decades, direct detection experiments have probed, with increasing sensitivities, the parameter space of weakly-interacting massive particles (WIMPs), a very well-motivated class of DM candidates (see, e.g., Ref.~\cite{Cirelli:2024ssz} for a recent review). These are produced in the early universe when they {\em freeze-out} of the thermal bath, in such a way that the resulting relic abundance matches the observed one if their thermally averaged annihilation cross-section is of the order of the electroweak scale. It is expected that WIMPs can scatter with the target material, leading to nuclear recoils (NR) in the keV range. There has not yet been any confirmed observation of WIMPs, which has led to stringent upper limits on their scattering cross-section off nuclei. Currently, large xenon-based experiments are leading the search for DM masses above $\sim 1$~GeV \cite{XENON:2025vwd,LZ:2024zvo,PandaX:2024qfu}.

More recently, direct detection  experiments have also begun to analyse potential DM signatures in electron recoils (ER) \cite{Essig:2011nj, Essig:2012yx}. Because of the very small mass of electrons (compared to nuclei), electron recoils can leave observable signals after the scattering of MeV scale DM particles (or even keV if the DM particle is absorbed). In parallel, some  detection techniques have specialised in lowering their experimental threshold, allowing them to explore low-mass DM candidates through nuclear recoils. As a result, the scattering of MeV scale DM particles with both electrons and nuclei is already severely constrained \cite{CRESST:2020wtj, CRESST:2019jnq, CRESST:2024cpr, SENSEI:2023zdf, SuperCDMS:2024yiv, SuperCDMS:2025dha, DarkSide:2022knj, DarkSide-50:2022qzh, PandaX:2022xqx, PandaX:2025rrz, XENON:2019gfn, XENON:2024znc, XENON:2026qow, DAMIC-M:2025luv, LZ:2022lsv, LZ:2023poo, LZ:2025igz}.

Interestingly, these probes of low-mass DM also test alternative production mechanisms, most notably {\em freeze-in}~\cite{McDonald:2001vt,Hall:2009bx,Chu:2011be, Hambye:2018dpi}, in which the coupling between the DM and Standard Model (SM) particles is so small that thermal equilibrium is not reached. In this paradigm, DM particles are gradually produced by annihilations or decays of other particles (normally SM ones) in the thermal bath. The couplings involved are very small and therefore DM particles never reach thermal equilibrium, their relic abundance gradually increasing until the temperature of the universe becomes so small that either DM particles cannot be created for kinematical reasons or that the abundance of the SM particles producing DM becomes very suppressed.

Freeze-in DM models are currently under scrutiny, given the spectacular latest results of DAMIC-M~\cite{DAMIC-M:2025luv} and PANDAX-4T~\cite{PandaX:2025rrz}, based on ER. It has been shown that they rule out a large range of viable DM masses where the correct abundance is achieved in the dark photon model, a scenario in which the force carrier of a new $U(1)$ gauge symmetry acts as the mediator between the DM and SM sector~\cite{Okun:1982xi, Holdom:1985ag}. Due to its simplicity, and consistency from the theoretical point of view, the dark photon scenario constitutes a very convenient benchmark model with which to gauge the reach of future detectors~\cite{2020Snowmass2021LetterOI}, such as DarkSide-50k~\cite{DarkSide-50:2025lns}, Oscura~\cite{Oscura:2023qik}, SuperCDMS SNOLAB~\cite{SuperCDMS:2022kse} or TESSERACT~\cite{TESSERACT:2025tfw}.

The current experimental situation makes it necessary to review the dark photon scenario to understand the circumstances under which current experimental bounds can be avoided, and to find other benchmark models for freeze-in DM that can serve as a reference for future experiments.

In this article, we pursue this avenue by focusing first on the regions of the parameter space where the dark fermion is a subdominant component of DM. We exploit the distinct scaling of the DM detection rate with the local relic abundance in freeze-in models to show that future experiments have a great potential to probe under-abundant DM,
similar to what has been observed for millicharged DM in Ref.~\cite{Iles:2024zka}.

We then consider anomaly-free $U(1)$ gauge groups~\cite{Heeba:2019jho, Chang:2019xva, Mohapatra:2019ysk, delaVega:2021wpx, Nath:2021uqb, Ghoshal:2022zwu} obtained from gauging combinations~\cite{Davidson:1978pm,Davidson:1979wr,Mohapatra:1980qe, Wetterich:1981bx, Foot:1990mn, He:1990pn,Foot:1990uf, Foot:1992ui} of the accidental global symmetries of the SM, $U(1)_B$ and $U(1)_{L_\alpha}$, with $\alpha=e,\mu,\tau$, in which new neutrino interactions may offer complementary avenues for probing physics beyond the SM (BSM). 
Since the freeze-in mechanism is dominated by the smallest mass scale or lowest temperature, DM production becomes largely controlled by infrared scales. Motivated by this, we further embed each of the $U(1)$ models analysed within a cosmological scenario featuring a low reheating temperature~\cite{Co:2015pka, Belanger:2018sti}. In this framework, the modified thermal history can significantly affect the freeze-in dynamics, allowing the observed relic abundance to be produced for larger values of the gauge coupling compared to the standard cosmological scenario~\cite{Bhattiprolu:2022sdd, Cosme:2023xpa, Gan:2023jbs, Silva-Malpartida:2023yks, Becker:2023tvd, Cosme:2024ndc, Belanger:2024yoj,Arcadi:2024obp,  Arcadi:2024wwg, Barman:2024nhr, Barman:2024lxy,Bernal:2024ndy, Boddy:2024vgt,Arias:2025tvd, Bernal:2025osg}.

There is a final ingredient of new physics in these scenarios: neutrinos have interactions through the new gauge bosons that might  increase the coherent elastic neutrino-nucleus scattering (\cevns) cross-section. This is a process that can be within the reach of current and future direct DM detection experiments from solar neutrinos. In fact, PandaX-4T~\cite{PandaX:2024muv}, XENONnT~\cite{XENON:2024hup}, and LZ~\cite{LZ:2025igz} have already claimed a potential first observation of $^8$B solar neutrinos through this process. Although \cevns is an obstacle to DM detection, often referred to as a neutrino fog~\cite{Billard:2013qya,OHare:2021utq}, it can also be an evidence of new physics \cite{Boehm:2018sux,Amaral:2020tga,Amaral:2021rzw,DeRomeri:2024iaw,DeRomeri:2024dbv,DeRomeri:2026prc}.

With this motivation, we explore the potential of direct detection experiments to discover new physics in $U(1)$ models, either through the observation of DM particles in the MeV range or new neutrino physics. To do this, we consider both nuclear and electronic recoils.

This article is organised as follows. In \cref{sec:models} we introduce the details of the different $U(1)$ models and introduce the relevant notation. In \cref{sec:freeze-in}, we review the basic principles of DM production through freeze-in, we explain how low reheating temperatures play a role in DM production and we apply it to the cases of the dark photon and the different $U(1)$ models, comparing the different production channels. In~\cref{sec:direct_detection}, we discuss the phenomenology of the freeze-in scenario in the different $U(1)$ models at direct detection experiments and new physics signals registered in their detectors. In \cref{sec:results}, we present the results of our analysis.
Finally, the conclusions are presented in \cref{sec:conclusion}.

\section{Light DM in Vector mediator models}
\label{sec:models}

In this work we study a class of light DM scenarios, where the dark sector interacts with the SM via a vector portal. In the minimal setup, this interaction arises through a gauge-invariant, renormalisable kinetic mixing term between the SM hypercharge $U(1)_Y$ and a new $U(1)_X$~\cite{Okun:1982xi,Holdom:1985ag}. The relevant Lagrangian in the gauge basis can be written as~\cite{Bauer:2022nwt},
\begin{align}
    \mathcal{L} =& - \frac{1}{4}\,
    (\hat B_{\mu\nu}, \hat W^3_{\mu\nu}, \hat X_{\mu\nu})
    \begin{pmatrix}
    1 & 0 & \epsilon_B \\
    0 & 1 & \epsilon_W \\
    \epsilon_B & \epsilon_W & 1
    \end{pmatrix}
    \begin{pmatrix}
    \hat B^{\mu\nu}\\
    \hat W^{3\mu\nu}\\
    \hat X^{\mu\nu}
    \end{pmatrix} 
    - \left( g'\, j^Y_\mu,\, g\, j^3_\mu,\, g_X \,j_\mu^X \right) 
    \begin{pmatrix}
    \hat B^{\mu}\\
    \hat W^{3\mu} \\ 
    \hat X^{\mu}
    \end{pmatrix} 
    \nonumber \\
    &+  \frac{1}{2} \, 
    (\hat B_{\mu}, \hat W^{3}_\mu, \hat X_{\mu})\, \frac{v^2}{4}
    \begin{pmatrix}
    g'^2 & - g' g & 0 \\
   - g' g & g^2 & 0 \\
    0 &  0 & \frac{4\,{m}_{X}^2}{v^2}
    \end{pmatrix}
    \begin{pmatrix}
    \hat B^{\mu}\\
    \hat W^{3\mu} \\ 
    \hat X^{\mu}
    \end{pmatrix}  \, , 
    \label{eq:loop_lag}
\end{align}
where the hatted fields $\hat B^{\mu}$ and $\hat W^{3\mu}$ are the neutral gauge bosons of the SM $U(1)_Y$ and $SU(2)_L$, while $\hat X^{\mu}$ is the \Ux gauge boson. Here, $\epsilon$ corresponds to the kinetic mixing parameter that couples the \Ux and $U(1)_Y$ bosons. We denote the coupling constants of $U(1)_Y$ and $SU(2)_L$ by $g'$ and $g$, respectively, and $j^Y_\mu$ and $j^3_\mu$ are their corresponding (neutral) gauge currents. Similarly, $g_X$ is the coupling constant and $j_\mu^X$ the dark fermionic current of the new \Ux, while $v$ denotes the electroweak (EW) vacuum expectation value (VEV) and $m_X$ is the tree-level \Ux gauge boson mass. For the purpose of this work, the origin of the vector boson mass is left unspecified, whether it results from the Stückelberg mechanism~\cite{Stueckelberg:1938hvi} or from the Brout-Englert-Higgs mechanism~\cite{Englert:1964et, Higgs:1964pj}.

To obtain the interactions of the physical vector bosons, one has to diagonalise both the kinetic mixing and mass matrix of the neutral gauge boson system. For a detailed description of the diagonalisation procedure see, e.g., Refs.~\cite{Babu:1997st,Bauer:2018onh,Alonso-Gonzalez:2025xqg}. However, in the double limit of small kinetic mixing, $\epsilon\ll1$, and light dark photon mass, $m_X^2/v^2 \ll 1$, (which we are considering in this work) we find for the relevant interactions of the mass eigenstate vector bosons 
\begin{align}
\label{eq:int_lag}
    \mathcal{L} \supset &- A^{\mu} (e\, j^\mathrm{EM}_\mu) - Z^\mu \left[ g_Z\, j^Z_\mu  - g_X\, \epsilon_Z \,j_\mu^X \right]  - A'^\mu \left[ g_X \,j_\mu^X  - e\, \epsilon_A\, j^\mathrm{EM}_\mu   \right] \,,
\end{align}
where $A_\mu$ is the SM photon, $Z_\mu$ the neutral SM $Z^0$-boson, and $A'_\mu$ the dark photon mass eigenstate. 
As pointed out in Ref.~\cite{Bauer:2022nwt}, the two mixing angles multiplying the electromagnetic and $Z^0$ current in~\cref{eq:int_lag} are in general given by
\begin{align}
    \begin{aligned}\label{eq:phys_mix}
\epsilon_A & =c_w \epsilon_B+s_w \epsilon_W\,, \\
\epsilon_Z & =-s_w \epsilon_B+c_w \epsilon_W \,,
\end{aligned}
\end{align}
with $\epsilon_B$ and $\epsilon_W$ referring to the fundamental mixing parameters of~\cref{eq:loop_lag}.

In the minimal dark photon model, we assume the dark sector to consist only of a vector-like DM fermion $\chi$ with charge $q_\chi$ under the new \Udark. In this work, however, we will also explore the freeze-in target for the minimal anomaly-free gauge groups $U(1)_{L_i-L_j}$, with $i,j=e,\mu,\tau$, and $U(1)_{B-L}$. The gauge current associated with the new $U(1)$ symmetry is extended by the corresponding charged SM fermions. Note that in order to cancel the gauge anomalies in $B-L$, we have to add three right-handed neutrino fields $\nu_R$. Since these fields are total SM singlets, they can have a Majorana mass term and a Dirac mass term with the active SM neutrinos. In the scope of this work, we assume a type-I seesaw mechanism, in which the heavy right-handed neutrinos completely decouple from the spectrum at low energies. The corresponding gauge currents are show in \cref{tab:gauge_current}.

\begin{table}[t!]
\centering
\begin{tabular}{l|l}
    Model & Dark gauge current \\ \hline
    \rule{0pt}{3ex}$U(1)_{D}$  & $j^D_\mu=q_\chi\, \bar\chi \gamma_\mu \chi$  \\ 
    \rule{0pt}{3ex}$U(1)_{L_i-L_j}$  & $j^{\, i-j}_\mu= j^D_\mu +  \bar{L}_i \gamma_\alpha L_i+\bar{\ell}_{iR} \gamma_\alpha \ell_{iR}-\bar{L}_{j} \gamma_\alpha L_j-\bar{\ell}_{jR} \gamma_\alpha \ell_{jR}$  \\
    \rule{0pt}{3ex}$U(1)_{B-L}$  & $j^{B-L}_\mu= j^D_\mu+ \frac{1}{3} \bar{Q} \gamma_\alpha Q+ \frac{1}{3}\bar{u}_{R} \gamma_\alpha u_{R} + \frac{1}{3}\bar{d}_{R} \gamma_\alpha d_{R}-\bar{L} \gamma_\alpha L - \bar{\ell}_{R} \gamma_\alpha \ell_{R} - \bar \nu_R\gamma_\mu \nu_R $  
\end{tabular}
\caption{Dark gauge currents for each model. Here, $L = (\nu_L,\ell_L)^T$ and $Q=(u_L,d_L)^T$ denote the left-handed lepton and quark doublets, while $u_R, d_R$ and $\ell_{R}$ are the corresponding right-handed fields.}
\label{tab:gauge_current}
\end{table}

In the minimal dark photon scenario, we just assume a fundamental tree-level kinetic mixing, $\epsilon_B$, and no non-abelian mixing, $\epsilon_W=0$. In this case, the two mixing parameters of~\cref{eq:phys_mix} obey the simple relation $\epsilon_Z=-\tan \theta_W\, \epsilon_A$.
However, special case of the $U(1)_{L_i-L_j}$ gauge groups both $\epsilon_B$ and $\epsilon_W$ are generated at the one-loop level. The general expressions for $\epsilon_A$ and $\epsilon_Z$ in this case are found in Eqs.~(19) and (20) of Ref.~\cite{Bauer:2022nwt}. In the low energy limit, these can be estimated as
\begin{align}\label{eq:eps_A}
    \epsilon_{A} &\approx\frac{e\, g_{X}}{6\pi^2}\ \log\left(\frac{m_i}{m_j}\right)  \,, \\
    \epsilon_{Z} &\approx\frac{g_Z\, g_{X}}{24\pi^2}\ \left[(1-4s_W^2)\,\log\left(\frac{m_i}{m_j
    }\right)\right]\,,\label{eq:eps_Z}
\end{align}
where we have neglected any neutrino mass effects.

\section{Freeze-in at low at a reheating temperature}
\label{sec:freeze-in}

In the context of these $U(1)$ scenarios, we assume that DM production in the early Universe proceeds via freeze-in from annihilations (and $Z^0$-boson decay within a specific DM mass range) of SM particles and, potentially, through dark photon self-annihilations (provided these reach the thermal equilibrium with the SM plasma)\footnote{We focus on the parameter regime $m_{A^\prime}<2m_\chi$, for which the decay $A'\to\bar\chi\chi$ is kinematically closed and therefore does not contribute to DM production. Beside this, the plasmon decay contribution becomes relevant for DM masses below the electron mass~\cite{Dvorkin:2019zdi}.}. The underlying assumption is that the gauge coupling $g_X$ or, in the case of $U(1)_D$, the product $\epsilon_D\,g_D$   is sufficiently small to guarantee that DM particles never attain thermal equilibrium with the plasma. 
The evolution of the number density of the DM particle is governed by the Boltzmann equation
\begin{align} \label{eq:BEn}
    \frac{dn_\chi}{dt} + 3 H n_\chi & =\, \left(n^{\rm eq}_\chi\right)^2\, \langle\sigma v\rangle,
\end{align}
where $n^{\rm eq}_\chi(T)$ indicates the equilibrium DM number density and $\langle\sigma v\rangle$ denotes the thermally averaged DM production cross section.

Introducing the DM yield, $Y_\chi=n_\chi/s$, with $s(T)$ being the SM entropy density,  the Boltzmann equation can be cast as
\begin{align} \label{eq:BEY}
    -\left(1+\frac{1}{3}\frac{d\ln\,g_{*s}}{d\ln\,T}\right)\frac{HT}{s}\frac{dY_\chi}{dT} &
     =\, \left(Y^{\rm eq}_\chi\right)^2\, \langle\sigma v\rangle,
\end{align}
where $H(T)$ is the Hubble expansion rate and $g_{*s}(T)$ corresponds to the number of relativistic degrees of freedom contributing to the SM entropy density. Assuming that the SM thermal bath is populated at a reheating temperature $T_{\rm rh}$, which is taken to be sufficiently close to the maximum temperature of the plasma~\cite{Chung:1998rq, Giudice:2000ex} so that a negligible amount of DM is present at $T_{\rm rh}$, the final yield is given by  
\begin{align}\label{eq:DMyield}
    Y_\chi(T_0)=&\int^{T_{\rm rh}}_{T_0} \,dT\, \frac{s}{HT}\,\left(1+\frac{1}{3}\frac{d\ln\,g_{*s}}{d\ln\,T}\right)^{-1}\,\left(Y^{\rm eq}_\chi\right)^2\, \langle\sigma v\rangle.
\end{align}
In this work, we rely on the {\tt F{\small REEZE}I{\small N}}~\cite{Bhattiprolu:2023akk, FreezeIn, Bhattiprolu:2024dmh}%
\footnote{This code incorporates the correction of the $Z$ boson coupling to the dark current 
from Ref.~\cite{Heeba:2023bik}.}
and {\tt micrOMEGAs}~\cite{Alguero:2023zol} packages to numerically compute $Y_\chi(T_0)$. 
To reproduce the entire observed DM relic abundance. $\Omega_{\rm DM} h^2 = 0.1198\pm 0.0012$~\cite{Planck:2018vyg}, the final yield must satisfy
\begin{equation}\label{eq:observed_DM}
    Y_\chi(T_0) = 4.35 \times 10^{-10}\left(\frac{{\rm GeV}}{m_\chi}\right). 
\end{equation}

Within the standard cosmological scenario, the reheating temperature is typically assumed to satisfy $T_{\rm rh}\gg m_\chi$. In this regime, the upper limit of the temperature integral in Eq.~\eqref{eq:DMyield} becomes irrelevant, implying that the final DM relic abundance is effectively insensitive to the initial conditions, in particular to the precise value of $T_{\rm rh}$. This regime is referred as infrared freeze-in~\cite{Hall:2009bx}, since the dominant contribution to the DM production takes place at temperatures of the order of the DM mass.

In contrast, if the reheating temperature satisfies $T_{\rm rh} < m_\chi$, DM production becomes kinematically suppressed. In this case, only SM particles in the exponentially suppressed tail of their thermal distributions possess sufficient energy to produce DM pairs. Consequently, the production rate is Boltzmann suppressed, leading to a significantly reduced DM yield. To compensate for this suppression and reproduce the observed relic abundance, larger values of the relevant couplings are generally required compared to those in the high-$T_{\rm rh}$ regime. 
As a byproduct, this shifts the viable parameter space toward stronger interactions, thereby altering the detection prospects of DM, as discussed in Sec.~\ref{sec:results}.

Before proceeding, let us comment on the compatibility of a low-reheating scenario with the generation of the baryon asymmetry. On the one hand, reheating temperatures well below the electroweak scale may disfavour conventional high-temperature mechanisms, such as standard electroweak baryogenesis
or thermal leptogenesis. In this sense, a future signal pointing to freeze-in production at a low $T_{\rm rh}$ could put such simple scenarios under tension~\cite{Belanger:2018sti}. On the other hand, several mechanisms capable of operating at low reheating temperatures have been discussed in the
literature~\cite{Davidson:2000dw, Giudice:2000ex, Allahverdi:2010im, Asaka:2019ocw, Nelson:2019fln, Elahi:2021jia, Jaeckel:2022osh}, as well as during an early matter-dominated phase~\cite{Bernal:2017zvx, Chen:2019etb, Bernal:2022pue, Chakraborty:2022gob}. 
A complete analysis of baryogenesis and DM production in a low-reheating or early matter-dominated cosmology is highly model dependent and lies beyond the scope of the present work. Consequently, the results presented here should be interpreted as benchmark freeze-in solutions in low-$T_{\rm rh}$ scenarios, under both the cosmological and particle physics assumptions specified above.

In the $U(1)_D$ case, the condition that the chemical equilibrium between the dark and visible sectors not be attained can be cast in terms of $\kappa\equiv \epsilon_D\,  g_D/e$ as $\kappa\lesssim1.2\times10^{-8}\left({m_\chi}/{{\rm MeV}}\right)^{1/2}$ \cite{Hambye:2019dwd}, while the condition $\epsilon_D\lesssim 1.2\times 10^{-9}\left({m_\chi}/{{\rm MeV}}\right)^{1/2}$ avoids the thermalisation of dark photons with the SM plasma~\cite{Hambye:2019dwd}. In this regime, DM particles are produced exclusively by SM particles, and therefore the relic abundance increases as $\kappa^2$. 
As a useful approximation, when the mediator mass is negligible compared to the DM mass, the contribution to the total yield for a $f\bar f\to\chi\chi$ channel, $Y_f$, can be found to be proportional to
\begin{equation}
    Y_f\sim\frac{M_P}{T^*} \frac{\kappa^2\,Q_f^2}{g_*^{3/2}(T^*)}\,,
    \label{eq:yield-approx}
\end{equation}
where  
$T^*={\rm max}(m_f,\,m_\chi)$ is the optimal temperature at which most of the DM production takes place,
$g_*(T^*)$ is the number of relativistic degrees of freedom at that temperature,
and $Q_f$ is the fermion electric charge.

\begin{figure}[t!]
    \centering
    \includegraphics[width=0.48\textwidth]{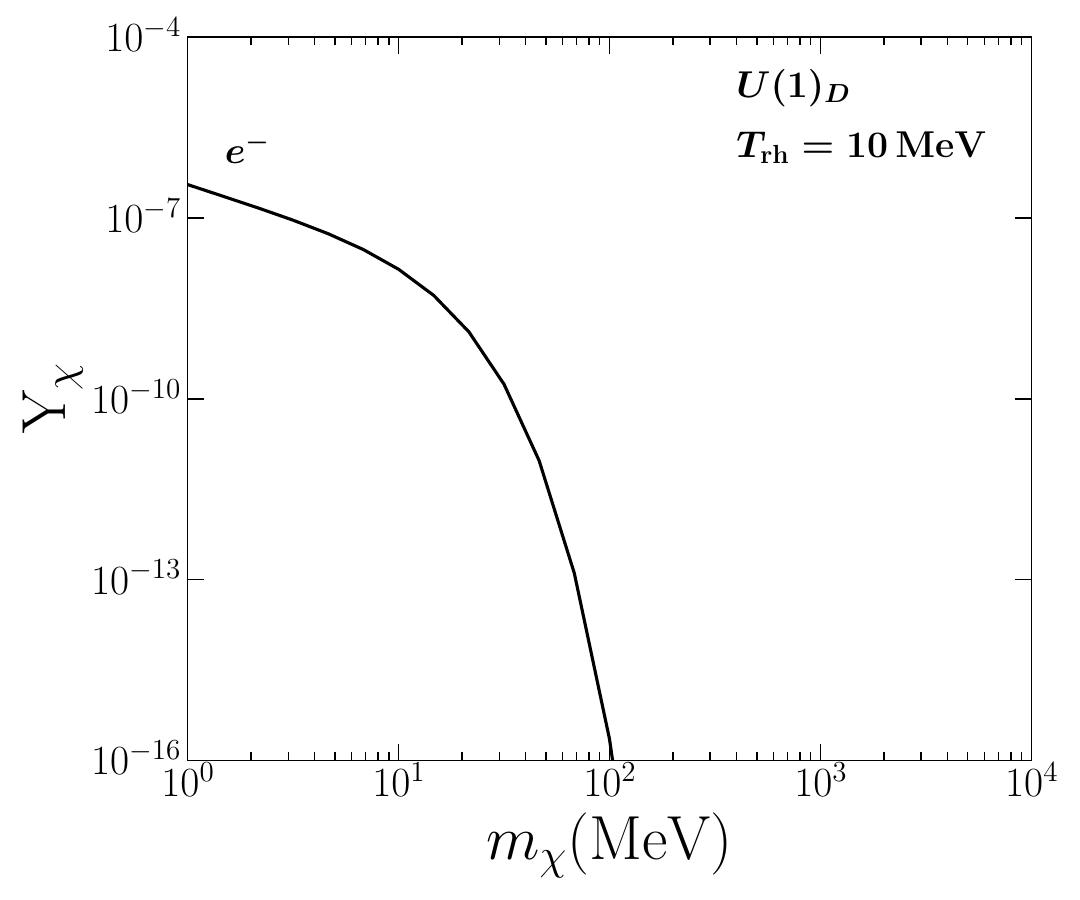}%
    \includegraphics[width=0.48\textwidth]{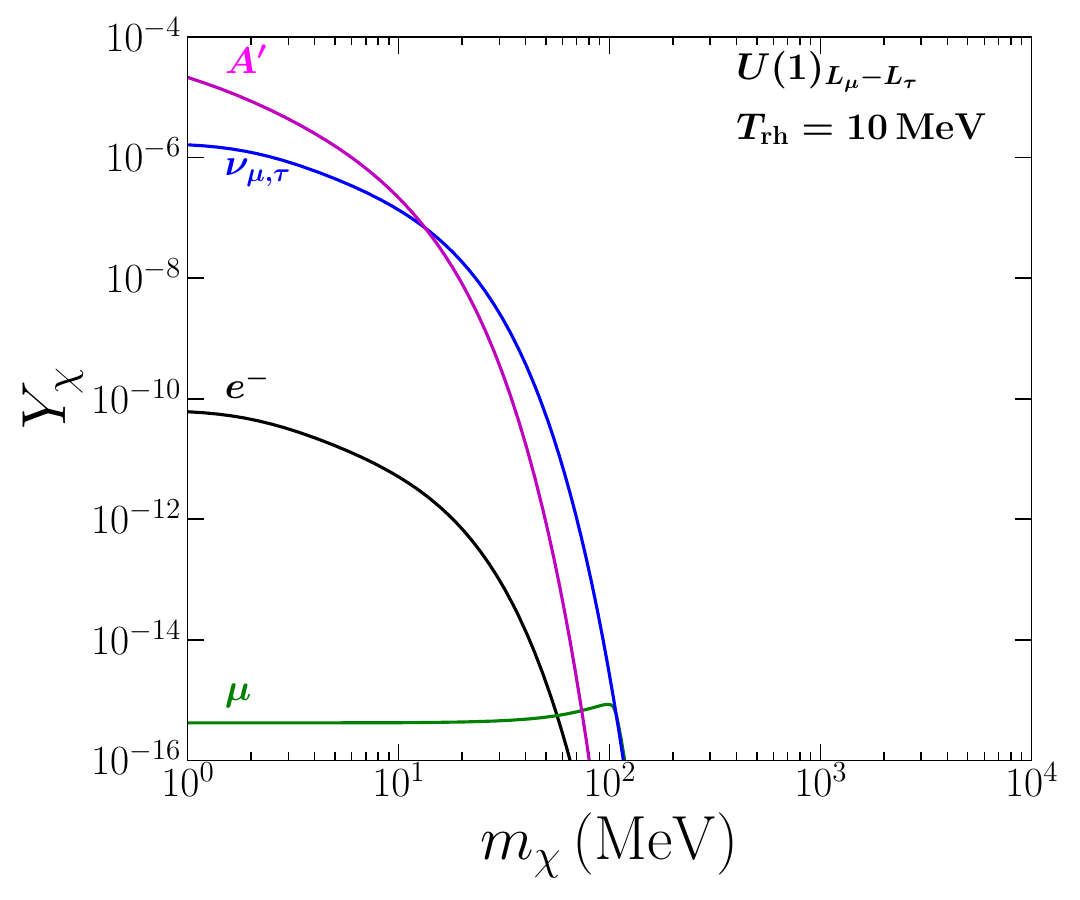}\\
    \includegraphics[width=0.48\textwidth]{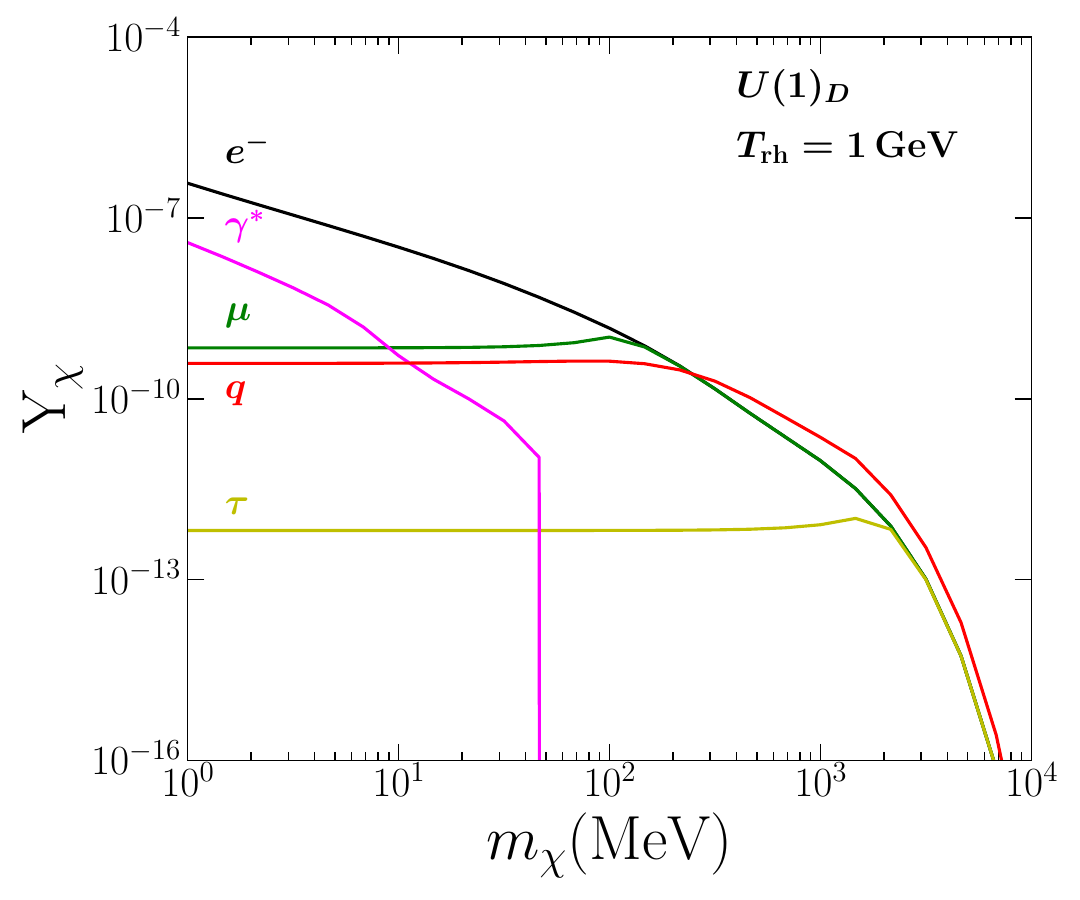}%
    \includegraphics[width=0.48\textwidth]{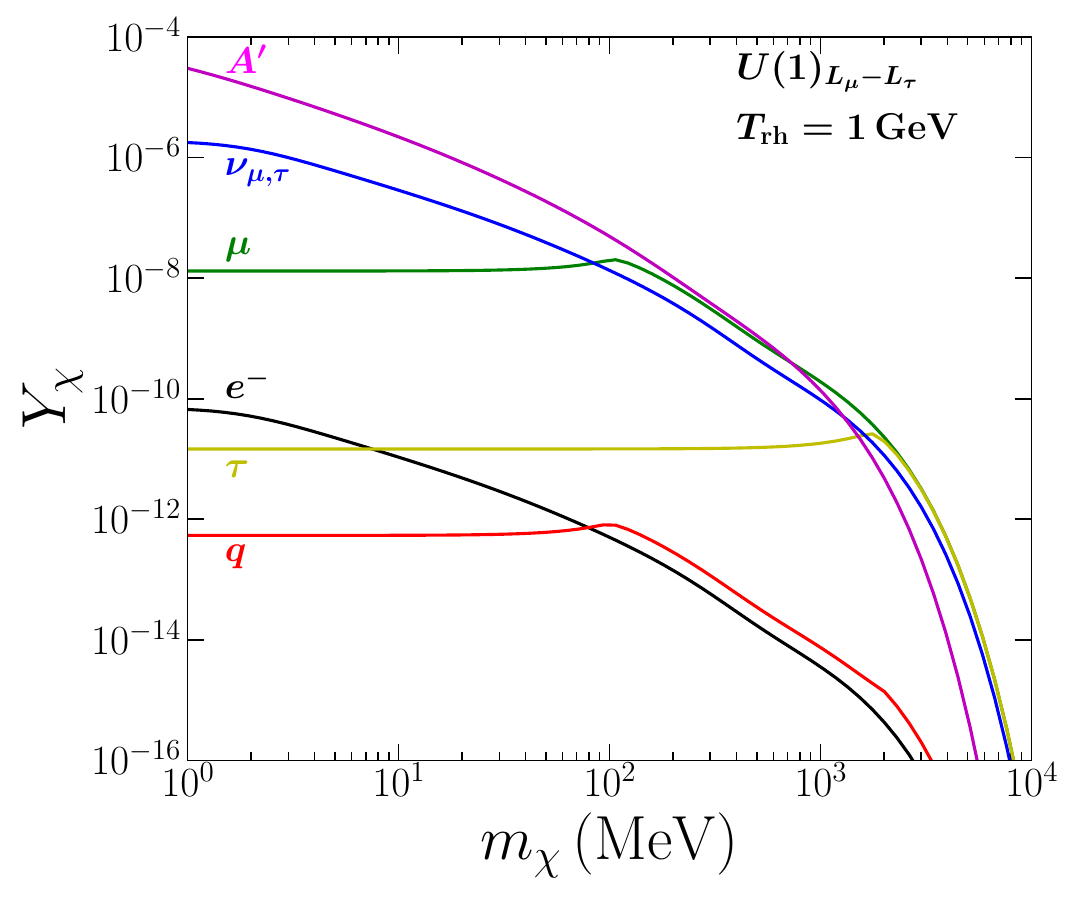} \\
    \includegraphics[width=0.48\textwidth]{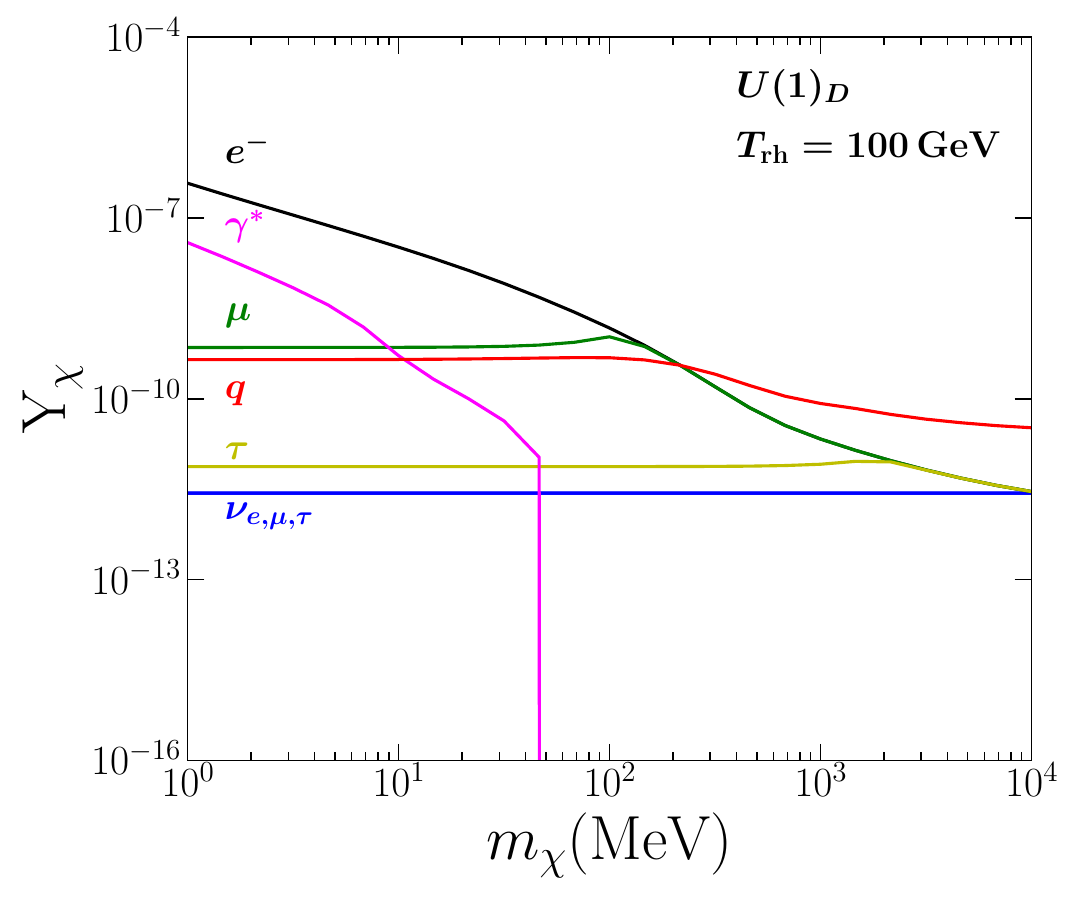}%
    \includegraphics[width=0.48\textwidth]{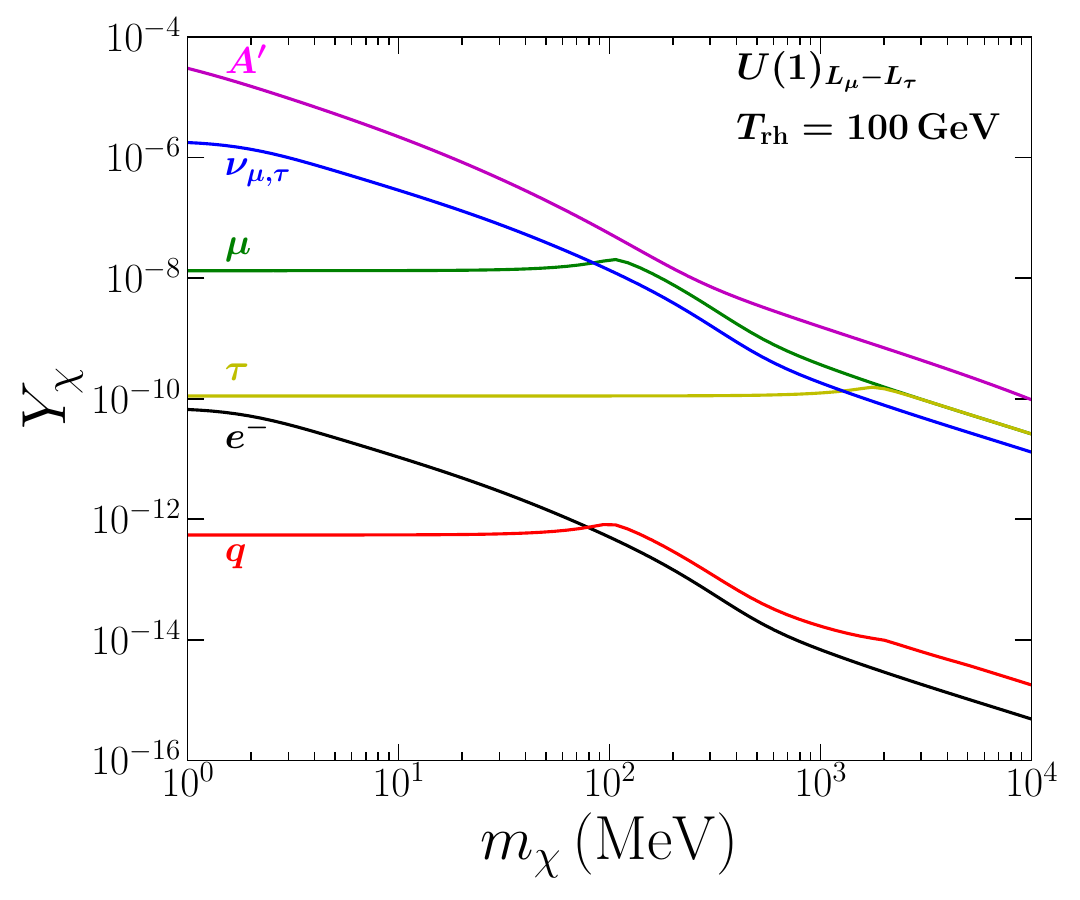}
    \caption{
    Contribution to the DM yield as function of the DM mass in the minimal \Udark (left) and $L_\mu-L_\tau$ (right) models for several  self-annihilation channel: quarks (red), SM plasmon/$A^\prime$ (magenta), $\nu_{\mu,\tau}$ (blue), $\mu$ (green), $\tau$ (yellow), electron (black). Top, middle and bottom panels correspond to a reheating temperature of 10 MeV, 1 GeV and 100 GeV, respectively. In the left (right) panel $\kappa=2\times10^{-11}$ ($g_{\mu\tau}^2/e=2\times10^{-11}$) has been fixed and in both panels $m_{A^\prime}=1.5\,m_\chi$.
    }
    \label{fig:yields-D-Lmu-Ltau}
\end{figure}

The left panels in \cref{fig:yields-D-Lmu-Ltau} show the contribution to the DM yield of each individual channel in \cref{eq:yield-approx} for a massive dark photon with $m_{A^\prime}=1.5\,m_\chi$. The bottom panel corresponds to $T_{\rm rh}=100$ GeV, while the middle and top panels correspond to $T_{\rm rh}=1$ GeV and 10 MeV, respectively.   Comparing the yields for the different \Trh, we can see the impact of Boltzmann suppression when the reheating temperature becomes lower than the DM mass. Two regimes arise depending on whether the DM is heavier or lighter than the annihilating particles. If the DM is heavier, the yield decreases with increasing DM mass, $Y_{i}\sim m_{\chi}^{-1}$. However, if the DM is lighter, the DM yield is inversely proportional to the mass of the annihilating SM fermion, $Y_{i}\sim m_{f}^{-1}$. This explains the behaviour of the yields from charged leptons, $e$, $\mu$ and $\tau$. Since they have the same charges under \Udark, the matrix elements in $\langle\sigma v\rangle$ are the same. For $m_\chi>m_\tau$, the contributions to the yield are equal and decrease as the DM increases. However, for DM masses smaller than the mass of each leptons, their contributions stabilise and are proportional to $m_f^{-1}$, explaining why $e^{-}e^{+}\rightarrow\bar\chi\chi$ dominates over muons and taus.\footnote{Quantum statistical and in-medium effects could in principle modify the freeze-in production rate~\cite{Heeba:2019jho, Bringmann:2021sth}, but we neglect them here. For instance, in a thermal plasma charged leptons acquire an effective mass correction $\Delta m_\ell^2 \simeq e^2 T^2/8$. For the reheating temperatures considered in this work, the resulting effective masses satisfy $m_{\ell,\rm eff}/T \lesssim 0.1$, so leptons remain relativistic and their equilibrium abundances—and hence the reaction densities—change only at the percent level. At the lowest reheating temperatures the muon and tau channels are already Boltzmann suppressed due to their vacuum masses.} In contrast, neutrinos are not charged under \Udark and they only couple to the DM via weak interactions mediated by the SM $Z^0$ boson.

On the other hand, since in the $L_i-L_j$ ($B-L$) models two generations of leptons (all the fermions) transform non-trivially under the corresponding symmetry, some important differences with respect to the dark photon case arise. For concreteness, we will focus on the \Umt model as an example, but the conclusions can be easily extrapolated to the other constructions. The right panels in Fig.~\ref{fig:yields-D-Lmu-Ltau} display the contributions to the DM yield as a function of the DM mass in this model for the same reheating temperatures as for the \Udark case.

First, the \Umt gauge boson has tree-level interactions with the second and third lepton generations. As a result, it thermalises with the SM plasma unless either extremely low values of the gauge coupling are considered ($g_{\mu\tau}\lesssim 4\times 10^{-9}$) or the mediator is sufficiently light ($m_{A^\prime}\lesssim 10$ MeV)~\cite{Escudero:2019gzq}. Therefore, in the mass regime $m_{A^\prime}<2m_\chi$, the annihilation process $A^\prime A^\prime\to\bar\chi\chi$ becomes the dominant channel for DM production via the freeze in mechanism. Second, the $\nu_\mu, \nu_\tau$ self-annihilation into DM provides a substantial contribution to the DM yield. Given that neutrinos are effectively massless at the relevant temperatures, these channels are indeed very efficient. This same qualitative behaviour occurs in the rest of the models analysed in this work. Finally, the contribution from $\bar q q\rightarrow\bar\chi\chi$ and $e^{-}e^{+}\rightarrow\bar\chi\chi$ are suppressed by a factor of $1/70^{2}$ relative to case of minimal dark photon.  This suppression arises because the coupling to quarks and electrons is generated at loop level and is therefore reduced by a factor of $1/70$ with respect to the tree-level coupling (see \cref{eq:eps_A}).

\section{Direct detection}
\label{sec:direct_detection}

Direct detection experiments can look for DM collisions on either nuclei or electrons. For several decades, nuclear recoils were considered the canonical signature, since the expected background can be very small if the experiment can discriminate between electron and nuclear recoils. Because of the experimental energy threshold and the nuclear mass, nuclear recoils allow to search for DM masses larger than $\sim 1$~GeV in liquid noble gas TPCs and $\sim 10$~MeV in solid-state detectors. In fact, there are ongoing attempts to lower the experimental thresholds to gain sensitivity to lighter DM particles. 
The DM nucleus elastic scattering cross section in the limit $m_{A^\prime}\ll m_{Z^0}$ reads~\cite{Evans:2017kti}
\begin{equation}\label{eq:sig_nuc}
    \sigma_{\chi n} =
    \frac{\mu_{\chi n}^2}{\pi}\,\frac{g_X^2\, f_{nX}^2}{m_{A^\prime}^4},
\end{equation}
with $\mu_{\chi n}=m_\chi m_n/(m_\chi+m_n)$  the nucleon-DM reduced mass, 
\begin{equation}
    f_{nX} = \frac{1}{A} \left[ Z\,(2 g_{uX} + g_{dX}) + (A - Z)\,(g_{uX} + 2 g_{dX}) \right],
\end{equation}
where $g_{uX}$ and $g_{dX}$ are the effective couplings of the $X$ boson to up and down quarks. They are given by
\begin{align}
    g_{qX}&=\,\begin{cases}
    e\,\epsilon_D\,Q_q,\,\,\,{\rm for}\,\,\, X=D;\\
    e\,\epsilon_A\,Q_q,\,\,\,{\rm for}\,\,\, X=L_i-L_j;\\
    g_{BL}\,B_q, \,\,\,{\rm for}\,\,\,  X=B-L,
    \end{cases}
\end{align}
where $\epsilon_A$ is given in Eq.~\eqref{eq:eps_A}.

More recently, experiments have also started to exploit the possibility that DM scatters off electrons. Since the electron is a much lighter target, it allows to probe DM particles with masses in the range of the MeV. The DM electron elastic scattering cross section reads~\cite{Essig:2015cda}
\begin{align}\label{eq:sig_e}
    \sigma_{\chi e}&=\,\frac{\mu_{\chi e}^2}{\pi}\frac{g_X^2\,g_{eX}^2}{\left(\alpha_{\rm EM}^2m_e^2+m^2_{A'}\right)^2}\,,
\end{align}
where $\mu_{\chi e}=m_\chi m_e/(m_\chi+m_e)$  is the electron-DM reduced mass and $g_{eX}$ is the coupling between the $X$ boson and the electron. This coupling takes a concrete form depending on the $X$ boson: 
\begin{align}
    g_{eX}&=\,\begin{cases}
    e\,\epsilon_D,\,\,\,{\rm for}\,\,\, U(1)_D;\\
    -g_{\mu e}, \,\,\,{\rm for}\,\,\,  U(1)_{L_\mu-L_e};\\
    g_{e\tau}, \,\,\,{\rm for}\,\,\,  U(1)_{L_e-L_\tau};\\
    e\,\epsilon_A, \,\,\,{\rm for}\,\,\, U(1)_{L_\mu-L_\tau};\\
    -g_{BL}, \,\,\,{\rm for}\,\,\,  U(1)_{B-L}.
    \end{cases}
\end{align}

If we consider the possibility that DM is multi-component, direct detection experiments would only be able to set constraints on $\xi \sigma_{\chi e}$ or $\xi \sigma_{\chi n}$, respectively, where $\rho_{\chi}=\xi\,\rho_{\rm CDM}$ is the local density of $\chi$, expressed as a fraction of the total DM local density, $\rho_{\rm CDM}=0.3$~GeV/cm$^3$~\cite{Weber:2009pt,Read:2014qva}. For consistency, we assume that the amount by which $\chi$ is underabundant in the Milky Way halo coincides with the amount by which it is underabundant in the universe, so that $\xi=\Omega_{\chi}/\Omega_{\rm DM}$.

In particular, in the case of the dark photon, the relic abundance is proportional to $\kappa^2$, and therefore the same holds for $\xi$. This means that $\xi\sigma_{\chi e}\propto \kappa^2(g_Xg_{eX})^2\propto\kappa^4$. 
Consequently, direct detection experiments quickly lose sensitivity when the DM is under-abundant. This is in stark contrast with how $\xi\sigma_{\chi e}$ scales in freeze-out scenarios, for which the relic density typically scales as the inverse of the coupling squared and the decrease in $\xi$ is compensated by a similar increase in $\sigma_{\chi e}$.

\subsection{New physics discovery limits}
\label{sec:nu-floor}

It is crucial to notice that in these scenarios, direct detection experiments are sensitive to two simultaneous signatures of new physics: the observation of a signal compatible with a DM particle or the detection of solar neutrinos with a flux that is inconsistent with the SM prediction. Both of these signatures can appear in nuclear and electron recoils. In order to classify the regions of the parameter space in which we would have such hints of new physics, in this article we have considered the following two conditions for each point in the parameter space, $\boldsymbol\theta$.
\begin{itemize}
\item[\textbf{(i)}] \textbf{Dark Matter discovery:}

If the DM spectrum is statistically separable from that of solar neutrinos (considering SM plus new physics contributions), then DM can be discovered for this parameter point. 
If the scattering cross section of the DM with nuclei or electrons is small enough, then the signal from solar neutrinos could surpass and conceal the DM events. Notice that this defines a \textit{new neutrino floor}, which can potentially differ from the SM one if the new neutrino interactions are important \cite{Boehm:2018sux,Sadhukhan:2020etu,DeRomeri:2025nkx}. 

\item[\textbf{(ii)}] \textbf{Neutrino discovery:}

If the enhanced neutrino spectrum due to the additional $A'$ interactions is statistically separable from that of the SM-only neutrinos, then new neutrino physics can be discovered for this parameter point. 
This defines a new neutrino physics discovery limit for each specific model. Note that this discovery limit is completely independent of the DM sector of these models and provides  a minimal BSM sensitivity of direct detection experiments to these $U(1)$ models.
\end{itemize}

Notice also that, in principle, we would have to apply the criteria above to both electron and nuclear recoils. As we will see in the following sections, in this article we will mainly be interested in nuclear recoils since the regions of the parameter space that would be accessible to electron recoils are extremely constrained by other experimental searches. Similarly, in these models the very relevant question arises of whether one can determine if a new physics signal in direct detection experiments is due to DM or neutrinos. We defer a more thorough analysis of this to future work.

In order to derive the two  discovery limits discussed above, we perform a set of hypothesis test via a maximum likelihood analysis. For this we perform two different tests; one, where we consider only the DM events as the signal and test them against the full neutrino background (i.e.~events induced both via SM interactions and the new $U(1)$ mediator) similar to the neutrino floor as derived in Ref.~\cite{Billard:2013qya}. And a second one, where we consider the signal to be composed of only the BSM neutrino  events above the SM expectation. For computing the expected number of solar neutrino events at the detector in all these models, we use the \snudd package~\cite{snudd2023,Amaral:2023tbs}.

We build our binned likelihood function for the model parameters $\boldsymbol{\theta}=(g_X, m_{A'})$ based on a Poisson probability function for the total number of events, and signal and background probability distribution functions as~\cite{Cowan:2010js},
\begin{align}\label{eq:likelihood}
    \mathscr{L}(\mu, \boldsymbol{\theta}, \boldsymbol{\phi}) = & \frac{e^{-\left(\mu\, s(\boldsymbol{\theta})+\sum_{j=1}^{n_\phi} b_j\right)}}{n!} \ \prod_{i=1}^N \left[\mu\, s(\boldsymbol{\theta})\,f(E_{R_i}|s(\boldsymbol{\theta})) + \sum_{j=1}^{n_\phi} b_j\,f(E_{R_i}|b_j)\right]^{n_i} \ \prod_{j=1}^{n_\phi} \  \mathscr{L}(\phi_j)\,,
\end{align}
where $n=\sum_i n_i$ are the total number of observed events, $N$ is the number of bins,  $s(\boldsymbol{\theta})$ and $f(E_{R_i}|s(\boldsymbol{\theta}))$ are the total number of signal events and unit-normalized signal probability distribution function (PDF) of the observed recoil energy $E_{R_i}$ for the model parameters $\boldsymbol{\theta}$. Similarly, $b_j$ and $f(E_{R_i}|b_j)$ are the total number of background (neutrino) events of flux component $j$ and background PDF of the recoil energy, $n_\phi$ is the number of neutrino flux components. The parameter $\mu$ denotes the signal strength. We denote the likelihoods for the neutrino flux normalizations as $\mathscr{L}(\phi_j)$, which we treat as nuisance parameters and model as Gaussian distributions with a standard deviations given by the different flux normalization uncertainties~\cite{Grevesse:1998bj,Asplund:2009fu,Vinyoles:2016djt,Evans:2016obt}.

In order to test whether the BSM signal exists in the data, we want to test it against the background-only hypothesis (ie.~the SM) and reject it. According to the Neyman-Pearson lemma, the test with the maximum discrimination power is then given by the log likelihood ratio test~\cite{Cowan:1998ji},
\begin{equation}
    -2 \ln \lambda(\boldsymbol{\theta}) = -2 \ln \frac{\mathscr{L}(1,\boldsymbol{\theta},  \boldsymbol{\hat\phi})}{\mathscr{L}(0,\boldsymbol{\theta}, \boldsymbol{\hat{\hat{\phi}}})}\,,
\end{equation}
where the hats symbol that we have profiled over the flux uncertainties to maximise the likelihoods.
Since under random fluctuations in the data $n_i$ the function $-2 \ln \lambda(\boldsymbol{\theta})$ asymptotically follows a $\chi^2$ distribution with the number of degrees of freedom $n$ equal to the number of model parameters $\boldsymbol{\theta}$, we can reject parameters $\boldsymbol{\theta}$ at the level of $1-p$ via~\cite{Kahlhoefer:2017ddj}, 
\begin{equation}
    1 - F_{\chi^2}(n; -2\ln\lambda(\boldsymbol{\theta})) < p\,,    
\end{equation}
where $F_{\chi^2}$ denotes the cumulative $\chi^2$ distribution with $n$ degrees of freedom. Since for a given mass $m_{A'}$ we are varying the gauge coupling $g_X$ of the gauge boson, we can reject the null hypothesis at the $2\, \sigma$ level via $-2 \ln \lambda(\boldsymbol{\theta}) > 4$ (for 1 degree of freedom).

For concreteness, in the remainder of this article, we will determine the DM and BSM discovery limits for a Si target with a benchmark exposure of 1~tonne-year and a minimum achievable energy threshold of 1~eV.

\section{Results}
\label{sec:results}

In this section, we discuss the results of our analysis of freeze-in production of a vector-like fermion, $\chi$, within the secluded \Udark dark photon model in~\cref{sec:dp_res}, as well as within the minimal anomaly-free gauge extensions \Uij and \UBL under which SM fields are also charged in~\cref{sec:gauge_res}.

\subsection{Dark Photon}
\label{sec:dp_res}

In the minimal kinetic-mixing dark photon case, the mass of the mediator is a free parameter. In the context of DM direct detection it is customary to consider two limiting cases of either a  light or heavy mediator, comparing its mass, $m_{A'}$, to the momentum exchange of the relevant processes. This results in two very different phenomenological regimes with profound implications for direct detection.

\begin{figure}[!t]
    \includegraphics[width=0.5\textwidth]{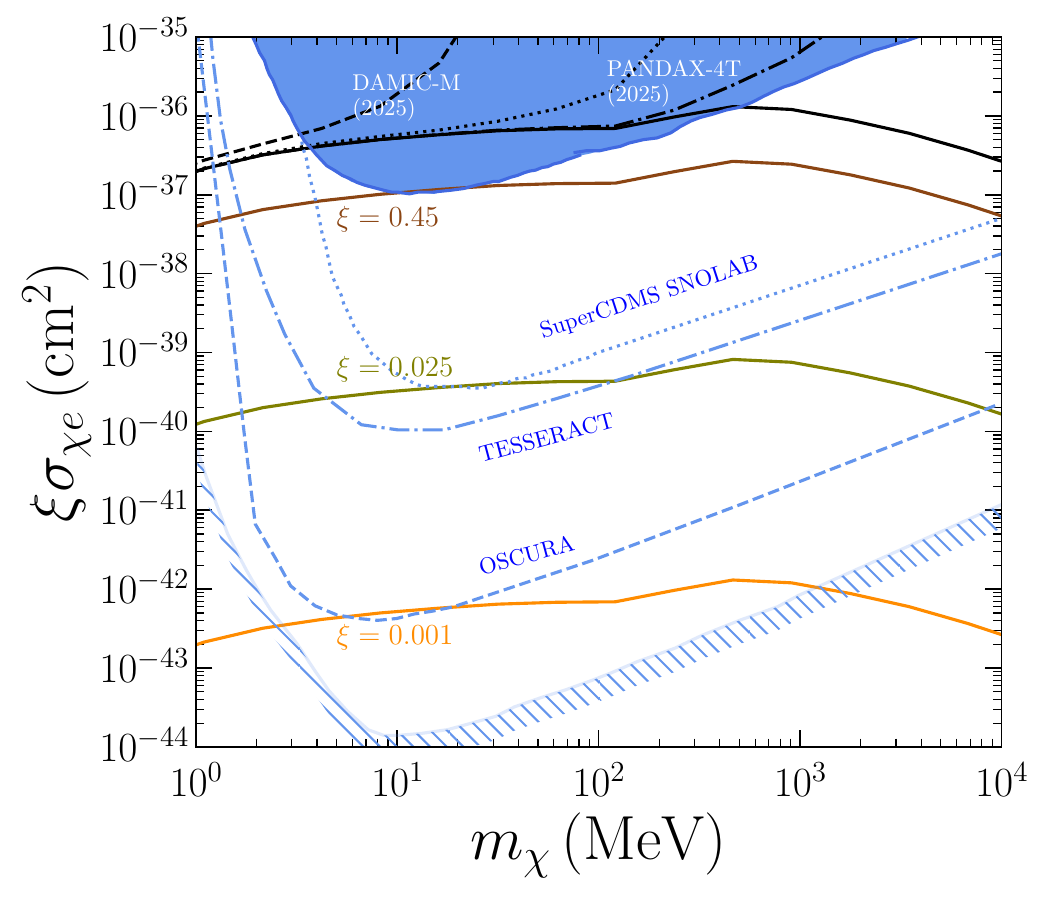}
    \includegraphics[width=0.5\textwidth]{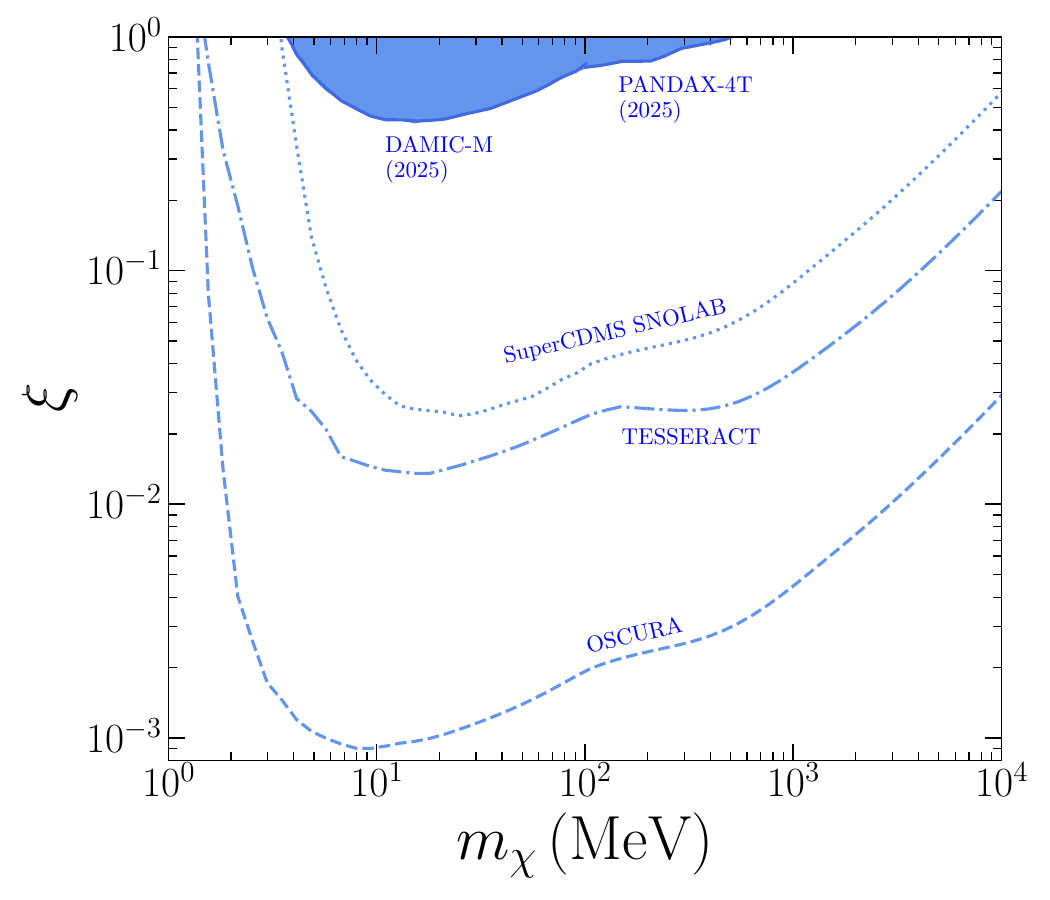}
    \caption{Constraints from electron recoil data of direct detection experiments on freeze-in DM in the ultra-light dark photon scenario.
    \textbf{Left:} Constraints on $\xi \sigma_{\chi e}$  as a function of the DM mass, $m_\chi$. The black solid line represents the freeze-in solution in a standard cosmology with high reheating temperature \Trh where $\chi$ accounts for all of the DM, $\Omega_\chi h^2=0.12$, whereas the 
    brown, green and orange
    lines correspond to an under-abundance of 
    $\xi=0.45$, $0.025$, and $0.001$, 
    respectively. The black dashed, dotted and dot-dashed lines are the freeze-in solutions where $\chi$ constitutes all the DM in non-standard cosmologies with low reheating temperatures \Trh of $10^{-2}$~GeV, $10^{-1}$~GeV, and $1$~GeV, respectively.
    The blue shaded areas correspond to the regions excluded by the DAMIC-M \cite{DAMIC-M:2025luv} and PandaX-4T \cite{PandaX:2025rrz} electron recoil data. The blue dotted, dot-dashed and dashed lines represent the future sensitivity of SuperCDMS, TESSERACT and OSCURA \cite{SuperCDMS:2022kse,2020Snowmass2021LetterOI,Oscura:2023qik},  respectively. 
    The region below the blue hatched line corresponds to the neutrino floor for electron recoils in a Si target~\cite{Carew:2023qrj}.  \textbf{Right:} Same experimental constraints and future sensitivities as on the left, but shown on the under-abundance fraction, $\xi$, as a function of the DM mass, $m_\chi$.}
    \label{fig:DP-directdetection}
\end{figure}

\subsubsection{Ultralight dark photons}
\label{sec:ultralight_dp}

We begin our discussion of freeze-in in the minimal secluded dark photon model with the limiting case of an ultra-light mediator scenario, $m_{A'}\ll m_{\chi}$. Since both the freeze-in abundance, $\Omega h^2_\chi$, and the electron scattering cross section times relative DM abundance, $\xi\sigma_{\chi e}$, have the same scaling with the coupling combination $\kappa=\epsilon_D\,  g_D/e$, we show our results for the freeze-in abundance of $\chi$ directly in the $\xi\sigma_{\chi e}$ versus $m_\chi$ plane.

In the left panel of \cref{fig:DP-directdetection}, the solid black line corresponds to the values of $\xi\sigma_{\chi e}$, for which freeze-in production of $\chi$ reproduces the total cold DM relic abundance in a standard cosmological scenario with high reheating temperature. As it has already been pointed out in the literature~\cite{Cheek:2025nul}, current experimental constraints from DAMIC-M \cite{DAMIC-M:2025luv} and PandaX-4T \cite{PandaX:2025rrz} (blue regions) exclude DM masses in the range 4 MeV $\lesssim m_\chi \lesssim$ 500 MeV. The predicted values of $\xi\sigma_{\chi e}$ for non-standard cosmologies with low reheating temperatures, $T_{rh}=10^{-2}$~GeV, $10^{-1}$~GeV, and $1$~GeV, are shown as the black dashed, dotted and dot-dashed lines, respectively. In these scenarios, all the parameter space with $m_\chi >4$ MeV is excluded. For reference, we also display the projected sensitivities for some selected future experiments, such as SuperCDMS (blue dotted) \cite{SuperCDMS:2022kse}, Tesseract (blue dot-dashed) \cite{2020Snowmass2021LetterOI} and Oscura (blues dashed) \cite{Oscura:2023qik}, which will significantly improve over the current bounds.

Note that due to the scaling of the relic abundance and scattering cross section under freeze-in, $\xi\sigma_{\chi e}\propto g_{X}^{2}\,g_{eX}^2\propto\kappa^2$, it is possible to avoid current experimental limits by slightly reducing the DM relic abundance. For example, in the left panel of \cref{fig:DP-directdetection}, the 
brown, green and orange
lines represent the solutions with an under-abundant $\chi$ with 
$\xi=0.45$, $0.025$, and $0.001$,
respectively. For reference, we also show in this plot the limitation that the neutrino fog for electron recoils in an Si target sets on this model~\cite{Carew:2023qrj}.

Given this scaling of $\xi\sigma_{\chi e}$ with the relic abundance, an interesting way to represent the sensitivity of direct detection experiments on the DM abundance in these models is 
shown in the right panel of \cref{fig:DP-directdetection}. Here, we can see how both DAMIC-M and PandaX-4T can only constrain regions where the abundance of $\chi$ is larger than 40\% of the total cold DM density. Future experiments will allow to test much smaller fractions: SuperCDMS SNOLAB will be able to probe this scenario down to $\xi=0.025$, while TESSERACT and OSCURA will reach $\xi\sim0.02$ and $\xi=0.001$, respectively.

\subsubsection{Massive dark photons}
\label{sec:mev_dp}

Let us now consider heavier mediators, with masses in the MeV - GeV range, comparable to the DM mass, $m_{A'}\sim m_\chi$. For freeze-in production of DM in the minimal dark photon model, it is usually enough to discuss the phenomenology in terms of the coupling combination $\kappa=\epsilon_D\,  g_D/e$, \cite{Hambye:2018dpi,Boddy:2024vgt}. However, to accurately assess the impact of the bounds on the (massive) dark photon mediator from accelerator experiments, astrophysics and cosmology on the relevant DM parameter space, we must specify the relative magnitudes of $\epsilon_D$ and $g_D$~\cite{Alenezi:2025kwl}. We have chosen to represent our results in terms of the kinetic mixing $\epsilon_D$ for a fixed value of $g_{D}=0.1$, which is safe from self-interaction bounds~\cite{Feng:2009hw}. In this case, freeze-in production of $\chi$ reproduces the observed DM relic abundance  for a value of $\epsilon_D\sim 10^{-10}$ for a standard cosmology with high reheating temperature~\cite{Caputo:2021eaa}.

\begin{figure}[t]
    \centering
    \includegraphics[width=0.5\linewidth]{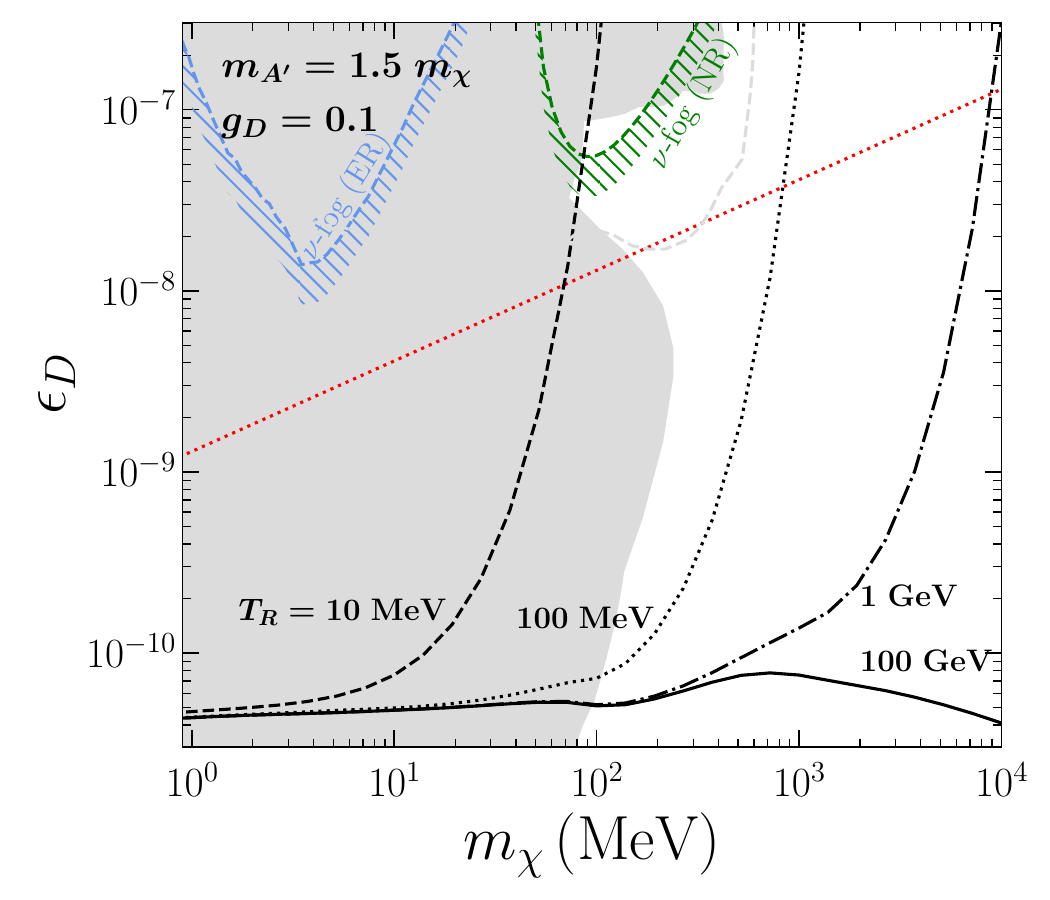}%
    \includegraphics[width=0.5\linewidth]{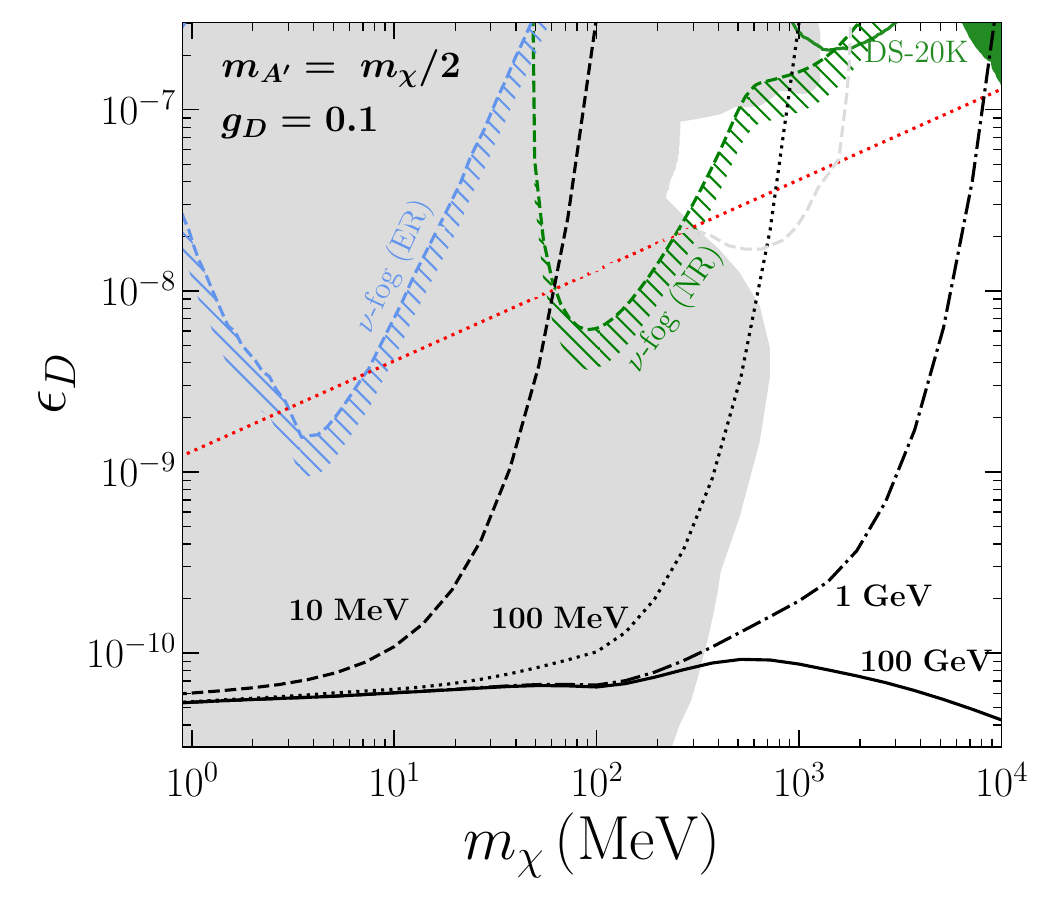}
    \includegraphics[width=0.5\linewidth]{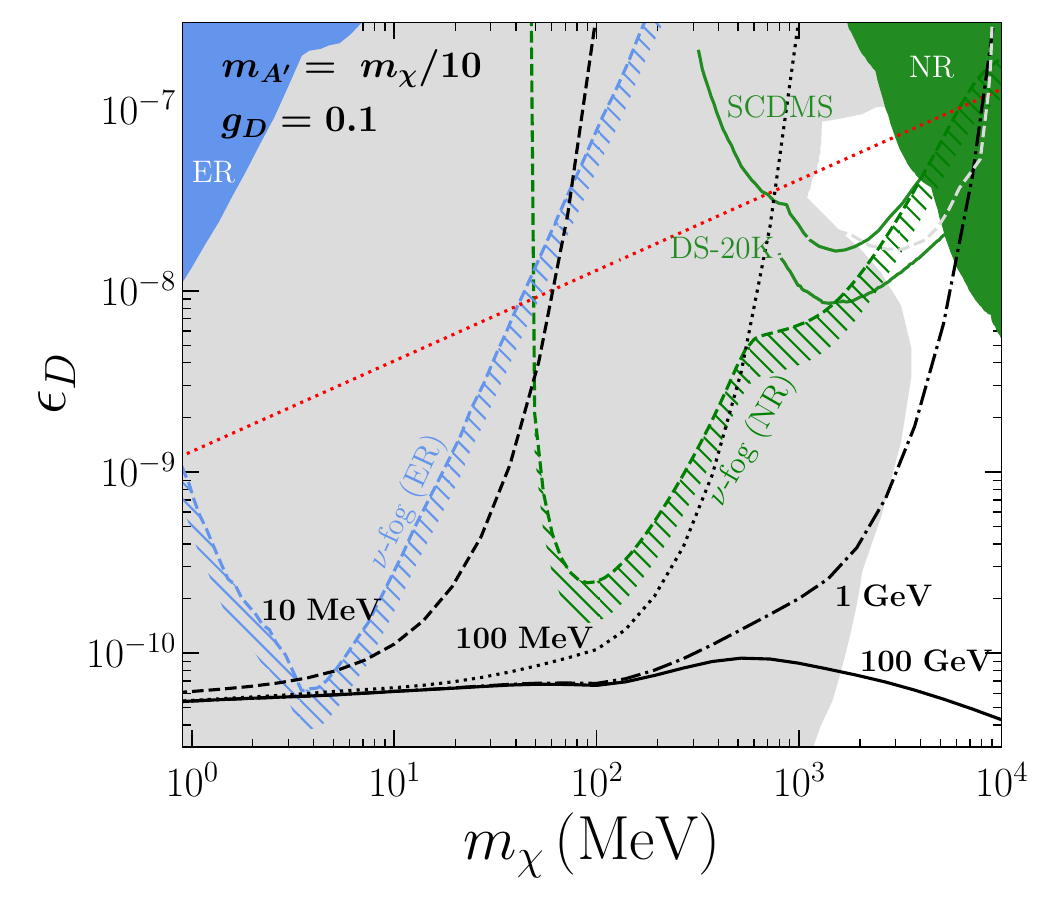}%
    \includegraphics[width=0.5\linewidth]{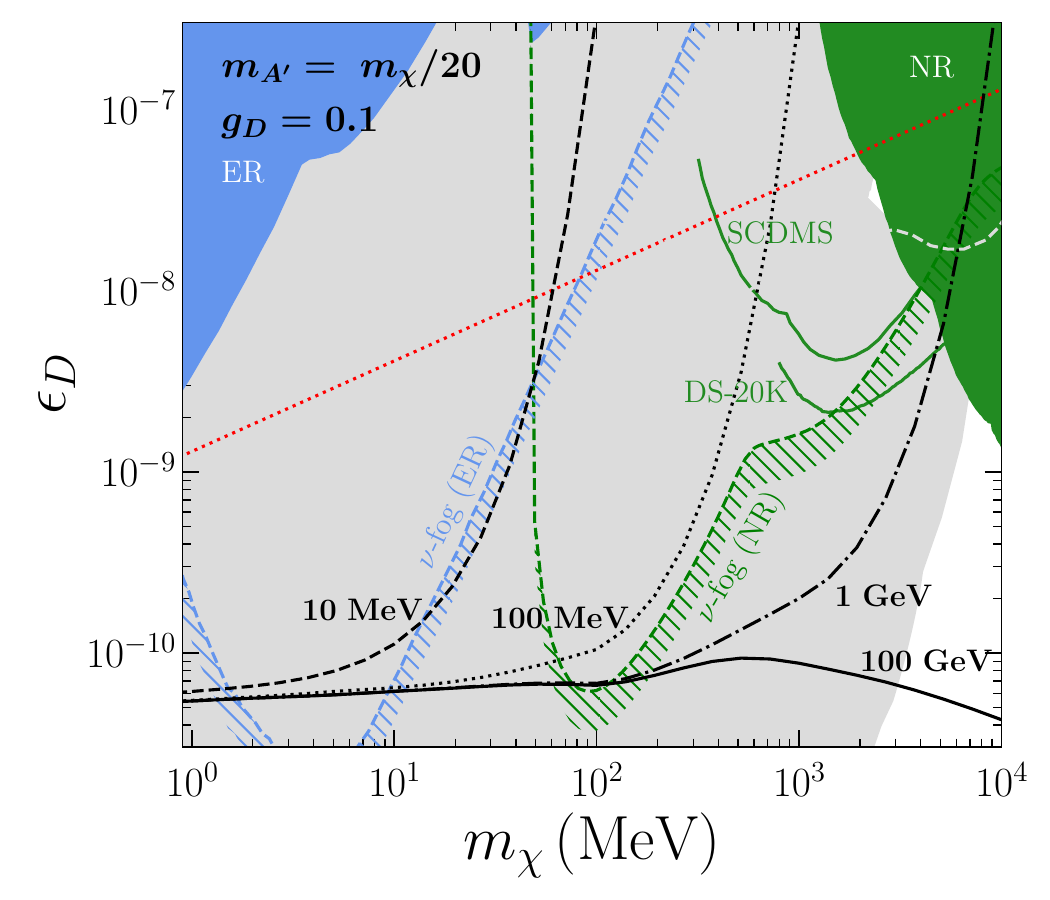}
    \caption{Black lines correspond to the regions in the $(m_{\chi},\epsilon_D)$ plane where the correct relic density is obtained via freeze-in for a reheating temperature of $T_{\mathrm{rh}}=10^2$~GeV (solid), $1$~GeV (dot-dashed), $10^{-1}$~GeV (dotted), and $10^{-2}$~GeV (dashed). We have chosen 
    a benchmark value of $g_D=0.1$ and four different values for the mediator and DM mass ratio, $m_{A'}/m_\chi=1.5$, $1/2$, $1/10$, and $1/20$. 
    The shaded blue areas are constrained via electron recoils in DAMIC-M \cite{DAMIC-M:2025luv}, DarkSide-50 \cite{DarkSide:2022knj}, PandaX-4T \cite{PandaX:2025rrz} and XENON1T from solar-reflected DM~\cite{Emken_2024}, while the green areas are constrained via nuclear recoils in CRESST \cite{CRESST:2020wtj}, DarkSide-50 \cite{DarkSide-50:2025lns}, LZ \cite{LZ:2025igz} and PandaX-4T~\cite{PandaX:2025rrz}. 
    The green solid lines show the projected sensitivities (from nuclear recoils) for SuperCDMS SNOLAB and DarkSide-20k \cite{SuperCDMS:2022kse,DarkSide-50:2025lns}. The hatched green (blue) lines represents the DM discovery limit in nuclear (electron) recoils in a Si target. The dotted red line denotes the limit up
    to which thermalisation of the dark photon can be neglected.}
    \label{fig:kDPM}
\end{figure}

For concreteness, we present our results in~\cref{fig:kDPM} for four different mediator-to-DM mass ratios, $m_{A'}/m_\chi$, of $3/2$ (top left), $1/2$ (top right), $1/10$ (bottom left) and $1/20$ (bottom right), respectively. For each of these cases, we show the freeze-in lines for the correct DM relic abundance for four different cosmological scenarios, with a reheating temperature \Trh of $100$ GeV (black solid), $1$ GeV (black dot-dashed), $100$ MeV (black dotted) and $10$ MeV (black dashed). Notice that the reheating temperature must satisfy that \Trh $>T_{BBN}\simeq 4~\mathrm{MeV}$ in order not to spoil the successful predictions of Big Bang nucleosynthesis (BBN)~\cite{de_Salas_2015}.  
Finally, the red dotted line corresponds to the thermalisation limit, above which the dark photon would thermalise with the SM and DM production via dark photon annihilations $A' A'\rightarrow \bar\chi\chi$ become relevant. 
We take this as the limit up to which the freeze-in predictions within our simple treatment remain valid. Above this line, a full treatment of the thermal DM production would require also the tracking of the thermal $A'$ population and assessing its impact on the DM abundance.

Since we consider scenarios of a visible $A'$ with $m_{A'}\lesssim 2 m_{\chi}$, there are rather strong bounds on the dark photon itself in this mass window, represented by the gray areas in~\cref{fig:kDPM}. For couplings of $\epsilon_D\gtrsim 10^{-7}$, these constraints are mostly due to searches for the $A'$ at beam dump experiments, like e.g.~E137~\cite{Bjorken:2009mm}, Charm~\cite{CHARM:1985anb}, NuCal~\cite{Blumlein:1990ay} and LSND~\cite{Batell:2009di}. In the region of parameter space relevant to freeze-in, $10^{-10}\lesssim\epsilon_D\lesssim10^{-7}$, strong bounds from astrophysics and cosmology arise, mostly due to supernova cooling or BBN~\cite{Caputo:2025avc}. Note that these constraints are mostly independent of the DM phenomenology. The constraints shown in gray are set on the mass of the mediator and, consequently, they shift to larger or smaller DM masses in~\cref{fig:kDPM} when the mediator-to-DM mass ratio, $m_{A'}/m_\chi$ increases or decreases, respectively.

The blue areas represent regions of parameter space that have been probed by direct detection experiments employing electron recoils ~\cite{DarkSide:2022knj,PandaX:2022xqx,Emken_2024}, whereas green areas correspond to those excluded by nuclear recoils~\cite{CRESST:2020wtj,DarkSide-50:2025lns,LZ:2025igz,PandaX:2025rrz}. It should be noted that, because the mediator mass enters in the denominator of the scattering cross sections via the propagator, the corresponding limits scale as $1/m_{A'}^4$ (see \cref{eq:sig_nuc,eq:sig_e}), and rescale accordingly as we change the mediator-to-DM mass ratios in~\cref{fig:kDPM}.  The green hatched region represents the DM discovery limit (conventionally referred to as the ``\textit{neutrino floor}") in nuclear recoils for a representative direct detection experiment employing a Si target (see \cref{sec:nu-floor} for details). For parameter points below this discovery limit, the BSM signal due to DM scattering is statistically not separable from the expected SM solar neutrino scattering signal. Similarly, the blue hatched line shows the corresponding sensitivity boundary for discriminating DM from neutrino signals in electron recoils.

As we can see from~\cref{fig:kDPM}, the freeze-in line for the correct relic abundance in a standard cosmology with high reheating temperature (black solid line) is excluded for mediator masses below $\sim100$ MeV. For higher masses, this still constitutes a valid solution for the DM abundance. However, it is well below the DM discovery limit in direct detection experiments and also well below the projected sensitivity of future fixed target experiments like SHiP. In contrast, 
in the low reheating scenarios, the value of the coupling along the relic abundance lines rapidly increases when \Trh$\sim m_\chi$.
This is due to the Boltzmann suppression that occurs for $m_{\chi}>$ \Trh, as explained in~\cref{sec:freeze-in}.

This effect could lead to observable DM signals in current and future direct detection experiments above the neutrino background. 
While for $m_{A'}=1.5~m_{\chi}$ ($m_{A'}=m_{\chi}/2$) all the allowed region of parameter space would yield signals buried in the neutrino floor (below the green hatched line), 
these scenarios could still be tested by searches for the mediator at SHiP~\cite{Zhou:2024aeu}.
However, for $m_{A'}=~m_{\chi}/10$ there is a potential window for DM masses of $0.6<m_{\chi}<3~\mathrm{GeV}$, where future direct detection experiments such as DarkSide-20k~\cite{DarkSide-50:2025lns} or SuperCDMS~\cite{SuperCDMS:2022kse} will be able to test relevant regions of the parameter space with signals above the neutrino floor.

\subsection{Leptophilic $U(1)$s}
\label{sec:gauge_res}

\begin{figure}[t]
    \centering
    \includegraphics[width=0.49\linewidth]{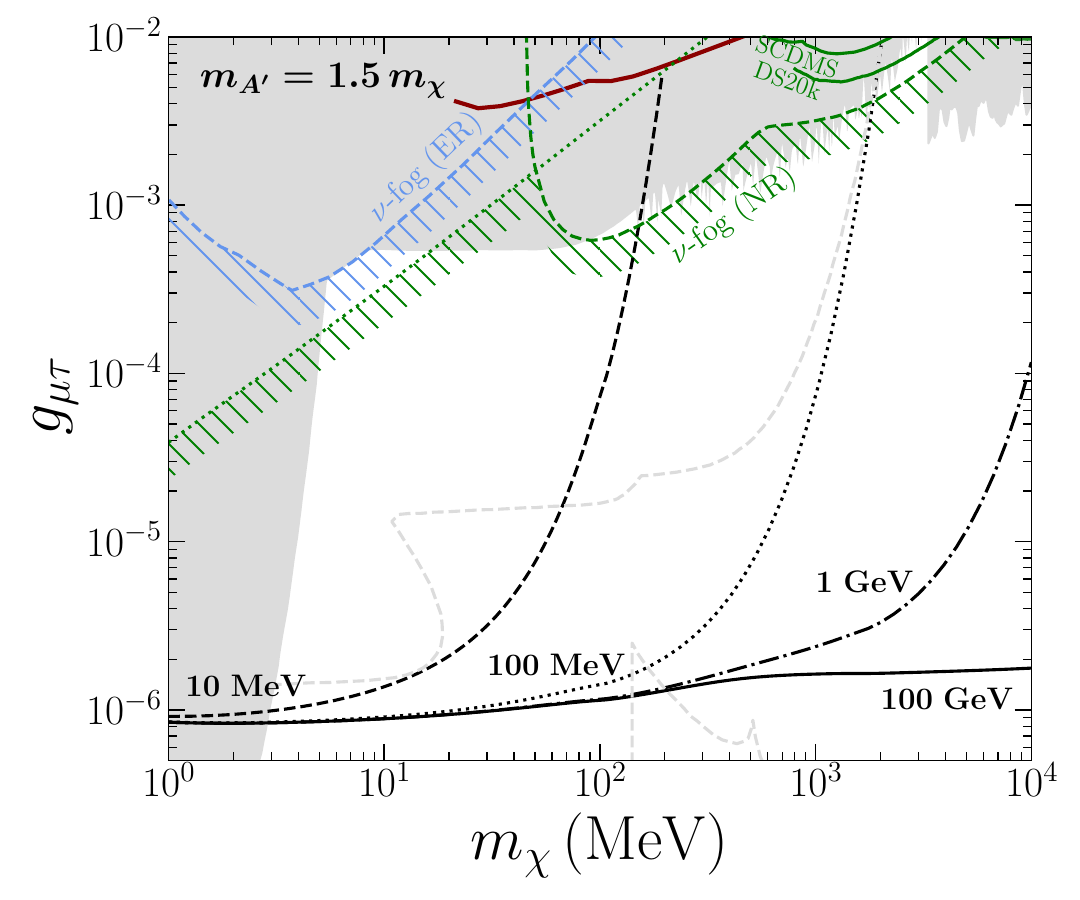}%
    \includegraphics[width=0.49\linewidth]{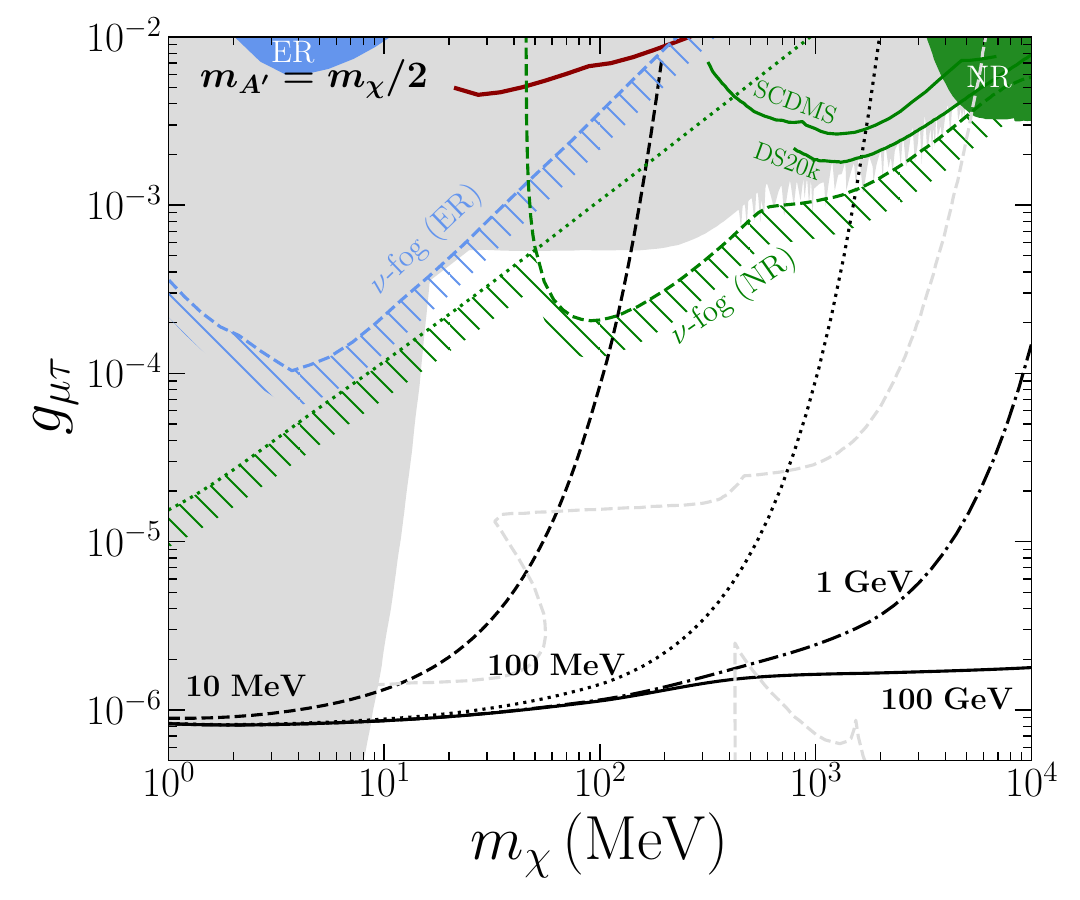}\\
    \includegraphics[width=0.49\linewidth]{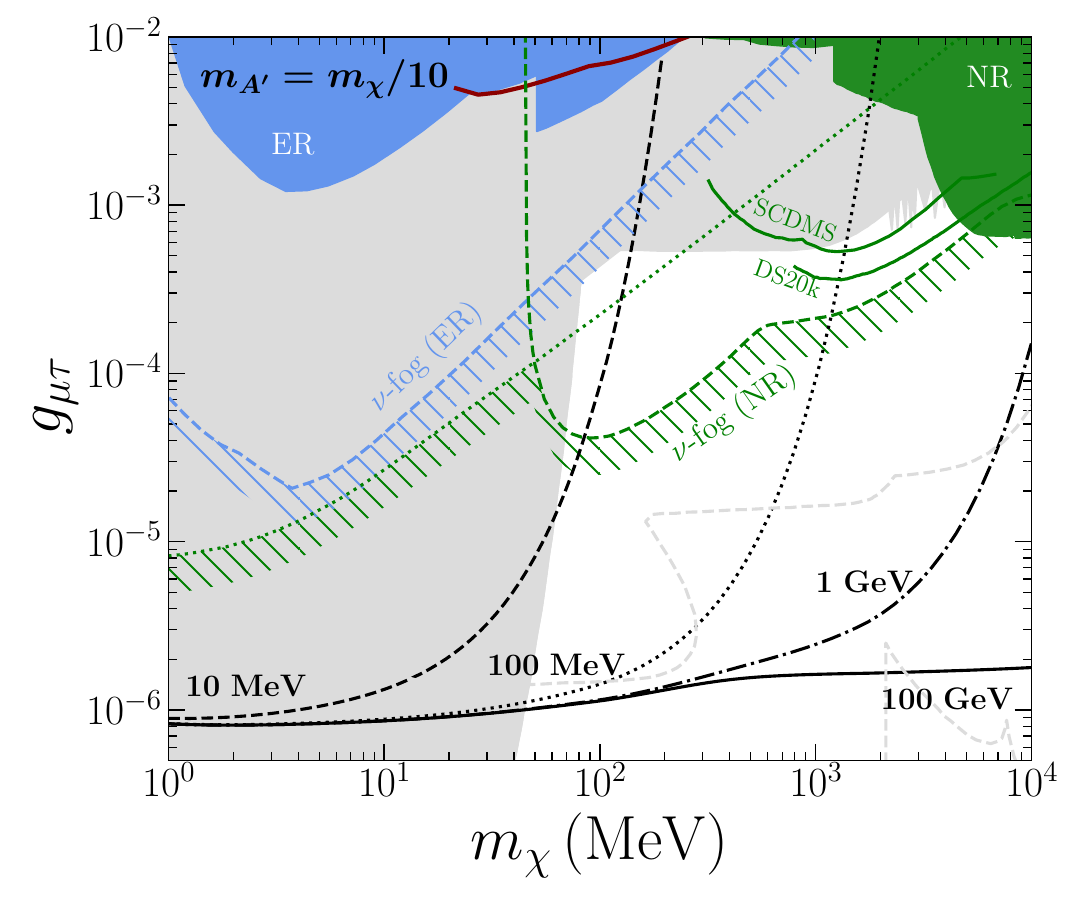}%
    \includegraphics[width=0.49\linewidth]{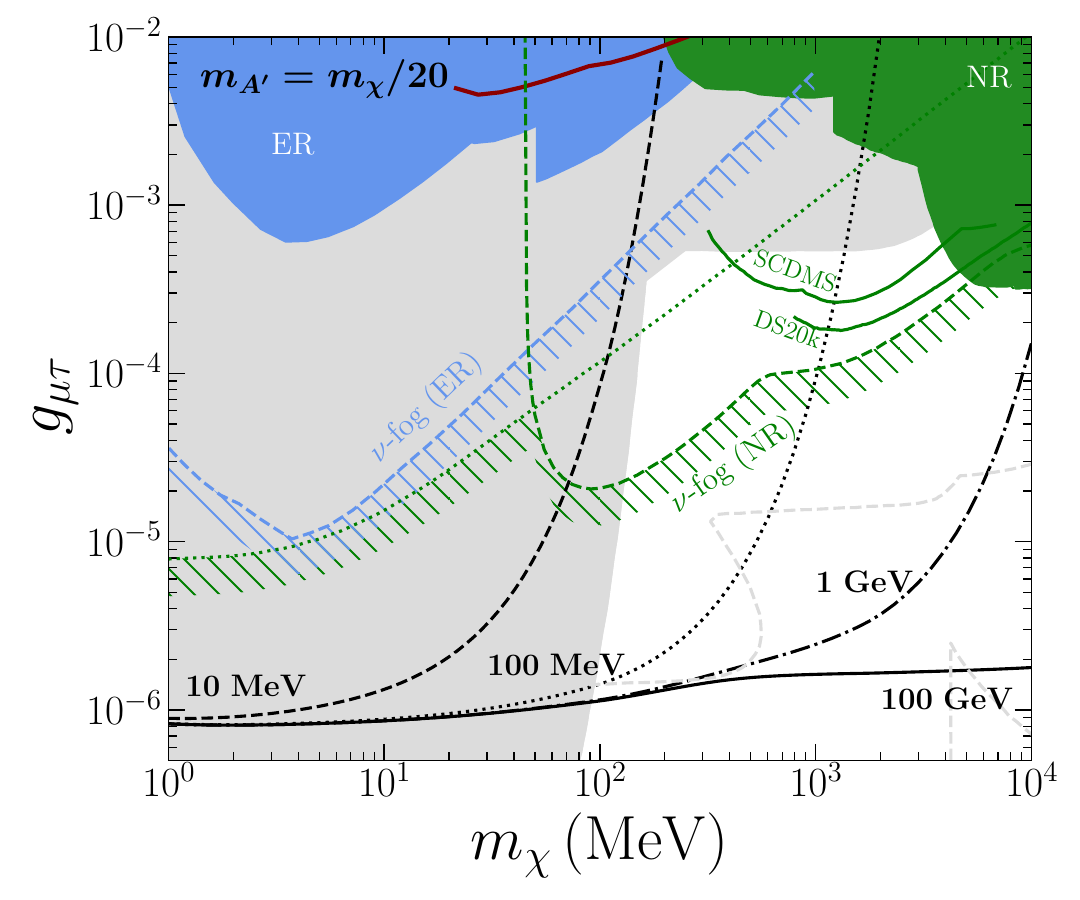}
    \caption{Parameter space relevant for freeze-in production of a vector-like fermion at low reheating temperature within \Umt. Results are shown for four different mediator-to-DM mass ratios of $m_{A'}/m_\chi = 3/2$ (\textbf{top left}), 1/2 (\textbf{top right}), 1/10 (\textbf{bottom left}), and 1/20 (\textbf{bottom right}). For each configurations, we show the relic density line for the observed DM abundance ($\Omega h =0.12$) for a reheating temperature of $T_R=10$ MeV (black dashed), 100 MeV (black dotted), 1 GeV (black dot-dashed), and 100 GeV (black solid). The dark red line illustrates the correct DM relic abundance from freeze-out production, as an indicator of when the DM thermalises with the SM plasma.}
    \label{fig:LMT}
\end{figure}

We now consider freeze-in production of the light DM particle, $\chi$, in the anomaly-free Abelian gauge models \Uij and \UBL (see Ref.~\cite{Han:2026ozp} for a recent analysis of thermal freeze-out in these models). 
The phenomenology of the associated $A'$ gauge bosons varies significantly from the simple \Udark dark photon discussed in~\cref{sec:dp_res}. The most significant difference is due to their tree-level couplings to neutrinos, which are especially relevant for light $A'$ with {MeV} masses. 
Thus, here we consider \Uij and \UBL as simple, prototypical examples
of the infinite family of accidentally anomaly-free $U(1)$ models with the SM field content only~\cite{Bauer:2020itv}. Hence, we expect qualitatively similar results for other anomaly-free constructions like $B-3L_i, B-2L_i-L_j, 2L_i-L_j-L_k$, etc., where $i,j,k={e,\mu,\tau}$.

In these models, strong cosmological constraints arise for gauge bosons lighter than $m_{A'}\lesssim 10$ MeV since they constitute extra relativistic degrees of freedom during the BBN era and the formation of the CMB, 
excluding gauge couplings as small as $g_X\lesssim 10^{-9}$~\cite{Escudero:2019gzq}. Therefore, in contrast to the minimal dark photon case, in this class of models there is no viable freeze-in option with ultralight mediators. For the same reason, strong constraints from BBN and CMB also exclude neutrino coupled fermionic DM below $m_\chi\sim10$ MeV~\cite{Boehm:2013jpa,Nollett:2014lwa}. To avoid these strong bounds, we will consider mediator and DM masses in the MeV-GeV range.

\begin{figure}[t]
\centering
    \begin{minipage}{0.48\textwidth}
    \centering
    \includegraphics[width=\textwidth]{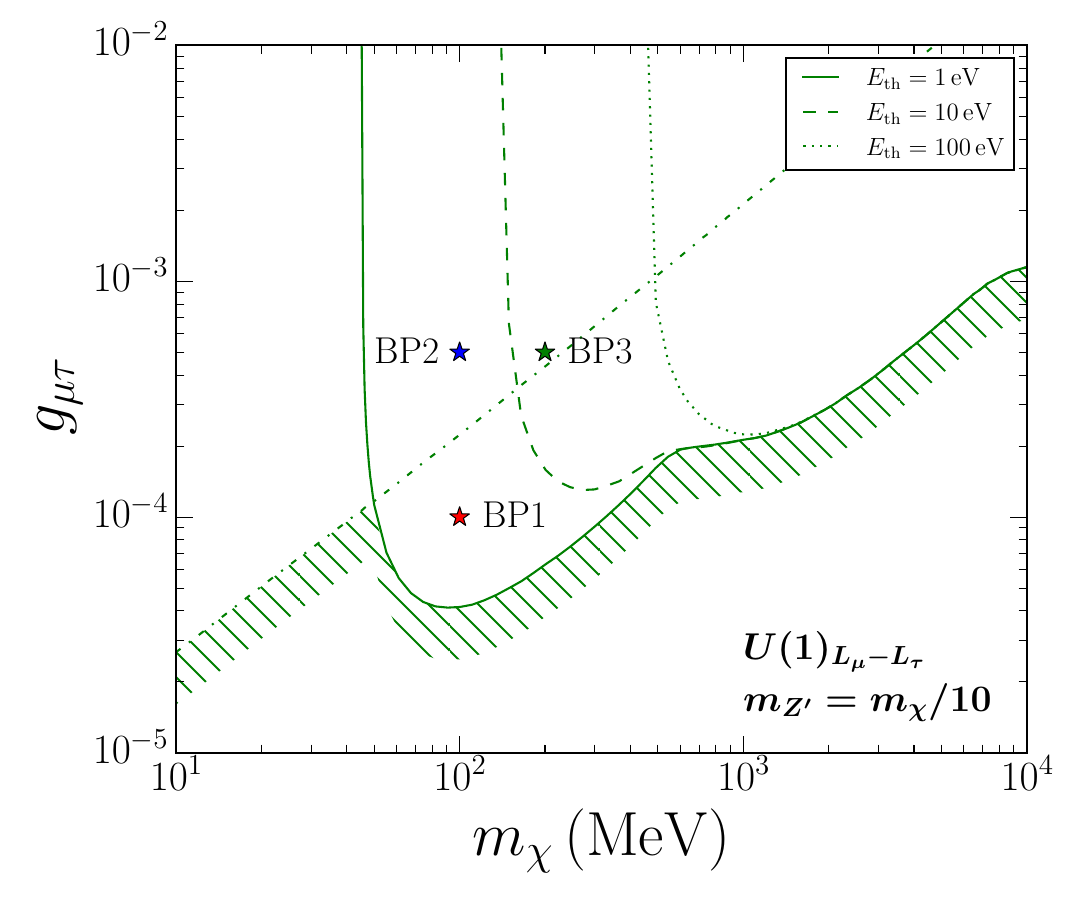}
\end{minipage}
\begin{minipage}{0.48\textwidth}
\centering
    \begin{tabular}{c | l}
     \rule{0pt}{3ex}{\textbf{BP1:}} & $m_\chi = 100$ MeV  \\[0.2cm] 
     & $g_{\mu\tau}=10^{-4}$  \\[0.1cm] \hline
    \rule{0pt}{3ex}{\textbf{BP2:}} & $m_\chi = 100$ MeV  \\[0.2cm] 
     & $g_{\mu\tau}=5\times 10^{-4}$ \\[0.1cm] \hline
    \rule{0pt}{3ex}{\textbf{BP3:}} & $m_\chi = 200$ MeV  \\[0.2cm] 
     & $g_{\mu\tau}=5\times 10^{-4}$ 
    \end{tabular} \vspace{.5cm}
    \end{minipage}
\caption{DM discovery limits in a \Umt model with $m_{A'}/m_\chi=1/10$ for a direct detection experiment employing Si with a nominal exposure of 1~tonne-years for 
varying energy thresholds $E_{\rm th}$ of 1 eV (solid), 10 eV (dashed) and 100 eV (dotted). While the mass reach of the DM discovery limit deteriorates significantly with increasing thresholds, the BSM neutrino discovery limit (dot-dashed) is essentially unaffected (at these low thresholds). 
The stars illustrate the three benchmark points defined on the right, for which we 
show the resulting DM and neutrino spectra in~\cref{fig:specs}.}
   \label{fig:thresholds} 
\end{figure}

\subsubsection{${\boldsymbol{L_\mu-L_\tau}}$}

We will begin by discussing the results for \Umt shown in~\cref{fig:LMT}. First, as illustrated by the gray areas, there are strong constraints from searches for the $A'$ mediator for gauge couplings of $g_{\mu\tau} \gtrsim 10^{-3}$, mostly from NA64$\mu$~\cite{NA64:2024klw}, White Dwarf cooling~\cite{Foldenauer:2024cdp}, Borexino~\cite{Bellini:2011rx,Borexino:2017rsf,Amaral:2020tga}, BaBar~\cite{BaBar:2016sci}, Belle-II~\cite{Belle-II:2024wtd} and ATLAS~\cite{ATLAS:2024uvu}.
For mediator masses below $m_{A'}\lesssim 10$ MeV, strong bounds from $\Delta N_\mathrm{eff}$ during BBN~\cite{Escudero:2019gzq} exclude all parameter space relevant to freeze-in production down to couplings of $g_{\mu\tau}\lesssim 10^{-9}$.

We can see that the limits from direct detection with ER cannot test parameter space beyond what is already excluded from searches of the mediator itself. However, the NR limits are excluding additional regions  for heavier DM particles with $m_\chi\sim 10$ GeV.
Nevertheless, there are still large parts of the parameter space available for $m_{A'}\gtrsim 10$ MeV and $g_{\mu\tau}\lesssim10^{-3}$, where freeze-in production of $\chi$ at low reheating temperatures of \Trh $\lesssim 1$ GeV  can reproduce the correct DM relic abundance (cf.~black lines).

However, for mediator masses larger than the DM mass, e.g., $m_{A'}/m_\chi = 3/2$ (top left panel), almost all of the viable parameter space lies below the discovery limits of DM direct detection experiments with both ER (blue hatched) and NR (green hatched). 
In turn, for $A'$  parametrically lighter than the DM, e.g.~for $m_{A'}/m_\chi = 1/2, 1/10$ and $1/20$, we can see that still large parts of the viable parameter space lie above the DM discovery limit in NR, where freeze-in can reproduce the correct relic DM abundance for low reheating temperatures of 10 MeV $\lesssim T_{\rm rh}\lesssim \mathcal{O}({\rm few})$  100 MeV. In fact, for  $m_{A'}/m_\chi = 1/10$ and $1/20$, the freeze-in abundances for DM masses of $m_\chi \gtrsim 400$ MeV with low reheating temperatures are within the reach of SuperCDMS SNOLAB~\cite{SuperCDMS:2022kse} and DarkSide-20k~\cite{DarkSide-50:2025lns}, as shown by the green lines.

As is shown by the gray dashed lines, future searches for the $A'$ mediator at NA64$\mu$~\cite{Gninenko:2014pea,Gninenko:2018tlp,NA64:2024nwj,NA64:2025ddk} or SHiP~\cite{Bauer:2018onh,Foldenauer:2018zrz,Albanese:2878604,Blinov:2025aha} will be able to independently probe large parts of the remaining parameter space relevant to freeze-in production for DM masses below a few GeV.%
\footnote{Note that in Ref.~\cite{Amaral:2021rzw} it has been argued that, in case of a positive detection, NA64$\mu$ could play a leading role in reconstructing the mass and coupling of the \Umt gauge boson.}

\bigskip

Interestingly,  for mediator masses of around $m_{A'}\gtrsim 10$ MeV, there is some region of parameter space (above the green dotted line) where the BSM neutrino signal alone can be distinguished from the SM neutrino signal, similar to what has been pointed out in Ref.~\cite{Amaral:2020tga}. We examine this behaviour in more detail for the example of a parametrically light \Umt boson with $m_{A'}/m_\chi = 1/10$. In~\cref{fig:thresholds} we show the DM and neutrino sensitivity limits for this example as a function of the experimental threshold $E_{\rm th}$.
We can see that the mass reach of the DM discovery limit (conventionally referred to as ``\textit{neutrino floor}'') crucially depends on the energy threshold of the specific DM experiment. We are considering a Si target, with a minimum energy threshold of $E_{\rm th} =1$ eV. Such a low threshold would allow to distinguish a DM signal for all coupling values of $g_{\mu\tau}$ above the solid green line in~\cref{fig:thresholds}, down to DM masses of $m_\chi\sim 50$ MeV. However, for thresholds of 10~eV (dashed green) and 100 eV (dotted green), the DM mass reach would deteriorate to $\sim150$~MeV and $\sim500$~MeV, respectively.

\begin{figure}[t]
    \centering
    \includegraphics[width=\textwidth]{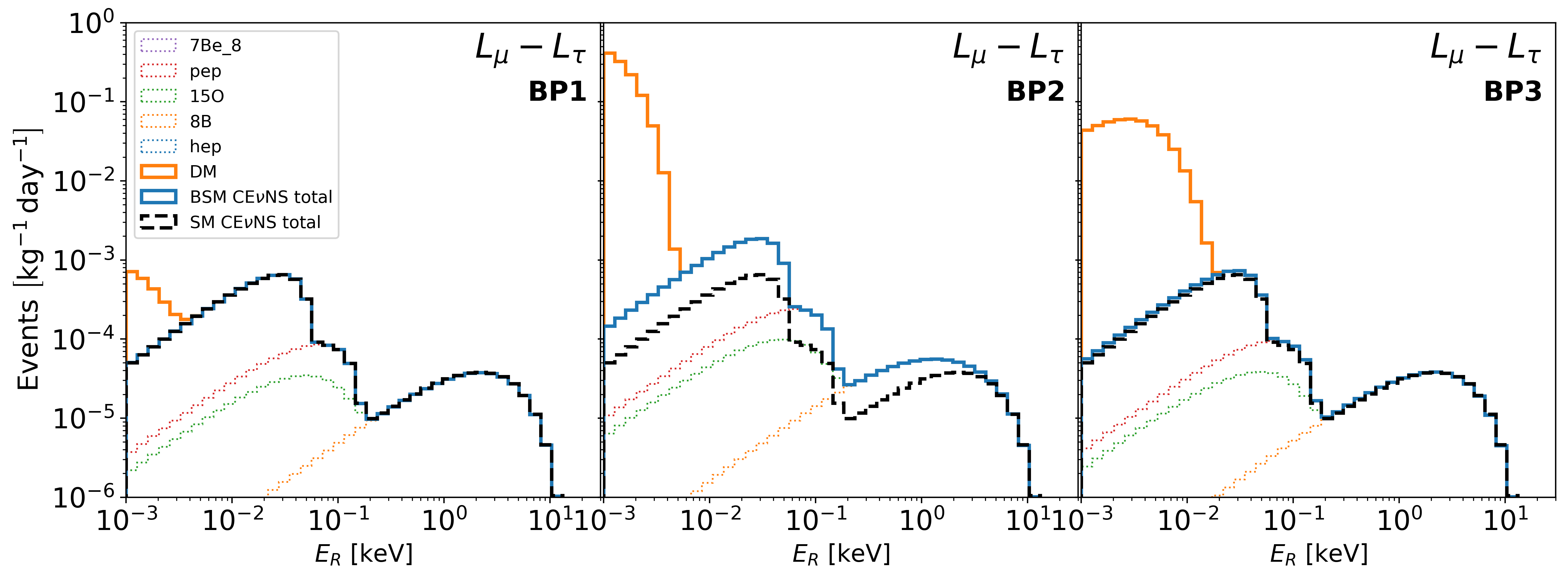}
    \caption{DM and neutrino spectra in \Umt for the three benchmark points defined in~\cref{fig:thresholds}. 
    We show the theoretical prediction for the expected events of DM (orange) and neutrinos (blue) in Si as a stacked histogram with $N=30$ bins and ignoring detector resolution and efficiency effects. 
    The black dashed line shows the SM neutrino spectrum for comparison.
    The dotted lines show the contribution of each of the neutrino flux components with the colour coding as defined in the legend.
    }
    \label{fig:specs}
\end{figure}

This can be readily understood by looking at the DM and neutrino recoil spectra in Si shown in~\cref{fig:specs} for the three benchmark points of~\cref{fig:thresholds}. Comparing, BP2 to BP3, we can see how the larger DM mass translates into a higher endpoint of the DM recoil spectrum (orange curves). Specifically, both DM recoil spectra of BP2 and BP3 would be detectable with a low energy threshold of $E_{\rm th}=1$ eV. However, all DM events of BP1 and BP2 would be too soft to be detected with the slightly higher threshold of $E_{\rm th}=10$ eV.  We can also see how for BP2 and BP3 the DM spectrum is clearly distinguishable from the (solar) neutrino spectrum (blue curve).

Next, let us consider the discrimination of the enhanced neutrino spectrum due to the $A'$ interactions (blue solid line) from the SM neutrino spectrum (black dashed) in~\cref{fig:specs}. We can see that for BP1 the SM and BSM spectra coincide and cannot be distinguished, while for BP2 the BSM spectrum is visibly enhanced almost across the entire energy range and can be clearly separated from the SM. For BP3, the BSM neutrino spectrum is much less markedly enhanced over the SM, but can still be statistically discriminated from the SM for the nominal exposure of 1~tonne-year that we consider.

Summing up this discussion, we note that direct detection experiments  might have the opportunity to observe both the neutrino and DM signal in such a leptophilic mediator model and discriminate it from the expected SM solar neutrino background.

\subsubsection{${\boldsymbol{L_e-L_\tau}}$ and ${\boldsymbol{L_\mu-L_e}}$}

\begin{figure}[t]
    \centering
    \includegraphics[width=0.49\linewidth]{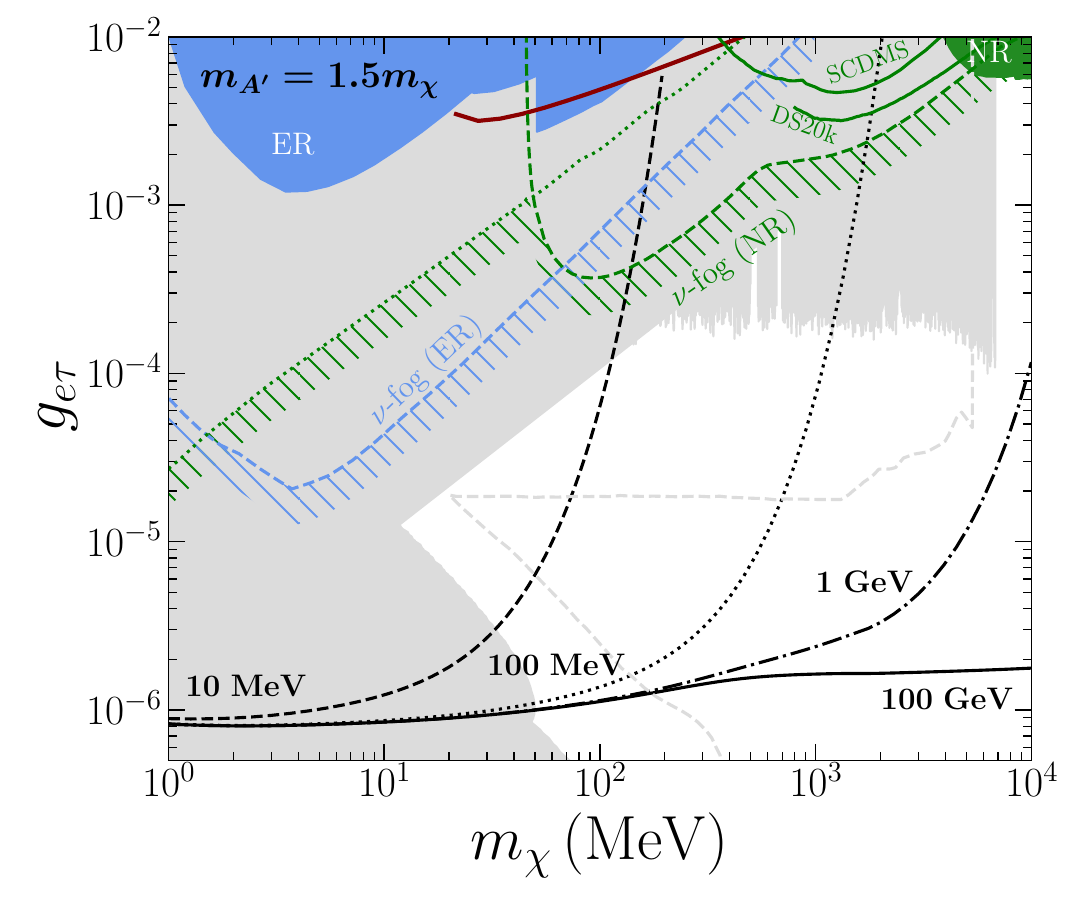}%
    \includegraphics[width=0.49\linewidth]{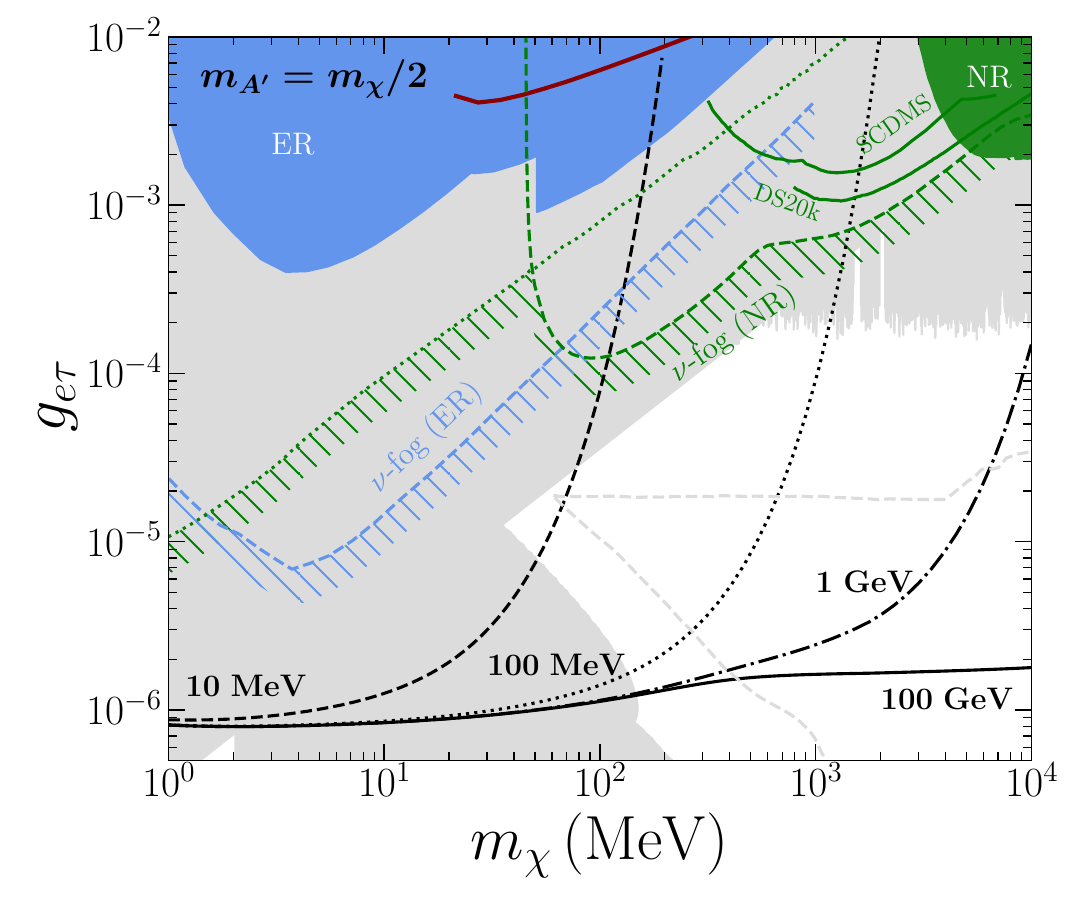}\\
    \includegraphics[width=0.49\linewidth]{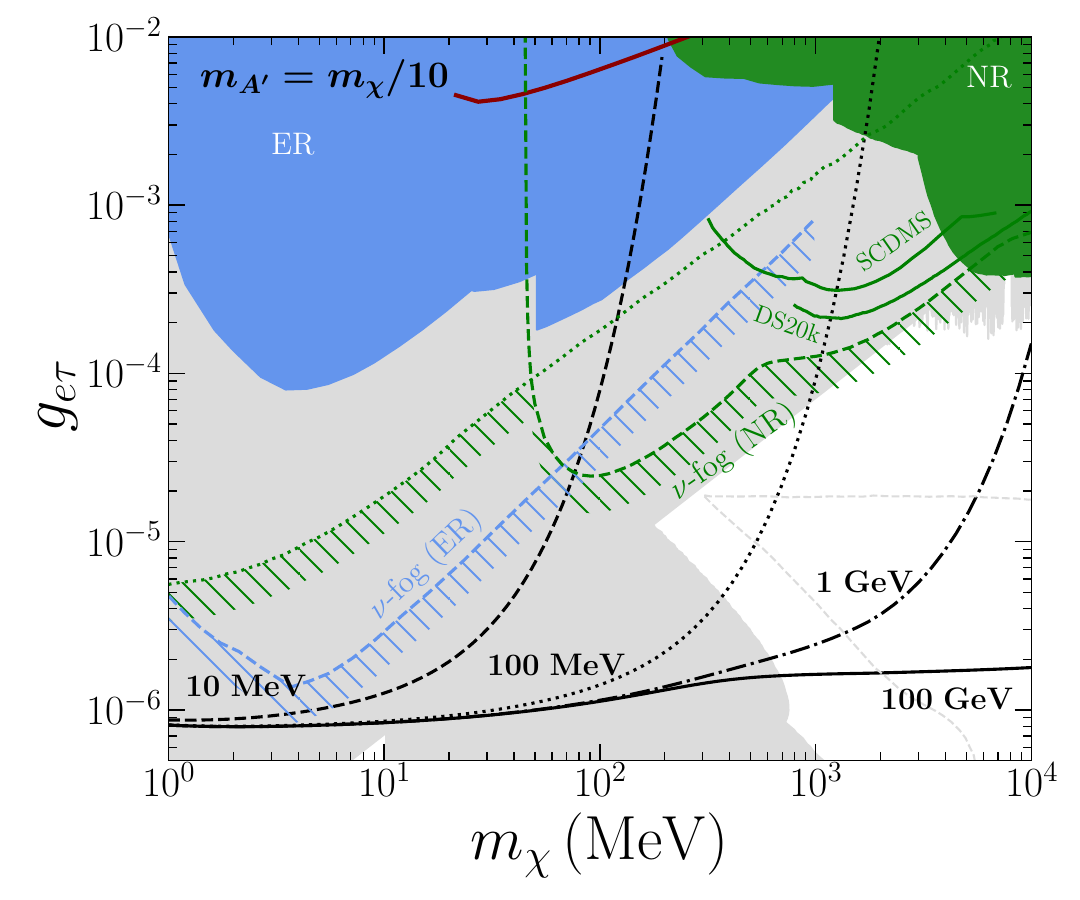}%
    \includegraphics[width=0.49\linewidth]{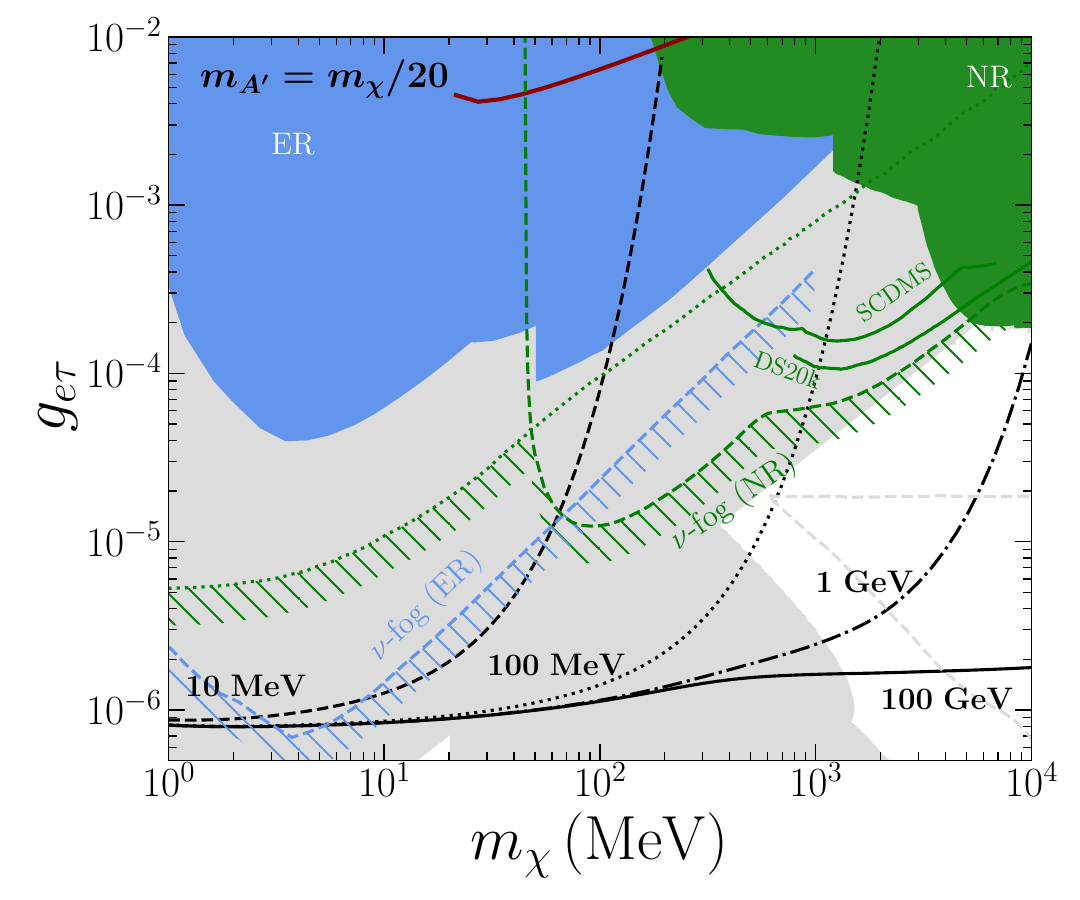}
    \caption{Parameter space for \Uet for the mass ratios and freeze-in temperatures as in~\cref{fig:LMT}.}
    \label{fig:LET}
\end{figure}

\begin{figure}[t!]
    \centering
    \includegraphics[width=0.49\linewidth]{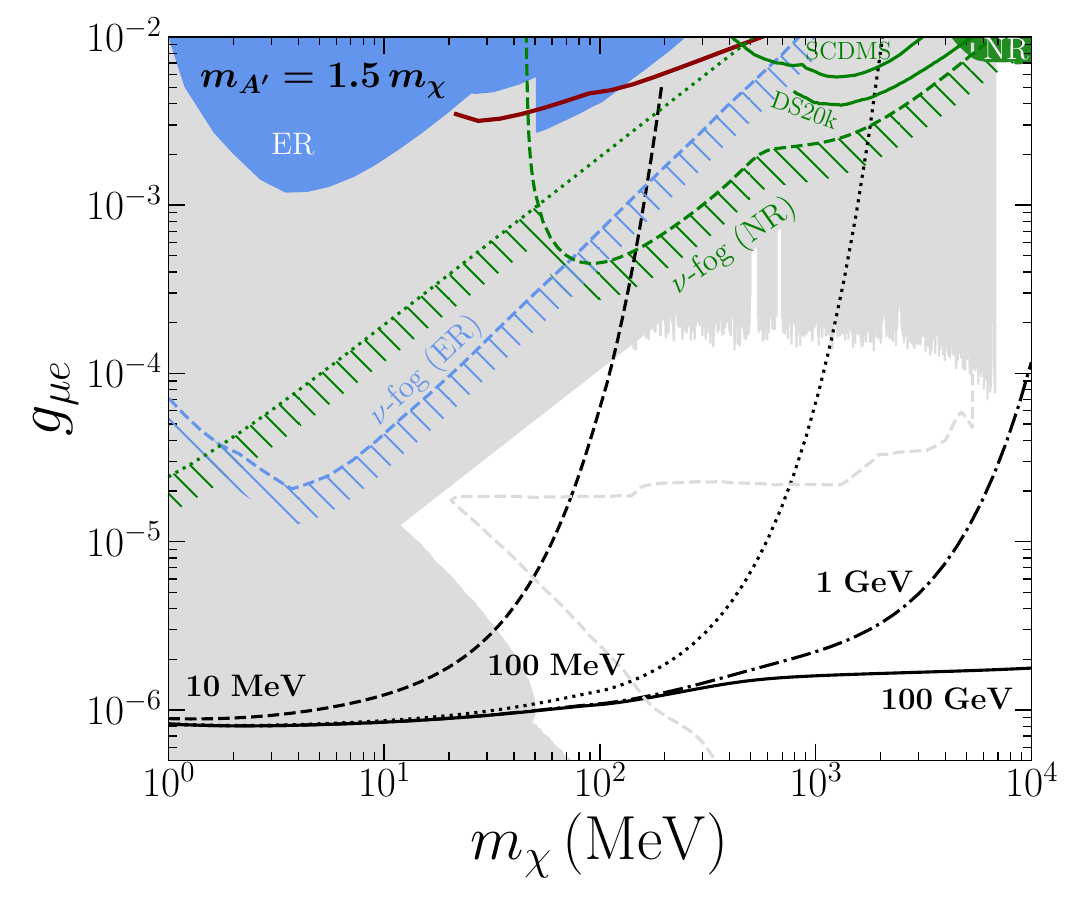}%
    \includegraphics[width=0.49\linewidth]{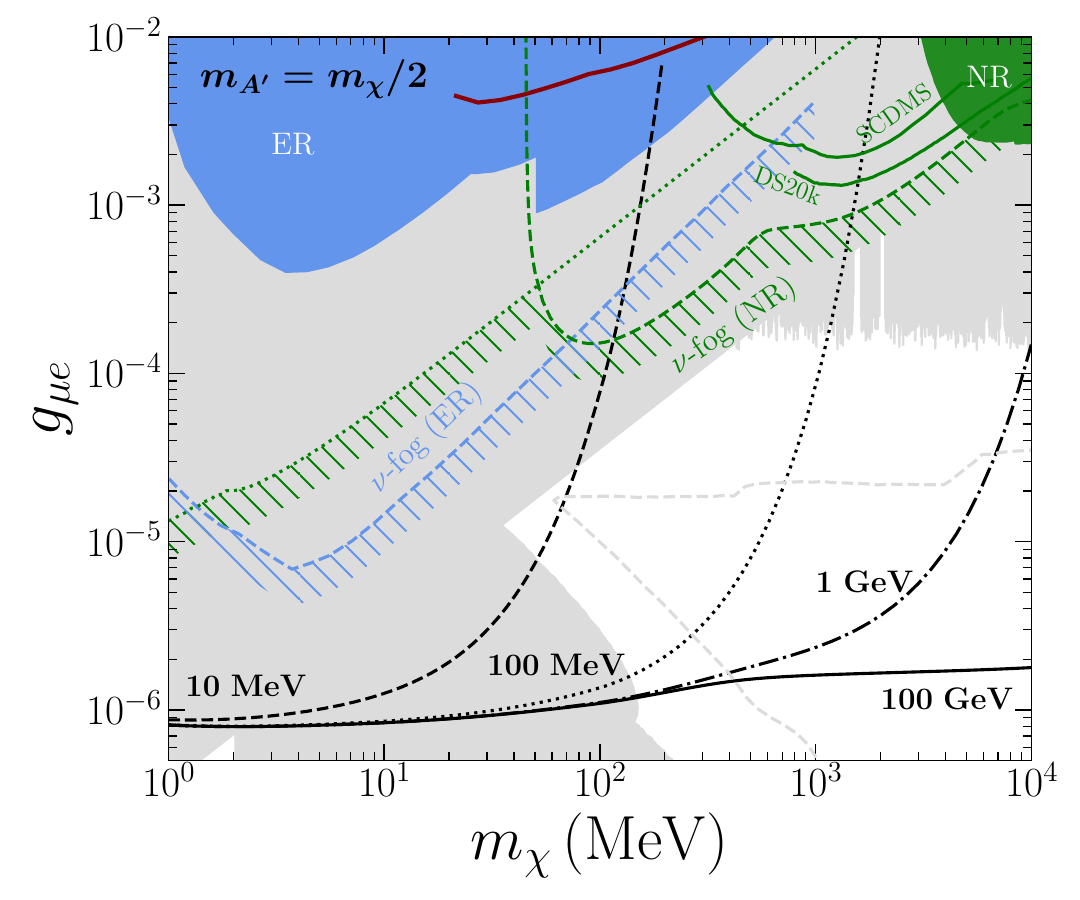}\\
    \includegraphics[width=0.49\linewidth]{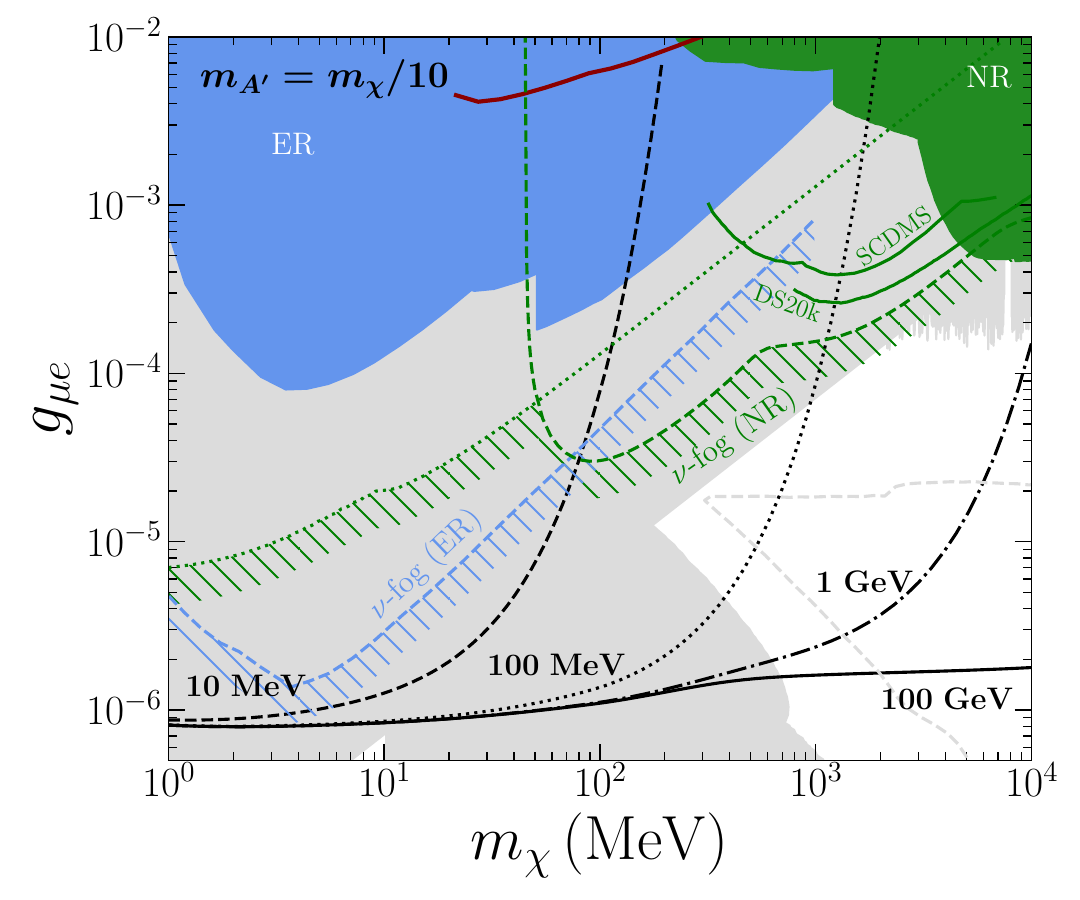}%
    \includegraphics[width=0.49\linewidth]{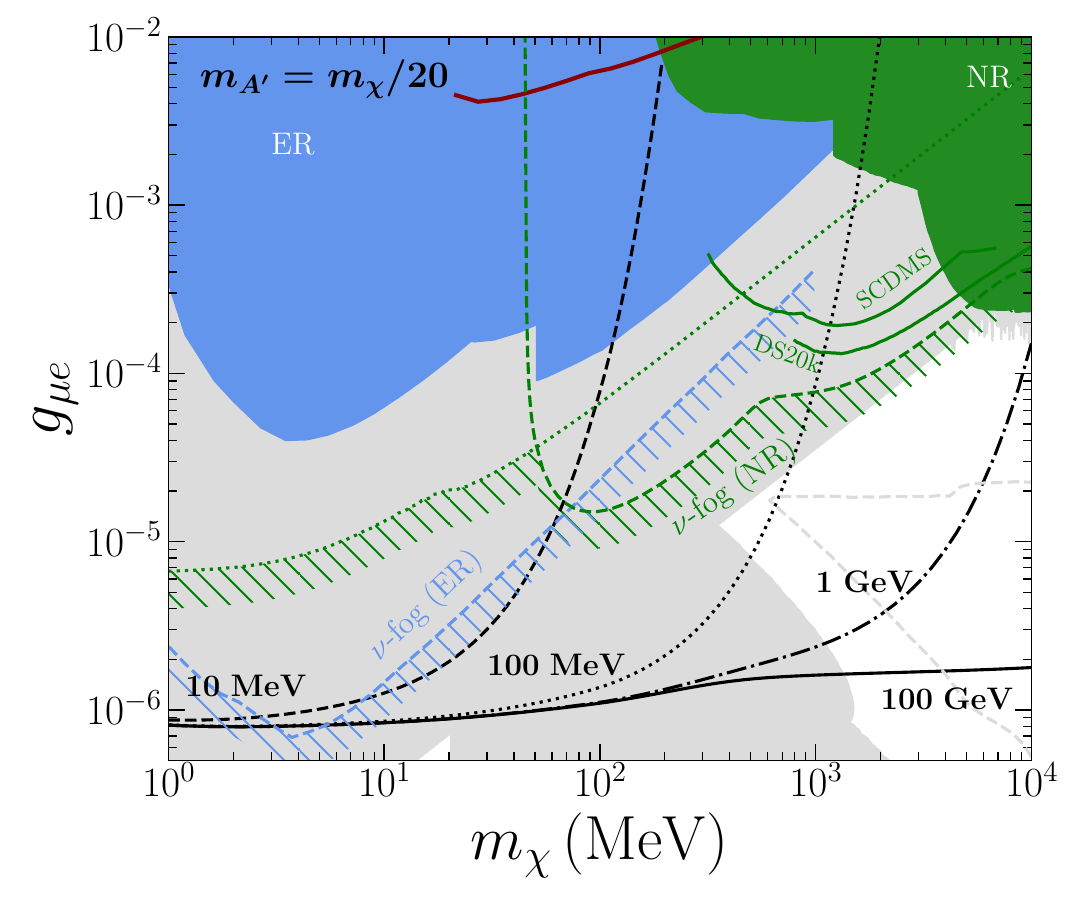}
    \caption{Parameter space for \Ume for the mass ratios and freeze-in temperatures as in~\cref{fig:LMT}.}
    \label{fig:LME}
\end{figure}

The most important feature of both \Uet and \Ume, distinguishing them from \Umt, is their coupling to first generation leptons, and therefore electrons. Comparing the constraints on the $A'$ mediator represented by the gray areas in~\cref{fig:LET,fig:LME} to those on the \Umt boson in~\cref{fig:LMT}, we can see that the $e^-$ and $\nu_e$ interactions result in much stronger bounds. Since their phenomenology is similar, we can discuss \Uet and \Ume simultaneously.

First of all, similarly to \Umt, strong BBN bounds rule out light mediators with $m_{A'}\lesssim 10$ MeV for couplings as small as $g_X\gtrsim 10 ^{-9}$~\cite{Escudero:2019gzq}. Furthermore, the electron interactions make the $A'$ subject to strong constraints from searches at electron beam dump and fixed target experiments, like E137, E141, E774~\cite{Bjorken:2009mm}, Orsay, and KEK~\cite{Andreas:2012mt}, excluding large parts of the parameter space for MeV mediators below $g_X\lesssim 10^{-5}$. Likewise, the \Ume and \Uet gauge bosons are subject to constraints coming from $e^+e^-$ colliders, like BaBar~\cite{BaBar:2014zli,BaBar:2017tiz}, excluding $A'$ up to masses of $m_{A'}\sim 10$ GeV for couplings above $g_X \gtrsim 10^{-4}$. Finally, the fact that the $A'$ have interactions with electron-flavoured neutrinos, $\nu_e$, leads to strong bounds from neutrino oscillation experiments~\cite{Heeck:2018nzc,Coloma:2020gfv}, especially COHERENT~\cite{Denton:2018xmq}, Super-Kamiokande~\cite{Wise:2018rnb} and Borexino~\cite{Coloma:2022umy}.

Similar to the case of \Umt, we can see in~\cref{fig:LET,fig:LME} that 
in the remaining viable parameter space of \Uet and \Ume the observed DM relic abundance can be reproduced via freeze-in of the dark fermion $\chi$  at low reheating temperatures of 10 MeV $\lesssim T_{\rm rh}\lesssim  \mathcal{O}(1)$ GeV. In contrast to the case of \Umt, direct detection searches are not sensitive to any of remaining viable parameter space. On the contrary, all of the viable freeze-in parameter space is in fact below the DM sensitivity line for direct detection experiments both for ER (blue hatched) and NR (green hatched) and will hence not lead to any testable signals above the SM solar neutrino background. Nevertheless, as illustrated by the projection shown as the gray dashed lines, part of this parameter can be tested by searches for the mediator itself at SHiP~\cite{Bauer:2018onh,Zhou:2024aeu}, NA64~\cite{Gninenko:2014pea} or Belle-II~\cite{Ferber:2015jzj,Belle-II:2018jsg}.

\subsubsection{${\boldsymbol{B-L}}$}

\begin{figure}[t]
    \centering
    \includegraphics[width=0.49\linewidth]{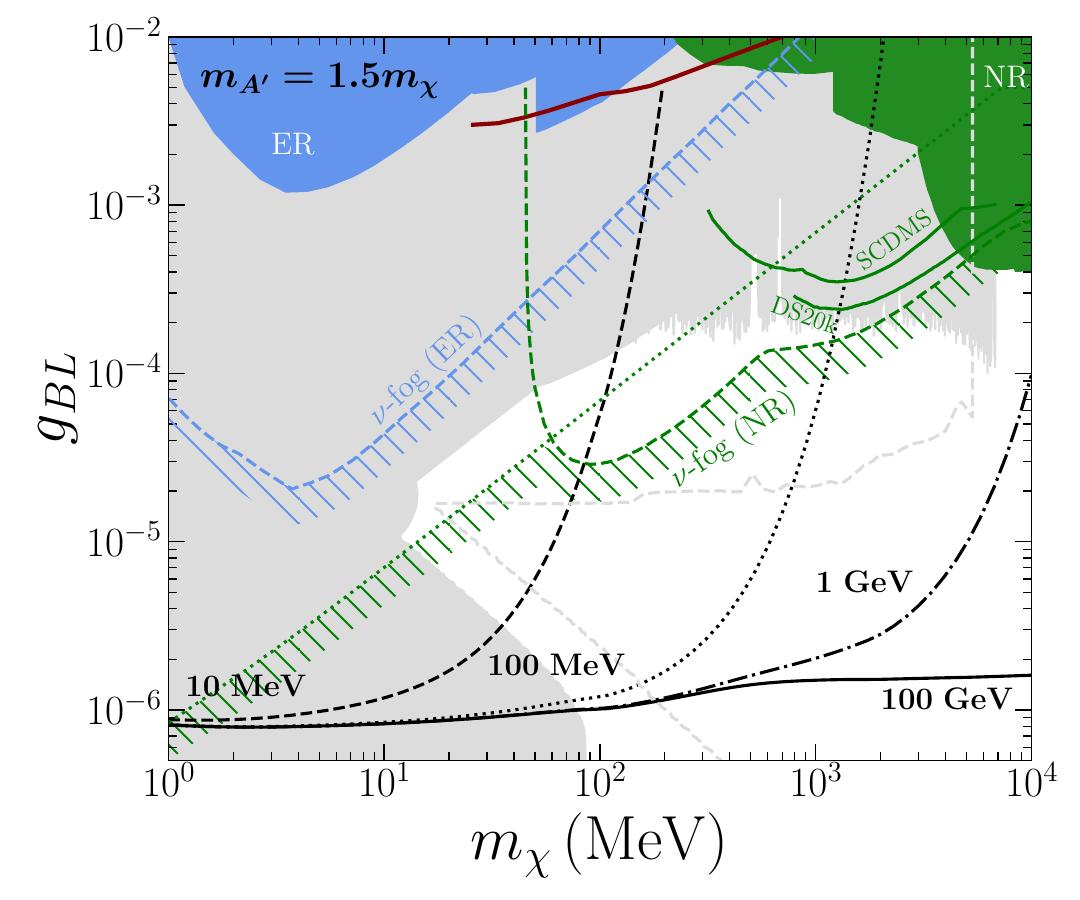}%
    \includegraphics[width=0.49\linewidth]{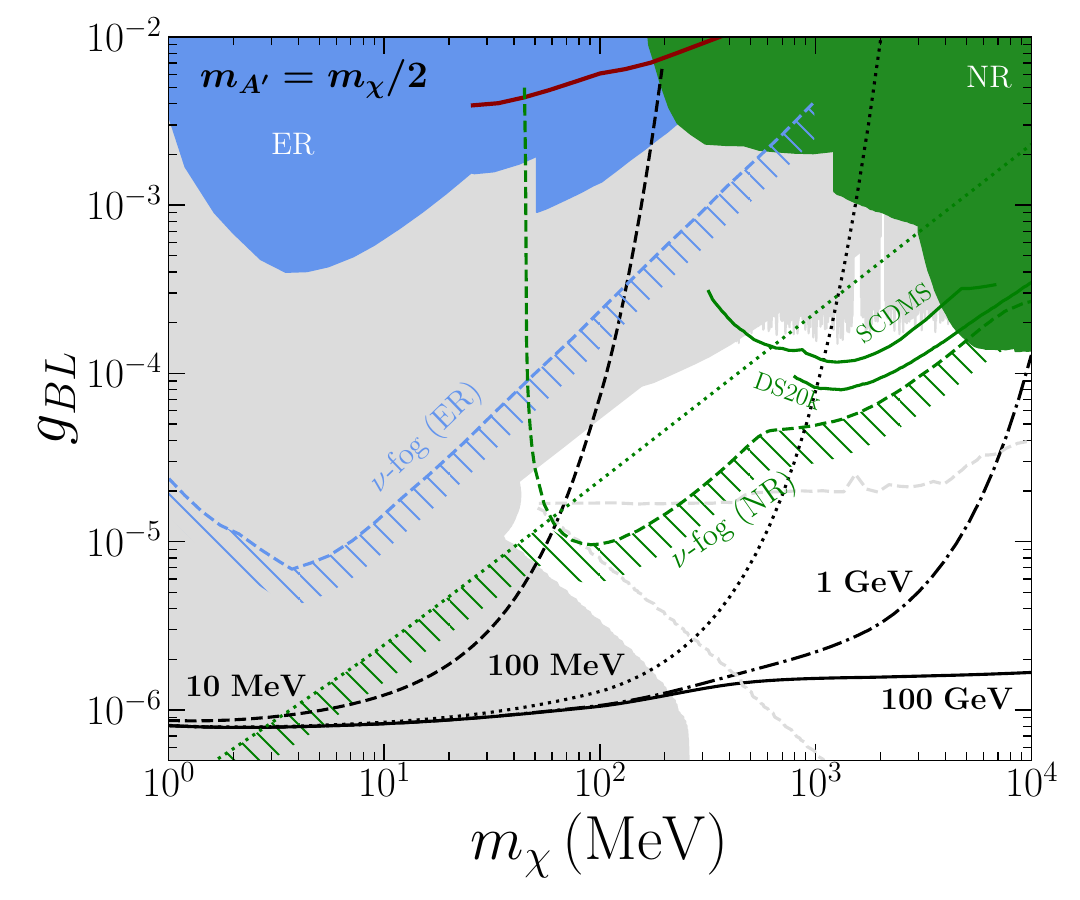}\\
    \includegraphics[width=0.49\linewidth]{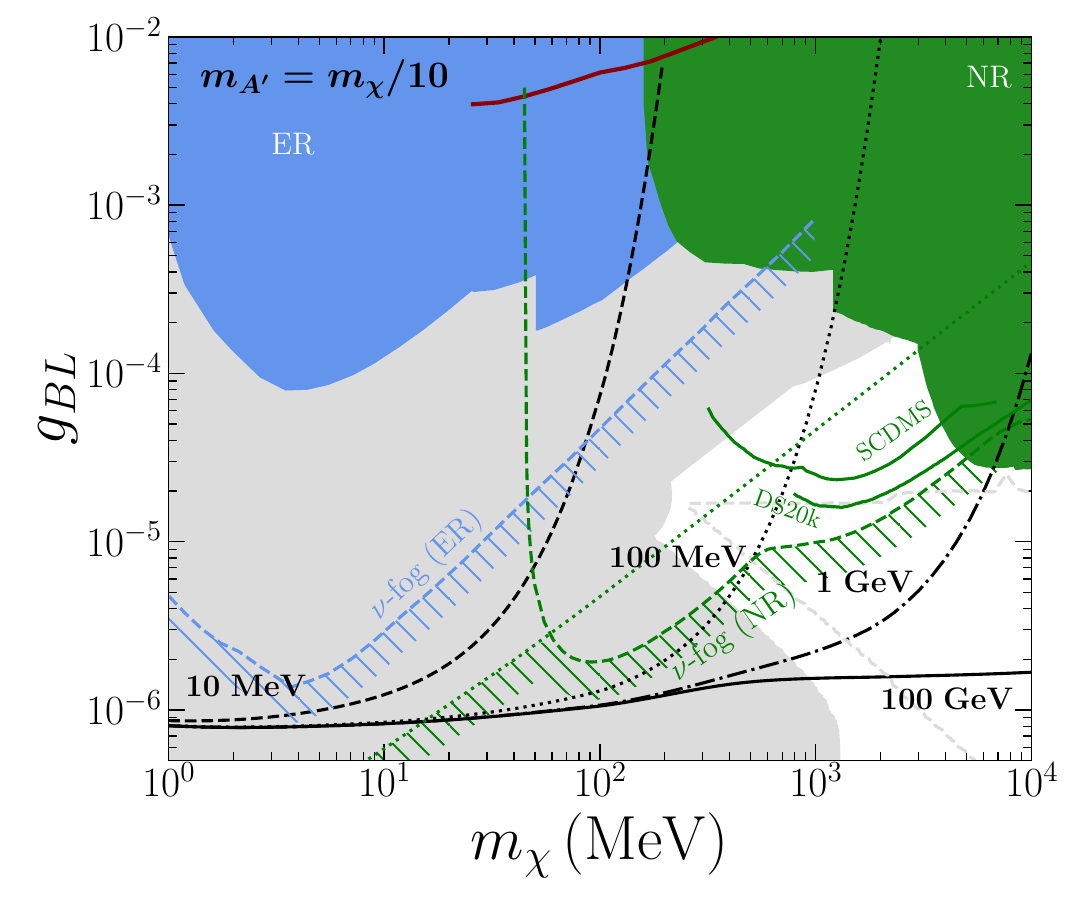}%
    \includegraphics[width=0.49\linewidth]{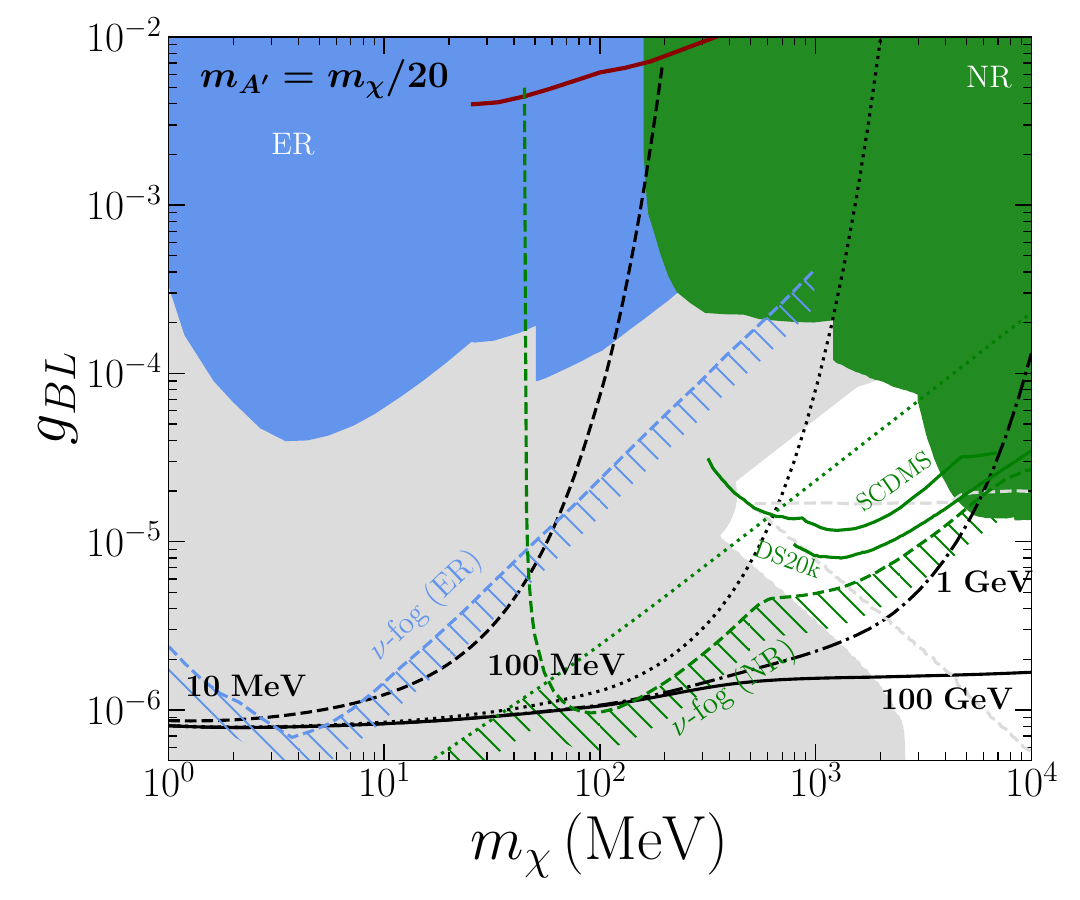}
    \caption{Parameter space for \UBL for the mass ratios and freeze-in temperatures as in~\cref{fig:LMT}.}
    \label{fig:BL}
\end{figure}

The \UBL gauge boson is subject to the most severe constraints, due to its couplings to all hadrons, charged leptons and neutrinos, as can be seen by the grey areas in~\cref{fig:BL}. As in the cases of \Ume and \Uet, the most stringent bounds on the \UBL gauge boson come from beam dump and fixed target experiments like E137, E141, E774, Orsay, NuCal/U70, FASER~\cite{Kling:2025udr}, neutrino interaction constraints from Borexino~\cite{Harnik:2012ni,Amaral:2020tga}, Texono~\cite{Bilmis:2015lja}, Dresden-II~\cite{Denton:2022nol}, as well as from $e^+e^-$ colliders like BaBar and Belle-II. In the future, large parts of the remaining parameter space will be tested by searches for the \UBL gauge boson at SHiP~\cite{Bauer:2018onh}, LHCb~\cite{Ilten:2015hya} or Belle-II~\cite{Ferber:2015jzj,Belle-II:2018jsg}.

Similar to what has been observed for freeze-in in \UBL with a standard cosmology~\cite{Heeba:2019jho,Eijima:2022dec,Fernandez-Martinez:2025qsw}, we find that at low reheating temperatures of 10 MeV $\lesssim T_{\rm rh}\lesssim  \mathcal{O}(1)$ GeV there exist large areas of available parameter space, where freeze-in can reproduce the observed DM relic abundance. These regions are a clear target for direct detection searches, as is illustrated by the relatively strong constraints that  NR searches already impose on the parameter space in~\cref{fig:BL}. Due to gauge couplings of the $A'$ to hadrons in \UBL, the NR signal for DM scattering is significantly enhanced compared to the kinetic mixing-induced interactions in purely leptonic $U(1)_{L_i-L_j}$ groups. Therefore, also the DM discovery limit in NR (green hatched) is shifted to lower values of the gauge coupling due to the increased sensitivity in this model. In fact, there are regions of available parameter space above the neutrino discovery limit (green dotted line), where direct detection experiments will be able to discriminate the \UBL-induced solar neutrino spectrum from  the SM expectation alone. Especially for the two cases of $m_{A'}/m_\chi=1/10$ and 1/20, large parts of these regions of parameter space are in reach of the future SuperCDMS SNOLAB~\cite{SuperCDMS:2022kse} and DarkSide-20k~\cite{DarkSide-50:2025lns} sensitivities illustrated by the solid green lines in~\cref{fig:BL}.

This makes direct detection experiments particularly well suited to test this scenario through future searches for both DM and neutrino signals.
In the case of a positive signal, identifying the underlying gauge structure would require exploiting the interplay of several complementary observables. While a single recoil excess would in general be degenerate, the combination of nuclear recoils, electron recoils can provide additional handles. The spectral shape of the DM signal is sensitive to the mass of the DM and mediator, while BSM neutrino signal depends only on the mediator.
The relative strength of the signal in the electron and nuclear channels probe the charge assignments of leptons and baryons under the new gauge symmetry. Moreover, comparing experiments with different targets and thresholds would help to disentangle spectral effects of the underlying interaction. We leave a dedicated study of parameter reconstruction and model discrimination for future work.

\section{Conclusions}
\label{sec:conclusion}

In this work, we have studied the potential of future direct detection experiments to detect new physics in a particular class of freeze-in DM scenarios, consisting of a light vector-like fermion, $\chi$, coupled to the SM sector via a new light vector mediator. We have considered the minimal kinetic mixing portal of a dark \Udark both for an ultra-light and a massive dark photon, $A'$, as well as the minimally anomaly-free gauge extensions of \Uij and \UBL. For each of these scenarios, we have determined the regions of the parameter space where the correct relic abundance can be obtained, taking into account non-standard cosmologies with low reheating temperatures. In the regions allowed by current experimental constraints, we have investigated the sensitivity of future direct detection experiments to either the DM particle or new physics in the neutrino sector.

\begin{itemize}
    \item 
    In the case of an ultra-light secluded \Udark mediator, current direct detection experiments like DAMIC-M and PandaX-4T already exclude freeze-in production at low reheating temperatures for DM masses above $m_{\chi} \gtrsim$ 4 MeV, while for high reheating temperatures DM masses of 4 MeV $\lesssim m_{\chi} \lesssim 50$ MeV are excluded.
    Exploiting the scaling of the detection rate for freeze-in DM with its relic abundance, we point out that current limits can in fact rule out scenarios where the dark fermion only constitutes a fraction (larger than 40\%) of the total cold DM.
    We then show that future experiments such as SuperCDMS SNOLAB and TESSERACT will be able to test DM fractions as low as 1\%
    while OSCURA could potentially reach 0.1\% 
    for MeV DM masses (see \cref{fig:DP-directdetection}). 
    
    \item 
    For massive dark photons in \Udark with MeV masses, strong constraints from astrophysics and cosmology already exclude large regions of parameter space where freeze-in with low-temperature reheating can reproduce the correct relic abundance. Nevertheless, there remain regions of parameter space for DM masses of 50 MeV $\lesssim m_{\chi}\lesssim$ 500 MeV in the case of $m_{A'}/m_\chi=1/10$,  where the freeze-in abundance is within the reach of SuperCDMS SNOLAB and DarkSide-20k. For small couplings, however, the predicted DM detection rate would be obscured by the neutrino fog, rendering it unobservable. The results are summarised in \cref{fig:kDPM}.
    
    \item 
    In the anomaly-free models \Uij and \UBL, the freeze-in production proceeds dominantly via $A'$  and neutrino annihilations. In these scenarios, freeze-in can reproduce the observed relic abundance for gauge couplings of $g_X\sim10^{-6}$ with high reheating temperatures, while for low reheating temperatures the correct DM abundance can be achieved at much higher couplings of up to $g_X\sim10^{-2}$, placing these scenarios in reach of future experimental searches. On the one hand, future searches for the mediator at beam dump and fixed target experiments, such as SHiP, NA64$\mu$ or Belle-II, will be sensitive to significant fractions of the parameter space relevant to low reheating temperature freeze-in. Nevertheless, especially in \Umt and \UBL, future direct detection experiments like SuperCDMS SNOLAB or DarkSide-20k will be sensitive to both the DM scattering and the enhanced solar neutrino scattering signal in nuclear recoils. Interestingly, the region of parameter space where these models could lead to appreciable signals in electron recoils is already firmly excluded by searches of the leptophilic mediator. The results are shown in \cref{fig:LMT,fig:LET,fig:LME,fig:BL}.
    
\end{itemize}

We conclude that freeze-in production of DM at low reheating temperatures within vector portal models leads to interesting phenomenology, putting these scenarios in reach of upcoming experimental searches both at beam dump and fixed target experiments, but most notably also for nuclear recoil searches at direct detection experiments. Given the prospects that direct detection can test both the DM signal of the low-reheating-temperature freeze-in as well as the BSM solar neutrino scattering signal in  leptophilic vector mediators like \Umt and \UBL, it will be crucial to develop strategies of how to discriminate those two types of signals in nuclear scattering.

\section*{Acknowledgements}

We thank  Enrique Fernández-Martínez, El\'ias L\'opez Asamar,  David Alonso-Gonz\'alez and Daniel Naredo-Tuero for helpful discussions. We are also thankful to  Andrea Caputo, Kevin Kelly and Felix Kling for providing the data points of Refs.~\cite{Caputo:2025avc,Blinov:2025aha,Kling:2025udr}.
The work of PF was supported by a fellowship from La Caixa Foundation (ID 100010434 with code LCF/BQ/PR25/12110019). 
OZ has been partially supported by Sostenibilidad-UdeA, the UdeA/CODI Grants 2022-52380 and  2024-76476.
DC, PF, and RLN acknowledge support from the Spanish Agencia Estatal de Investigaci\'on through the grants PID2024-155874NB-C22 (TheDeAs) and CEX2020-001007-S, funded by 
MCIN/AEI/10.13039/501100011033.
DC also acknowledges the kind hospitality of the Universidad de Antioquia and the Universidad Pedag\'ogica y Tecnol\'ogica de Colombia during the initial stages of this project. 
This work is partially funded by the European Commission – NextGenerationEU, through Momentum CSIC Programme: Develop Your Digital Talent. We acknowledge High Performance Computing support by Emilio Ambite, staff hired under the Generation D initiative, promoted by Red.es, an organisation attached to the Spanish Ministry for Digital Transformation and the Civil Service, for the attraction and retention of talent through grants and training contracts, financed by the Recovery, Transformation and Resilience Plan through the EU’s Next Generation funds.

\bibliographystyle{JHEP}
\bibliography{biblio}

\providecommand{\href}[2]{#2}\begingroup\raggedright\begin{thebibliography}{100}

\bibitem{Cirelli:2024ssz}
M.~Cirelli, A.~Strumia and J.~Zupan, \emph{{Dark Matter}},
  \href{https://arxiv.org/abs/2406.01705}{{\ttfamily 2406.01705}}.

\bibitem{XENON:2025vwd}
{\scshape XENON} collaboration, \emph{{WIMP Dark Matter Search Using a 3.1
  Tonne-Year Exposure of the XENONnT Experiment}},
  \href{https://doi.org/10.1103/msw4-t342}{\emph{Phys. Rev. Lett.} {\bfseries
  135} (2025) 221003} [\href{https://arxiv.org/abs/2502.18005}{{\ttfamily
  2502.18005}}].

\bibitem{LZ:2024zvo}
{\scshape LZ} collaboration, \emph{{Dark Matter Search Results from
  4.2{\,}{\,}Tonne-Years of Exposure of the LUX-ZEPLIN (LZ) Experiment}},
  \href{https://doi.org/10.1103/4dyc-z8zf}{\emph{Phys. Rev. Lett.} {\bfseries
  135} (2025) 011802} [\href{https://arxiv.org/abs/2410.17036}{{\ttfamily
  2410.17036}}].

\bibitem{PandaX:2024qfu}
{\scshape PandaX} collaboration, \emph{{Dark Matter Search Results from
  1.54{\,}{\,}Tonne{\textperiodcentered}Year Exposure of PandaX-4T}},
  \href{https://doi.org/10.1103/PhysRevLett.134.011805}{\emph{Phys. Rev. Lett.}
  {\bfseries 134} (2025) 011805}
  [\href{https://arxiv.org/abs/2408.00664}{{\ttfamily 2408.00664}}].

\bibitem{Essig:2011nj}
R.~Essig, J.~Mardon and T.~Volansky, \emph{{Direct Detection of Sub-GeV Dark
  Matter}}, \href{https://doi.org/10.1103/PhysRevD.85.076007}{\emph{Phys. Rev.
  D} {\bfseries 85} (2012) 076007}
  [\href{https://arxiv.org/abs/1108.5383}{{\ttfamily 1108.5383}}].

\bibitem{Essig:2012yx}
R.~Essig, A.~Manalaysay, J.~Mardon, P.~Sorensen and T.~Volansky, \emph{{First
  Direct Detection Limits on sub-GeV Dark Matter from XENON10}},
  \href{https://doi.org/10.1103/PhysRevLett.109.021301}{\emph{Phys. Rev. Lett.}
  {\bfseries 109} (2012) 021301}
  [\href{https://arxiv.org/abs/1206.2644}{{\ttfamily 1206.2644}}].

\bibitem{CRESST:2020wtj}
{\scshape CRESST} collaboration, \emph{{Searches for Light Dark Matter with the
  CRESST-III Experiment}},
  \href{https://doi.org/10.1007/s10909-020-02343-3}{\emph{J. Low Temp. Phys.}
  {\bfseries 199} (2020) 547}.

\bibitem{CRESST:2019jnq}
{\scshape CRESST} collaboration, \emph{{First results from the CRESST-III
  low-mass dark matter program}},
  \href{https://doi.org/10.1103/PhysRevD.100.102002}{\emph{Phys. Rev. D}
  {\bfseries 100} (2019) 102002}
  [\href{https://arxiv.org/abs/1904.00498}{{\ttfamily 1904.00498}}].

\bibitem{CRESST:2024cpr}
{\scshape CRESST} collaboration, \emph{{First observation of single photons in
  a CRESST detector and new dark matter exclusion limits}},
  \href{https://doi.org/10.1103/PhysRevD.110.083038}{\emph{Phys. Rev. D}
  {\bfseries 110} (2024) 083038}
  [\href{https://arxiv.org/abs/2405.06527}{{\ttfamily 2405.06527}}].

\bibitem{SENSEI:2023zdf}
{\scshape SENSEI} collaboration, \emph{{First Direct-Detection Results on
  Sub-GeV Dark Matter Using the SENSEI Detector at SNOLAB}},
  \href{https://doi.org/10.1103/PhysRevLett.134.011804}{\emph{Phys. Rev. Lett.}
  {\bfseries 134} (2025) 011804}
  [\href{https://arxiv.org/abs/2312.13342}{{\ttfamily 2312.13342}}].

\bibitem{SuperCDMS:2024yiv}
{\scshape SuperCDMS} collaboration, \emph{{Light dark matter constraints from
  SuperCDMS HVeV detectors operated underground with an anticoincidence event
  selection}}, \href{https://doi.org/10.1103/PhysRevD.111.012006}{\emph{Phys.
  Rev. D} {\bfseries 111} (2025) 012006}
  [\href{https://arxiv.org/abs/2407.08085}{{\ttfamily 2407.08085}}].

\bibitem{SuperCDMS:2025dha}
{\scshape SuperCDMS} collaboration, \emph{{Search for low-mass electron-recoil
  dark matter using a single-charge sensitive SuperCDMS-HVeV detector}},
  \href{https://doi.org/10.1103/5lnp-6mng}{\emph{Phys. Rev. D} {\bfseries 113}
  (2026) 032001} [\href{https://arxiv.org/abs/2509.03608}{{\ttfamily
  2509.03608}}].

\bibitem{DarkSide:2022knj}
{\scshape DarkSide} collaboration, \emph{{Search for Dark Matter Particle
  Interactions with Electron Final States with DarkSide-50}},
  \href{https://doi.org/10.1103/PhysRevLett.130.101002}{\emph{Phys. Rev. Lett.}
  {\bfseries 130} (2023) 101002}
  [\href{https://arxiv.org/abs/2207.11968}{{\ttfamily 2207.11968}}].

\bibitem{DarkSide-50:2022qzh}
{\scshape DarkSide-50} collaboration, \emph{{Search for low-mass dark matter
  WIMPs with 12~ton-day exposure of DarkSide-50}},
  \href{https://doi.org/10.1103/PhysRevD.107.063001}{\emph{Phys. Rev. D}
  {\bfseries 107} (2023) 063001}
  [\href{https://arxiv.org/abs/2207.11966}{{\ttfamily 2207.11966}}].

\bibitem{PandaX:2022xqx}
{\scshape PandaX} collaboration, \emph{{Search for Light Dark Matter with
  Ionization Signals in the PandaX-4T Experiment}},
  \href{https://doi.org/10.1103/PhysRevLett.130.261001}{\emph{Phys. Rev. Lett.}
  {\bfseries 130} (2023) 261001}
  [\href{https://arxiv.org/abs/2212.10067}{{\ttfamily 2212.10067}}].

\bibitem{PandaX:2025rrz}
{\scshape PandaX} collaboration, \emph{{Search for Light Dark Matter with 259
  Days of Data in PandaX-4T}},
  \href{https://doi.org/10.1103/rtnh-jn8s}{\emph{Phys. Rev. Lett.} {\bfseries
  135} (2025) 211001} [\href{https://arxiv.org/abs/2507.11930}{{\ttfamily
  2507.11930}}].

\bibitem{XENON:2019gfn}
{\scshape XENON} collaboration, \emph{{Light Dark Matter Search with Ionization
  Signals in XENON1T}},
  \href{https://doi.org/10.1103/PhysRevLett.123.251801}{\emph{Phys. Rev. Lett.}
  {\bfseries 123} (2019) 251801}
  [\href{https://arxiv.org/abs/1907.11485}{{\ttfamily 1907.11485}}].

\bibitem{XENON:2024znc}
{\scshape XENON} collaboration, \emph{{Search for Light Dark Matter in
  Low-Energy Ionization Signals from XENONnT}},
  \href{https://doi.org/10.1103/PhysRevLett.134.161004}{\emph{Phys. Rev. Lett.}
  {\bfseries 134} (2025) 161004}
  [\href{https://arxiv.org/abs/2411.15289}{{\ttfamily 2411.15289}}].

\bibitem{XENON:2026qow}
{\scshape XENON} collaboration, \emph{{Light Dark Matter Search with 7.8
  Tonne-Year of Ionization-Only Data in XENONnT}},
  \href{https://arxiv.org/abs/2601.11296}{{\ttfamily 2601.11296}}.

\bibitem{DAMIC-M:2025luv}
{\scshape DAMIC-M} collaboration, \emph{{Probing Benchmark Models of
  Hidden-Sector Dark Matter with DAMIC-M}},
  \href{https://doi.org/10.1103/2tcc-bqck}{\emph{Phys. Rev. Lett.} {\bfseries
  135} (2025) 071002} [\href{https://arxiv.org/abs/2503.14617}{{\ttfamily
  2503.14617}}].

\bibitem{LZ:2022lsv}
{\scshape LZ} collaboration, \emph{{First Dark Matter Search Results from the
  LUX-ZEPLIN (LZ) Experiment}},
  \href{https://doi.org/10.1103/PhysRevLett.131.041002}{\emph{Phys. Rev. Lett.}
  {\bfseries 131} (2023) 041002}
  [\href{https://arxiv.org/abs/2207.03764}{{\ttfamily 2207.03764}}].

\bibitem{LZ:2023poo}
{\scshape LZ} collaboration, \emph{{Search for new physics in low-energy
  electron recoils from the first LZ exposure}},
  \href{https://doi.org/10.1103/PhysRevD.108.072006}{\emph{Phys. Rev. D}
  {\bfseries 108} (2023) 072006}
  [\href{https://arxiv.org/abs/2307.15753}{{\ttfamily 2307.15753}}].

\bibitem{LZ:2025igz}
{\scshape LZ} collaboration, \emph{{Searches for Light Dark Matter and Evidence
  of Coherent Elastic Neutrino-Nucleus Scattering of Solar Neutrinos with the
  LUX-ZEPLIN (LZ) Experiment}},
  \href{https://arxiv.org/abs/2512.08065}{{\ttfamily 2512.08065}}.

\bibitem{McDonald:2001vt}
J.~McDonald, \emph{{Thermally generated gauge singlet scalars as
  selfinteracting dark matter}},
  \href{https://doi.org/10.1103/PhysRevLett.88.091304}{\emph{Phys. Rev. Lett.}
  {\bfseries 88} (2002) 091304}
  [\href{https://arxiv.org/abs/hep-ph/0106249}{{\ttfamily hep-ph/0106249}}].

\bibitem{Hall:2009bx}
L.J.~Hall, K.~Jedamzik, J.~March-Russell and S.M.~West, \emph{{Freeze-In
  Production of FIMP Dark Matter}},
  \href{https://doi.org/10.1007/JHEP03(2010)080}{\emph{JHEP} {\bfseries 03}
  (2010) 080} [\href{https://arxiv.org/abs/0911.1120}{{\ttfamily 0911.1120}}].

\bibitem{Chu:2011be}
X.~Chu, T.~Hambye and M.H.G.~Tytgat, \emph{{The Four Basic Ways of Creating
  Dark Matter Through a Portal}},
  \href{https://doi.org/10.1088/1475-7516/2012/05/034}{\emph{JCAP} {\bfseries
  05} (2012) 034} [\href{https://arxiv.org/abs/1112.0493}{{\ttfamily
  1112.0493}}].

\bibitem{Hambye:2018dpi}
T.~Hambye, M.H.G.~Tytgat, J.~Vandecasteele and L.~Vanderheyden, \emph{{Dark
  matter direct detection is testing freeze-in}},
  \href{https://doi.org/10.1103/PhysRevD.98.075017}{\emph{Phys. Rev. D}
  {\bfseries 98} (2018) 075017}
  [\href{https://arxiv.org/abs/1807.05022}{{\ttfamily 1807.05022}}].

\bibitem{Okun:1982xi}
L.B.~Okun, \emph{{LIMITS OF ELECTRODYNAMICS: PARAPHOTONS?}}, {\emph{Sov. Phys.
  JETP} {\bfseries 56} (1982) 502}.

\bibitem{Holdom:1985ag}
B.~Holdom, \emph{{Two U(1)'s and Epsilon Charge Shifts}},
  \href{https://doi.org/10.1016/0370-2693(86)91377-8}{\emph{Phys. Lett. B}
  {\bfseries 166} (1986) 196}.

\bibitem{2020Snowmass2021LetterOI}
\emph{Snowmass2021-letter of interest the tesseract dark matter project},
  2020,
  \href{https://api.semanticscholar.org/CorpusID:227194749}{https://api.semanticscholar.org/CorpusID:227194749}.

\bibitem{DarkSide-50:2025lns}
{\scshape DarkSide-50, DarkSide-20k} collaboration, \emph{{Sensitivity to
  low-mass WIMPs with an improved liquid argon ionization response model within
  the DarkSide programme}},  \href{https://arxiv.org/abs/2511.13629}{{\ttfamily
  2511.13629}}.

\bibitem{Oscura:2023qik}
{\scshape Oscura} collaboration, \emph{{Skipper-CCD sensors for the Oscura
  experiment: requirements and preliminary tests}},
  \href{https://doi.org/10.1088/1748-0221/18/08/P08016}{\emph{JINST} {\bfseries
  18} (2023) P08016} [\href{https://arxiv.org/abs/2304.04401}{{\ttfamily
  2304.04401}}].

\bibitem{SuperCDMS:2022kse}
{\scshape SuperCDMS} collaboration, \emph{{A Strategy for Low-Mass Dark Matter
  Searches with Cryogenic Detectors in the SuperCDMS SNOLAB Facility}},  in
  \emph{{Snowmass 2021}}, 3, 2022
  [\href{https://arxiv.org/abs/2203.08463}{{\ttfamily 2203.08463}}].

\bibitem{TESSERACT:2025tfw}
{\scshape TESSERACT} collaboration, \emph{{First Limits on Light Dark Matter
  Interactions in a Low Threshold Two-Channel Athermal Phonon Detector from the
  TESSERACT Collaboration}},
  \href{https://doi.org/10.1103/hsrl-crvf}{\emph{Phys. Rev. Lett.} {\bfseries
  135} (2025) 161002} [\href{https://arxiv.org/abs/2503.03683}{{\ttfamily
  2503.03683}}].

\bibitem{Iles:2024zka}
E.~Iles, S.~Heeba and K.~Schutz, \emph{{Dark Matter Direct Detection
  Experiments Are Sensitive to the Millicharged Background}},
  \href{https://doi.org/10.1103/PhysRevLett.134.121002}{\emph{Phys. Rev. Lett.}
  {\bfseries 134} (2025) 121002}
  [\href{https://arxiv.org/abs/2407.21096}{{\ttfamily 2407.21096}}].

\bibitem{Heeba:2019jho}
S.~Heeba and F.~Kahlhoefer, \emph{{Probing the freeze-in mechanism in dark
  matter models with U(1)' gauge extensions}},
  \href{https://doi.org/10.1103/PhysRevD.101.035043}{\emph{Phys. Rev. D}
  {\bfseries 101} (2020) 035043}
  [\href{https://arxiv.org/abs/1908.09834}{{\ttfamily 1908.09834}}].

\bibitem{Chang:2019xva}
J.H.~Chang, R.~Essig and A.~Reinert, \emph{{Light(ly)-coupled Dark Matter in
  the keV Range: Freeze-In and Constraints}},
  \href{https://doi.org/10.1007/JHEP03(2021)141}{\emph{JHEP} {\bfseries 03}
  (2021) 141} [\href{https://arxiv.org/abs/1911.03389}{{\ttfamily
  1911.03389}}].

\bibitem{Mohapatra:2019ysk}
R.N.~Mohapatra and N.~Okada, \emph{{Dark Matter Constraints on Low Mass and
  Weakly Coupled B-L Gauge Boson}},
  \href{https://doi.org/10.1103/PhysRevD.102.035028}{\emph{Phys. Rev. D}
  {\bfseries 102} (2020) 035028}
  [\href{https://arxiv.org/abs/1908.11325}{{\ttfamily 1908.11325}}].

\bibitem{delaVega:2021wpx}
L.M.G.~de~la Vega, L.J.~Flores, N.~Nath and E.~Peinado, \emph{{Complementarity
  between dark matter direct searches and CE{\ensuremath{\nu}}NS experiments in
  U(1)' models}}, \href{https://doi.org/10.1007/JHEP09(2021)146}{\emph{JHEP}
  {\bfseries 09} (2021) 146}
  [\href{https://arxiv.org/abs/2107.04037}{{\ttfamily 2107.04037}}].

\bibitem{Nath:2021uqb}
N.~Nath, N.~Okada, S.~Okada, D.~Raut and Q.~Shafi, \emph{{Light $Z^\prime $ and
  Dirac fermion dark matter in the $B-L$ model}},
  \href{https://doi.org/10.1140/epjc/s10052-022-10801-3}{\emph{Eur. Phys. J. C}
  {\bfseries 82} (2022) 864}
  [\href{https://arxiv.org/abs/2112.08960}{{\ttfamily 2112.08960}}].

\bibitem{Ghoshal:2022zwu}
A.~Ghoshal, N.~Okada and A.~Paul, \emph{{eV Hubble scale inflation with a
  radiative plateau: Very light inflaton, reheating, and dark matter in B-L
  extensions}}, \href{https://doi.org/10.1103/PhysRevD.106.095021}{\emph{Phys.
  Rev. D} {\bfseries 106} (2022) 095021}
  [\href{https://arxiv.org/abs/2203.03670}{{\ttfamily 2203.03670}}].

\bibitem{Davidson:1978pm}
A.~Davidson, \emph{{$B-L$ as the fourth color within an $\mathrm{SU}(2)_L
  \times \mathrm{U}(1)_R \times \mathrm{U}(1)$ model}},
  \href{https://doi.org/10.1103/PhysRevD.20.776}{\emph{Phys. Rev. D} {\bfseries
  20} (1979) 776}.

\bibitem{Davidson:1979wr}
A.~Davidson, M.~Koca and K.C.~Wali, \emph{{U(1) as the Minimal Horizontal Gauge
  Symmetry}}, \href{https://doi.org/10.1103/PhysRevLett.43.92}{\emph{Phys. Rev.
  Lett.} {\bfseries 43} (1979) 92}.

\bibitem{Mohapatra:1980qe}
R.N.~Mohapatra and R.E.~Marshak, \emph{{Local B-L Symmetry of Electroweak
  Interactions, Majorana Neutrinos and Neutron Oscillations}},
  \href{https://doi.org/10.1103/PhysRevLett.44.1316}{\emph{Phys. Rev. Lett.}
  {\bfseries 44} (1980) 1316}.

\bibitem{Wetterich:1981bx}
C.~Wetterich, \emph{{Neutrino Masses and the Scale of B-L Violation}},
  \href{https://doi.org/10.1016/0550-3213(81)90279-0}{\emph{Nucl. Phys. B}
  {\bfseries 187} (1981) 343}.

\bibitem{Foot:1990mn}
R.~Foot, \emph{{New Physics From Electric Charge Quantization?}},
  \href{https://doi.org/10.1142/S0217732391000543}{\emph{Mod. Phys. Lett. A}
  {\bfseries 6} (1991) 527}.

\bibitem{He:1990pn}
X.G.~He, G.C.~Joshi, H.~Lew and R.R.~Volkas, \emph{{NEW Z-prime
  PHENOMENOLOGY}}, \href{https://doi.org/10.1103/PhysRevD.43.R22}{\emph{Phys.
  Rev. D} {\bfseries 43} (1991) 22}.

\bibitem{Foot:1990uf}
R.~Foot, G.C.~Joshi, H.~Lew and R.R.~Volkas, \emph{{Charge quantization in the
  standard model and some of its extensions}},
  \href{https://doi.org/10.1142/S0217732390003176}{\emph{Mod. Phys. Lett. A}
  {\bfseries 5} (1990) 2721}.

\bibitem{Foot:1992ui}
R.~Foot, H.~Lew and R.R.~Volkas, \emph{{Electric charge quantization}},
  \href{https://doi.org/10.1088/0954-3899/19/3/005}{\emph{J. Phys. G}
  {\bfseries 19} (1993) 361}
  [\href{https://arxiv.org/abs/hep-ph/9209259}{{\ttfamily hep-ph/9209259}}].

\bibitem{Co:2015pka}
R.T.~Co, F.~D'Eramo, L.J.~Hall and D.~Pappadopulo, \emph{{Freeze-In Dark Matter
  with Displaced Signatures at Colliders}},
  \href{https://doi.org/10.1088/1475-7516/2015/12/024}{\emph{JCAP} {\bfseries
  12} (2015) 024} [\href{https://arxiv.org/abs/1506.07532}{{\ttfamily
  1506.07532}}].

\bibitem{Belanger:2018sti}
G.~B{\'e}langer et~al., \emph{{LHC-friendly minimal freeze-in models}},
  \href{https://doi.org/10.1007/JHEP02(2019)186}{\emph{JHEP} {\bfseries 02}
  (2019) 186} [\href{https://arxiv.org/abs/1811.05478}{{\ttfamily
  1811.05478}}].

\bibitem{Bhattiprolu:2022sdd}
P.N.~Bhattiprolu, G.~Elor, R.~McGehee and A.~Pierce, \emph{{Freezing-in
  hadrophilic dark matter at low reheating temperatures}},
  \href{https://doi.org/10.1007/JHEP01(2023)128}{\emph{JHEP} {\bfseries 01}
  (2023) 128} [\href{https://arxiv.org/abs/2210.15653}{{\ttfamily
  2210.15653}}].

\bibitem{Cosme:2023xpa}
C.~Cosme, F.~Costa and O.~Lebedev, \emph{{Freeze-in at stronger coupling}},
  \href{https://doi.org/10.1103/PhysRevD.109.075038}{\emph{Phys. Rev. D}
  {\bfseries 109} (2024) 075038}
  [\href{https://arxiv.org/abs/2306.13061}{{\ttfamily 2306.13061}}].

\bibitem{Gan:2023jbs}
X.~Gan and Y.-D.~Tsai, \emph{{Cosmic millicharge background and reheating
  probes}}, \href{https://doi.org/10.1007/JHEP07(2025)094}{\emph{JHEP}
  {\bfseries 07} (2025) 094}
  [\href{https://arxiv.org/abs/2308.07951}{{\ttfamily 2308.07951}}].

\bibitem{Silva-Malpartida:2023yks}
J.~Silva-Malpartida, N.~Bernal, J.~Jones-P{\'e}rez and R.A.~Lineros,
  \emph{{From WIMPs to FIMPs with low~reheating~temperatures}},
  \href{https://doi.org/10.1088/1475-7516/2023/09/015}{\emph{JCAP} {\bfseries
  09} (2023) 015} [\href{https://arxiv.org/abs/2306.14943}{{\ttfamily
  2306.14943}}].

\bibitem{Becker:2023tvd}
M.~Becker, E.~Copello, J.~Harz, J.~Lang and Y.~Xu, \emph{{Confronting dark
  matter freeze-in during reheating with constraints from inflation}},
  \href{https://doi.org/10.1088/1475-7516/2024/01/053}{\emph{JCAP} {\bfseries
  01} (2024) 053} [\href{https://arxiv.org/abs/2306.17238}{{\ttfamily
  2306.17238}}].

\bibitem{Cosme:2024ndc}
C.~Cosme, F.~Costa and O.~Lebedev, \emph{{Temperature evolution in the Early
  Universe and freeze-in at stronger coupling}},
  \href{https://doi.org/10.1088/1475-7516/2024/06/031}{\emph{JCAP} {\bfseries
  06} (2024) 031} [\href{https://arxiv.org/abs/2402.04743}{{\ttfamily
  2402.04743}}].

\bibitem{Belanger:2024yoj}
G.~B{\'e}langer, N.~Bernal and A.~Pukhov, \emph{{Z'-mediated dark matter with
  low-temperature reheating}},
  \href{https://doi.org/10.1007/JHEP03(2025)079}{\emph{JHEP} {\bfseries 03}
  (2025) 079} [\href{https://arxiv.org/abs/2412.12303}{{\ttfamily
  2412.12303}}].

\bibitem{Arcadi:2024obp}
G.~Arcadi, D.~Cabo-Almeida and O.~Lebedev, \emph{{Z'-mediated dark matter
  freeze-in at stronger coupling}},
  \href{https://doi.org/10.1016/j.physletb.2025.139268}{\emph{Phys. Lett. B}
  {\bfseries 861} (2025) 139268}
  [\href{https://arxiv.org/abs/2409.02191}{{\ttfamily 2409.02191}}].

\bibitem{Arcadi:2024wwg}
G.~Arcadi, F.~Costa, A.~Goudelis and O.~Lebedev, \emph{{Higgs portal dark
  matter freeze-in at stronger coupling: observational benchmarks}},
  \href{https://doi.org/10.1007/JHEP07(2024)044}{\emph{JHEP} {\bfseries 07}
  (2024) 044} [\href{https://arxiv.org/abs/2405.03760}{{\ttfamily
  2405.03760}}].

\bibitem{Barman:2024nhr}
B.~Barman, S.~Bhattacharya, S.~Jahedi, D.~Pradhan and A.~Sarkar, \emph{{Lepton
  collider as a window to reheating via freezing in dark matter detection. Part
  I}}, \href{https://doi.org/10.1016/j.physletb.2025.139863}{\emph{Phys. Lett.
  B} {\bfseries 869} (2025) 139863}
  [\href{https://arxiv.org/abs/2406.11963}{{\ttfamily 2406.11963}}].

\bibitem{Barman:2024lxy}
B.~Barman, A.~Das and S.~Mandal, \emph{{Dark matter-electron scattering and
  freeze-in scenarios in the light of Z' mediation}},
  \href{https://doi.org/10.1103/PhysRevD.110.055029}{\emph{Phys. Rev. D}
  {\bfseries 110} (2024) 055029}
  [\href{https://arxiv.org/abs/2407.00969}{{\ttfamily 2407.00969}}].

\bibitem{Bernal:2024ndy}
N.~Bernal, C.S.~Fong and {\'O}.~Zapata, \emph{{Probing low-reheating scenarios
  with minimal freeze-in dark matter}},
  \href{https://doi.org/10.1007/JHEP02(2025)161}{\emph{JHEP} {\bfseries 02}
  (2025) 161} [\href{https://arxiv.org/abs/2412.04550}{{\ttfamily
  2412.04550}}].

\bibitem{Boddy:2024vgt}
K.K.~Boddy, K.~Freese, G.~Montefalcone and B.~Shams Es~Haghi, \emph{{Minimal
  dark matter freeze-in with low reheating temperatures and implications for
  direct detection}},
  \href{https://doi.org/10.1103/PhysRevD.111.063537}{\emph{Phys. Rev. D}
  {\bfseries 111} (2025) 063537}
  [\href{https://arxiv.org/abs/2405.06226}{{\ttfamily 2405.06226}}].

\bibitem{Arias:2025tvd}
P.~Arias, B.~D{\'\i}az~S{\'a}ez, L.~Duarte, J.~Jones-P{\'e}rez, W.~Rodriguez
  and D.Z.~Herrera, \emph{{Probing displaced (dark)photons from low reheating
  freeze-in at the LHC}},
  \href{https://doi.org/10.1007/JHEP01(2026)135}{\emph{JHEP} {\bfseries 01}
  (2026) 135} [\href{https://arxiv.org/abs/2507.15930}{{\ttfamily
  2507.15930}}].

\bibitem{Bernal:2025osg}
N.~Bernal, E.~Cervantes, K.~Deka and A.~Hryczuk, \emph{{Freezing-in cannibals
  with low-reheating temperature}},
  \href{https://doi.org/10.1007/JHEP09(2025)083}{\emph{JHEP} {\bfseries 09}
  (2025) 083} [\href{https://arxiv.org/abs/2506.09155}{{\ttfamily
  2506.09155}}].

\bibitem{PandaX:2024muv}
{\scshape PandaX} collaboration, \emph{{First Indication of Solar B8 Neutrinos
  through Coherent Elastic Neutrino-Nucleus Scattering in PandaX-4T}},
  \href{https://doi.org/10.1103/PhysRevLett.133.191001}{\emph{Phys. Rev. Lett.}
  {\bfseries 133} (2024) 191001}
  [\href{https://arxiv.org/abs/2407.10892}{{\ttfamily 2407.10892}}].

\bibitem{XENON:2024hup}
{\scshape XENON} collaboration, \emph{{First Search for Light Dark Matter in
  the Neutrino Fog with XENONnT}},
  \href{https://doi.org/10.1103/PhysRevLett.134.111802}{\emph{Phys. Rev. Lett.}
  {\bfseries 134} (2025) 111802}
  [\href{https://arxiv.org/abs/2409.17868}{{\ttfamily 2409.17868}}].

\bibitem{Billard:2013qya}
J.~Billard, L.~Strigari and E.~Figueroa-Feliciano, \emph{{Implication of
  neutrino backgrounds on the reach of next generation dark matter direct
  detection experiments}},
  \href{https://doi.org/10.1103/PhysRevD.89.023524}{\emph{Phys. Rev. D}
  {\bfseries 89} (2014) 023524}
  [\href{https://arxiv.org/abs/1307.5458}{{\ttfamily 1307.5458}}].

\bibitem{OHare:2021utq}
C.A.J.~O'Hare, \emph{{New Definition of the Neutrino Floor for Direct Dark
  Matter Searches}},
  \href{https://doi.org/10.1103/PhysRevLett.127.251802}{\emph{Phys. Rev. Lett.}
  {\bfseries 127} (2021) 251802}
  [\href{https://arxiv.org/abs/2109.03116}{{\ttfamily 2109.03116}}].

\bibitem{Boehm:2018sux}
C.~B{\oe}hm, D.G.~Cerde{\~n}o, P.A.N.~Machado, A.~Olivares-Del~Campo,
  E.~Perdomo and E.~Reid, \emph{{How high is the neutrino floor?}},
  \href{https://doi.org/10.1088/1475-7516/2019/01/043}{\emph{JCAP} {\bfseries
  01} (2019) 043} [\href{https://arxiv.org/abs/1809.06385}{{\ttfamily
  1809.06385}}].

\bibitem{Amaral:2020tga}
D.W.P.d.~Amaral, D.G.~Cerde{\~n}o, P.~Foldenauer and E.~Reid, \emph{{Solar
  neutrino probes of the muon anomalous magnetic moment in the gauged $
  \mathrm{U}{(1)}_{L_{\mu }-{L}_{\tau }} $}},
  \href{https://doi.org/10.1007/JHEP12(2020)155}{\emph{JHEP} {\bfseries 12}
  (2020) 155} [\href{https://arxiv.org/abs/2006.11225}{{\ttfamily
  2006.11225}}].

\bibitem{Amaral:2021rzw}
D.W.P.~Amaral, D.G.~Cerde{\~n}o, A.~Cheek and P.~Foldenauer, \emph{{Confirming
  $U(1)_{L_\mu -L_{\tau }}$ as a solution for $(g-2)_\mu $ with neutrinos}},
  \href{https://doi.org/10.1140/epjc/s10052-021-09670-z}{\emph{Eur. Phys. J. C}
  {\bfseries 81} (2021) 861}
  [\href{https://arxiv.org/abs/2104.03297}{{\ttfamily 2104.03297}}].

\bibitem{DeRomeri:2024iaw}
V.~De~Romeri, D.K.~Papoulias and C.A.~Ternes, \emph{{Bounds on new neutrino
  interactions from the first CE{\ensuremath{\nu}}NS data at direct detection
  experiments}},
  \href{https://doi.org/10.1088/1475-7516/2025/05/012}{\emph{JCAP} {\bfseries
  05} (2025) 012} [\href{https://arxiv.org/abs/2411.11749}{{\ttfamily
  2411.11749}}].

\bibitem{DeRomeri:2024dbv}
V.~De~Romeri, D.K.~Papoulias and C.A.~Ternes, \emph{{Light vector mediators at
  direct detection experiments}},
  \href{https://doi.org/10.1007/JHEP05(2024)165}{\emph{JHEP} {\bfseries 05}
  (2024) 165} [\href{https://arxiv.org/abs/2402.05506}{{\ttfamily
  2402.05506}}].

\bibitem{DeRomeri:2026prc}
V.~De~Romeri, D.K.~Papoulias, F.~Pompa, G.~Sanchez~Garcia and C.A.~Ternes,
  \emph{{Testing light and heavy vector mediators with solar CE$\nu$NS
  measurements}},  \href{https://arxiv.org/abs/2603.00554}{{\ttfamily
  2603.00554}}.

\bibitem{Bauer:2022nwt}
M.~Bauer and P.~Foldenauer, \emph{{Consistent Theory of Kinetic Mixing and the
  Higgs Low-Energy Theorem}},
  \href{https://doi.org/10.1103/PhysRevLett.129.171801}{\emph{Phys. Rev. Lett.}
  {\bfseries 129} (2022) 171801}
  [\href{https://arxiv.org/abs/2207.00023}{{\ttfamily 2207.00023}}].

\bibitem{Stueckelberg:1938hvi}
E.C.G.~Stueckelberg, \emph{{Interaction energy in electrodynamics and in the
  field theory of nuclear forces}},
  \href{https://doi.org/10.5169/seals-110852}{\emph{Helv. Phys. Acta}
  {\bfseries 11} (1938) 225}.

\bibitem{Englert:1964et}
F.~Englert and R.~Brout, \emph{{Broken Symmetry and the Mass of Gauge Vector
  Mesons}}, \href{https://doi.org/10.1103/PhysRevLett.13.321}{\emph{Phys. Rev.
  Lett.} {\bfseries 13} (1964) 321}.

\bibitem{Higgs:1964pj}
P.W.~Higgs, \emph{{Broken Symmetries and the Masses of Gauge Bosons}},
  \href{https://doi.org/10.1103/PhysRevLett.13.508}{\emph{Phys. Rev. Lett.}
  {\bfseries 13} (1964) 508}.

\bibitem{Babu:1997st}
K.S.~Babu, C.F.~Kolda and J.~March-Russell, \emph{{Implications of generalized
  Z - Z-prime mixing}},
  \href{https://doi.org/10.1103/PhysRevD.57.6788}{\emph{Phys. Rev. D}
  {\bfseries 57} (1998) 6788}
  [\href{https://arxiv.org/abs/hep-ph/9710441}{{\ttfamily hep-ph/9710441}}].

\bibitem{Bauer:2018onh}
M.~Bauer, P.~Foldenauer and J.~Jaeckel, \emph{{Hunting All the Hidden
  Photons}}, \href{https://doi.org/10.1007/JHEP07(2018)094}{\emph{JHEP}
  {\bfseries 07} (2018) 094}
  [\href{https://arxiv.org/abs/1803.05466}{{\ttfamily 1803.05466}}].

\bibitem{Alonso-Gonzalez:2025xqg}
D.~Alonso-Gonz{\'a}lez, D.~Cerde{\~n}o, P.~Foldenauer and J.M.~No,
  \emph{{GeV-scale thermal dark matter from dark photons: tightly constrained,
  yet allowed}},  \href{https://arxiv.org/abs/2507.11376}{{\ttfamily
  2507.11376}}.

\bibitem{Dvorkin:2019zdi}
C.~Dvorkin, T.~Lin and K.~Schutz, \emph{{Making dark matter out of light:
  freeze-in from plasma effects}},
  \href{https://doi.org/10.1103/PhysRevD.99.115009}{\emph{Phys. Rev. D}
  {\bfseries 99} (2019) 115009}
  [\href{https://arxiv.org/abs/1902.08623}{{\ttfamily 1902.08623}}].

\bibitem{Chung:1998rq}
D.J.H.~Chung, E.W.~Kolb and A.~Riotto, \emph{{Production of massive particles
  during reheating}},
  \href{https://doi.org/10.1103/PhysRevD.60.063504}{\emph{Phys. Rev. D}
  {\bfseries 60} (1999) 063504}
  [\href{https://arxiv.org/abs/hep-ph/9809453}{{\ttfamily hep-ph/9809453}}].

\bibitem{Giudice:2000ex}
G.F.~Giudice, E.W.~Kolb and A.~Riotto, \emph{{Largest temperature of the
  radiation era and its cosmological implications}},
  \href{https://doi.org/10.1103/PhysRevD.64.023508}{\emph{Phys. Rev. D}
  {\bfseries 64} (2001) 023508}
  [\href{https://arxiv.org/abs/hep-ph/0005123}{{\ttfamily hep-ph/0005123}}].

\bibitem{Bhattiprolu:2023akk}
P.N.~Bhattiprolu, R.~McGehee and A.~Pierce, \emph{{Dark sink enhances the
  direct detection of freeze-in dark matter}},
  \href{https://doi.org/10.1103/PhysRevD.110.L031702}{\emph{Phys. Rev. D}
  {\bfseries 110} (2024) L031702}
  [\href{https://arxiv.org/abs/2312.14152}{{\ttfamily 2312.14152}}].

\bibitem{FreezeIn}
\url{https://github.com/prudhvibhattiprolu/FreezeIn}.

\bibitem{Bhattiprolu:2024dmh}
P.N.~Bhattiprolu, R.~McGehee, E.~Petrosky and A.~Pierce, \emph{{Sub-MeV dark
  sink dark matter}},
  \href{https://doi.org/10.1103/PhysRevD.111.035027}{\emph{Phys. Rev. D}
  {\bfseries 111} (2025) 035027}
  [\href{https://arxiv.org/abs/2408.07744}{{\ttfamily 2408.07744}}].

\bibitem{Heeba:2023bik}
S.~Heeba, T.~Lin and K.~Schutz, \emph{{Inelastic freeze-in}},
  \href{https://doi.org/10.1103/PhysRevD.108.095016}{\emph{Phys. Rev. D}
  {\bfseries 108} (2023) 095016}
  [\href{https://arxiv.org/abs/2304.06072}{{\ttfamily 2304.06072}}].

\bibitem{Alguero:2023zol}
G.~Alguero, G.~Belanger, F.~Boudjema, S.~Chakraborti, A.~Goudelis, S.~Kraml
  et~al., \emph{{micrOMEGAs 6.0: N-component dark matter}},
  \href{https://doi.org/10.1016/j.cpc.2024.109133}{\emph{Comput. Phys. Commun.}
  {\bfseries 299} (2024) 109133}
  [\href{https://arxiv.org/abs/2312.14894}{{\ttfamily 2312.14894}}].

\bibitem{Planck:2018vyg}
{\scshape Planck} collaboration, \emph{{Planck 2018 results. VI. Cosmological
  parameters}},
  \href{https://doi.org/10.1051/0004-6361/201833910}{\emph{Astron. Astrophys.}
  {\bfseries 641} (2020) A6}
  [\href{https://arxiv.org/abs/1807.06209}{{\ttfamily 1807.06209}}].

\bibitem{Davidson:2000dw}
S.~Davidson, M.~Losada and A.~Riotto, \emph{{A New perspective on
  baryogenesis}},
  \href{https://doi.org/10.1103/PhysRevLett.84.4284}{\emph{Phys. Rev. Lett.}
  {\bfseries 84} (2000) 4284}
  [\href{https://arxiv.org/abs/hep-ph/0001301}{{\ttfamily hep-ph/0001301}}].

\bibitem{Allahverdi:2010im}
R.~Allahverdi, B.~Dutta and K.~Sinha, \emph{{Baryogenesis and Late-Decaying
  Moduli}}, \href{https://doi.org/10.1103/PhysRevD.82.035004}{\emph{Phys. Rev.
  D} {\bfseries 82} (2010) 035004}
  [\href{https://arxiv.org/abs/1005.2804}{{\ttfamily 1005.2804}}].

\bibitem{Asaka:2019ocw}
T.~Asaka, H.~Ishida and W.~Yin, \emph{{Direct baryogenesis in the broken
  phase}}, \href{https://doi.org/10.1007/JHEP07(2020)174}{\emph{JHEP}
  {\bfseries 07} (2020) 174}
  [\href{https://arxiv.org/abs/1912.08797}{{\ttfamily 1912.08797}}].

\bibitem{Nelson:2019fln}
A.E.~Nelson and H.~Xiao, \emph{{Baryogenesis from B Meson Oscillations}},
  \href{https://doi.org/10.1103/PhysRevD.100.075002}{\emph{Phys. Rev. D}
  {\bfseries 100} (2019) 075002}
  [\href{https://arxiv.org/abs/1901.08141}{{\ttfamily 1901.08141}}].

\bibitem{Elahi:2021jia}
F.~Elahi, G.~Elor and R.~McGehee, \emph{{Charged B mesogenesis}},
  \href{https://doi.org/10.1103/PhysRevD.105.055024}{\emph{Phys. Rev. D}
  {\bfseries 105} (2022) 055024}
  [\href{https://arxiv.org/abs/2109.09751}{{\ttfamily 2109.09751}}].

\bibitem{Jaeckel:2022osh}
J.~Jaeckel and W.~Yin, \emph{{High energy sphalerons for baryogenesis at low
  temperatures}},
  \href{https://doi.org/10.1103/PhysRevD.107.015001}{\emph{Phys. Rev. D}
  {\bfseries 107} (2023) 015001}
  [\href{https://arxiv.org/abs/2206.06376}{{\ttfamily 2206.06376}}].

\bibitem{Bernal:2017zvx}
N.~Bernal and C.S.~Fong, \emph{{Hot Leptogenesis from Thermal Dark Matter}},
  \href{https://doi.org/10.1088/1475-7516/2017/10/042}{\emph{JCAP} {\bfseries
  10} (2017) 042} [\href{https://arxiv.org/abs/1707.02988}{{\ttfamily
  1707.02988}}].

\bibitem{Chen:2019etb}
S.-L.~Chen, A.~Dutta~Banik and Z.-K.~Liu, \emph{{Leptogenesis in fast expanding
  Universe}}, \href{https://doi.org/10.1088/1475-7516/2020/03/009}{\emph{JCAP}
  {\bfseries 03} (2020) 009}
  [\href{https://arxiv.org/abs/1912.07185}{{\ttfamily 1912.07185}}].

\bibitem{Bernal:2022pue}
N.~Bernal, C.S.~Fong, Y.F.~Pérez-González and J.~Turner, \emph{{Rescuing
  high-scale leptogenesis using primordial black holes}},
  \href{https://doi.org/10.1103/PhysRevD.106.035019}{\emph{Phys. Rev. D}
  {\bfseries 106} (2022) 035019}
  [\href{https://arxiv.org/abs/2203.08823}{{\ttfamily 2203.08823}}].

\bibitem{Chakraborty:2022gob}
M.~Chakraborty and S.~Roy, \emph{{Baryon asymmetry and lower bound on right
  handed neutrino mass in fast expanding Universe: an analytical approach}},
  \href{https://doi.org/10.1088/1475-7516/2022/11/053}{\emph{JCAP} {\bfseries
  11} (2022) 053} [\href{https://arxiv.org/abs/2208.04046}{{\ttfamily
  2208.04046}}].

\bibitem{Hambye:2019dwd}
T.~Hambye, M.H.G.~Tytgat, J.~Vandecasteele and L.~Vanderheyden, \emph{{Dark
  matter from dark photons: a taxonomy of dark matter production}},
  \href{https://doi.org/10.1103/PhysRevD.100.095018}{\emph{Phys. Rev. D}
  {\bfseries 100} (2019) 095018}
  [\href{https://arxiv.org/abs/1908.09864}{{\ttfamily 1908.09864}}].

\bibitem{Bringmann:2021sth}
T.~Bringmann, S.~Heeba, F.~Kahlhoefer and K.~Vangsnes, \emph{{Freezing-in a hot
  bath: resonances, medium effects and phase transitions}},
  \href{https://doi.org/10.1007/JHEP02(2022)110}{\emph{JHEP} {\bfseries 02}
  (2022) 110} [\href{https://arxiv.org/abs/2111.14871}{{\ttfamily
  2111.14871}}].

\bibitem{Escudero:2019gzq}
M.~Escudero, D.~Hooper, G.~Krnjaic and M.~Pierre, \emph{{Cosmology with A Very
  Light L$_{\mu}$ {\ensuremath{-}} L$_{\tau}$ Gauge Boson}},
  \href{https://doi.org/10.1007/JHEP03(2019)071}{\emph{JHEP} {\bfseries 03}
  (2019) 071} [\href{https://arxiv.org/abs/1901.02010}{{\ttfamily
  1901.02010}}].

\bibitem{Evans:2017kti}
J.A.~Evans, S.~Gori and J.~Shelton, \emph{{Looking for the WIMP Next Door}},
  \href{https://doi.org/10.1007/JHEP02(2018)100}{\emph{JHEP} {\bfseries 02}
  (2018) 100} [\href{https://arxiv.org/abs/1712.03974}{{\ttfamily
  1712.03974}}].

\bibitem{Essig:2015cda}
R.~Essig, M.~Fernandez-Serra, J.~Mardon, A.~Soto, T.~Volansky and T.-T.~Yu,
  \emph{{Direct Detection of sub-GeV Dark Matter with Semiconductor Targets}},
  \href{https://doi.org/10.1007/JHEP05(2016)046}{\emph{JHEP} {\bfseries 05}
  (2016) 046} [\href{https://arxiv.org/abs/1509.01598}{{\ttfamily
  1509.01598}}].

\bibitem{Weber:2009pt}
M.~Weber and W.~de~Boer, \emph{{Determination of the Local Dark Matter Density
  in our Galaxy}},
  \href{https://doi.org/10.1051/0004-6361/200913381}{\emph{Astron. Astrophys.}
  {\bfseries 509} (2010) A25}
  [\href{https://arxiv.org/abs/0910.4272}{{\ttfamily 0910.4272}}].

\bibitem{Read:2014qva}
J.I.~Read, \emph{{The Local Dark Matter Density}},
  \href{https://doi.org/10.1088/0954-3899/41/6/063101}{\emph{J. Phys. G}
  {\bfseries 41} (2014) 063101}
  [\href{https://arxiv.org/abs/1404.1938}{{\ttfamily 1404.1938}}].

\bibitem{Sadhukhan:2020etu}
S.~Sadhukhan and M.P.~Singh, \emph{{Neutrino floor in leptophilic $U(1)$
  models: Modification in $U(1)_{L_{\mu}-L_{\tau}}$}},
  \href{https://doi.org/10.1103/PhysRevD.103.015015}{\emph{Phys. Rev. D}
  {\bfseries 103} (2021) 015015}
  [\href{https://arxiv.org/abs/2006.05981}{{\ttfamily 2006.05981}}].

\bibitem{DeRomeri:2025nkx}
V.~De~Romeri, A.~Majumdar, D.K.~Papoulias and R.~Srivastava, \emph{{New light
  mediators and the neutrino fog: Implications from XENONnT nuclear recoil
  data}},  \href{https://arxiv.org/abs/2512.08853}{{\ttfamily 2512.08853}}.

\bibitem{snudd2023}
D.~Amaral, D.~Cerde{\~n}o, A.~Cheek and P.~Foldenauer, ``{SNuDD} [computer
  software].'' \url{https://github.com/snudd/snudd.git}, 2023.

\bibitem{Amaral:2023tbs}
D.W.P.~Amaral, D.~Cerde{\~n}o, A.~Cheek and P.~Foldenauer, \emph{{A direct
  detection view of the neutrino NSI landscape}},
  \href{https://doi.org/10.1007/JHEP07(2023)071}{\emph{JHEP} {\bfseries 07}
  (2023) 071} [\href{https://arxiv.org/abs/2302.12846}{{\ttfamily
  2302.12846}}].

\bibitem{Cowan:2010js}
G.~Cowan, K.~Cranmer, E.~Gross and O.~Vitells, \emph{{Asymptotic formulae for
  likelihood-based tests of new physics}},
  \href{https://doi.org/10.1140/epjc/s10052-011-1554-0}{\emph{Eur. Phys. J. C}
  {\bfseries 71} (2011) 1554}
  [\href{https://arxiv.org/abs/1007.1727}{{\ttfamily 1007.1727}}].

\bibitem{Grevesse:1998bj}
N.~Grevesse and A.J.~Sauval, \emph{{Standard Solar Composition}},
  \href{https://doi.org/10.1023/A:1005161325181}{\emph{Space Sci. Rev.}
  {\bfseries 85} (1998) 161}.

\bibitem{Asplund:2009fu}
M.~Asplund, N.~Grevesse, A.J.~Sauval and P.~Scott, \emph{{The chemical
  composition of the Sun}},
  \href{https://doi.org/10.1146/annurev.astro.46.060407.145222}{\emph{Ann. Rev.
  Astron. Astrophys.} {\bfseries 47} (2009) 481}
  [\href{https://arxiv.org/abs/0909.0948}{{\ttfamily 0909.0948}}].

\bibitem{Vinyoles:2016djt}
N.~Vinyoles, A.M.~Serenelli, F.L.~Villante, S.~Basu, J.~Bergstr{\"o}m,
  M.C.~Gonzalez-Garcia et~al., \emph{{A new Generation of Standard Solar
  Models}}, \href{https://doi.org/10.3847/1538-4357/835/2/202}{\emph{Astrophys.
  J.} {\bfseries 835} (2017) 202}
  [\href{https://arxiv.org/abs/1611.09867}{{\ttfamily 1611.09867}}].

\bibitem{Evans:2016obt}
J.~Evans, D.G.~Gamez, S.D.~Porzio, S.~S{\"o}ldner-Rembold and S.~Wren,
  \emph{{Uncertainties in Atmospheric Muon-Neutrino Fluxes Arising from
  Cosmic-Ray Primaries}},
  \href{https://doi.org/10.1103/PhysRevD.95.023012}{\emph{Phys. Rev. D}
  {\bfseries 95} (2017) 023012}
  [\href{https://arxiv.org/abs/1612.03219}{{\ttfamily 1612.03219}}].

\bibitem{Cowan:1998ji}
G.~Cowan, \emph{{Statistical data analysis}}, Oxford University Press (1998).

\bibitem{Kahlhoefer:2017ddj}
F.~Kahlhoefer, S.~Kulkarni and S.~Wild, \emph{{Exploring light mediators with
  low-threshold direct detection experiments}},
  \href{https://doi.org/10.1088/1475-7516/2017/11/016}{\emph{JCAP} {\bfseries
  11} (2017) 016} [\href{https://arxiv.org/abs/1707.08571}{{\ttfamily
  1707.08571}}].

\bibitem{Carew:2023qrj}
B.~Carew, A.R.~Caddell, T.N.~Maity and C.A.J.~O'Hare, \emph{{Neutrino fog for
  dark matter-electron scattering experiments}},
  \href{https://doi.org/10.1103/PhysRevD.109.083016}{\emph{Phys. Rev. D}
  {\bfseries 109} (2024) 083016}
  [\href{https://arxiv.org/abs/2312.04303}{{\ttfamily 2312.04303}}].

\bibitem{Cheek:2025nul}
A.~Cheek, P.~Figueroa, G.~Herrera and I.M.~Shoemaker, \emph{{Sub-GeV Dark
  Matter Under Pressure from Direct Detection}},
  \href{https://arxiv.org/abs/2507.15956}{{\ttfamily 2507.15956}}.

\bibitem{Alenezi:2025kwl}
A.~Alenezi, C.~Cesarotti, S.~Gori and J.~Shelton, \emph{{Discovery Prospects
  for a Minimal Dark Matter Model at Cosmic and Intensity Frontier
  Experiments}},  \href{https://arxiv.org/abs/2504.00077}{{\ttfamily
  2504.00077}}.

\bibitem{Feng:2009hw}
J.L.~Feng, M.~Kaplinghat and H.-B.~Yu, \emph{{Halo Shape and Relic Density
  Exclusions of Sommerfeld-Enhanced Dark Matter Explanations of Cosmic Ray
  Excesses}}, \href{https://doi.org/10.1103/PhysRevLett.104.151301}{\emph{Phys.
  Rev. Lett.} {\bfseries 104} (2010) 151301}
  [\href{https://arxiv.org/abs/0911.0422}{{\ttfamily 0911.0422}}].

\bibitem{Caputo:2021eaa}
A.~Caputo, A.J.~Millar, C.A.J.~O'Hare and E.~Vitagliano, \emph{{Dark photon
  limits: A handbook}},
  \href{https://doi.org/10.1103/PhysRevD.104.095029}{\emph{Phys. Rev. D}
  {\bfseries 104} (2021) 095029}
  [\href{https://arxiv.org/abs/2105.04565}{{\ttfamily 2105.04565}}].

\bibitem{Emken_2024}
T.~Emken, R.~Essig and H.~Xu, \emph{Solar reflection of dark matter with
  dark-photon mediators},
  \href{https://doi.org/10.1088/1475-7516/2024/07/023}{\emph{Journal of
  Cosmology and Astroparticle Physics} {\bfseries 2024} (2024) 023}.

\bibitem{de_Salas_2015}
P.~de~Salas, M.~Lattanzi, G.~Mangano, G.~Miele, S.~Pastor and O.~Pisanti,
  \emph{Bounds on very low reheating scenarios after planck},
  \href{https://doi.org/10.1103/physrevd.92.123534}{\emph{Physical Review D}
  {\bfseries 92} (2015) }.

\bibitem{Bjorken:2009mm}
J.D.~Bjorken, R.~Essig, P.~Schuster and N.~Toro, \emph{{New Fixed-Target
  Experiments to Search for Dark Gauge Forces}},
  \href{https://doi.org/10.1103/PhysRevD.80.075018}{\emph{Phys. Rev. D}
  {\bfseries 80} (2009) 075018}
  [\href{https://arxiv.org/abs/0906.0580}{{\ttfamily 0906.0580}}].

\bibitem{CHARM:1985anb}
{\scshape CHARM} collaboration, \emph{{Search for Axion Like Particle
  Production in 400-{GeV} Proton - Copper Interactions}},
  \href{https://doi.org/10.1016/0370-2693(85)90400-9}{\emph{Phys. Lett. B}
  {\bfseries 157} (1985) 458}.

\bibitem{Blumlein:1990ay}
J.~Blumlein et~al., \emph{{Limits on neutral light scalar and pseudoscalar
  particles in a proton beam dump experiment}},
  \href{https://doi.org/10.1007/BF01548556}{\emph{Z. Phys. C} {\bfseries 51}
  (1991) 341}.

\bibitem{Batell:2009di}
B.~Batell, M.~Pospelov and A.~Ritz, \emph{{Exploring Portals to a Hidden Sector
  Through Fixed Targets}},
  \href{https://doi.org/10.1103/PhysRevD.80.095024}{\emph{Phys. Rev. D}
  {\bfseries 80} (2009) 095024}
  [\href{https://arxiv.org/abs/0906.5614}{{\ttfamily 0906.5614}}].

\bibitem{Caputo:2025avc}
A.~Caputo, J.~Park and S.~Yun, \emph{{The Heavy Dark Photon Handbook:
  Cosmological and Astrophysical Bounds}},
  \href{https://arxiv.org/abs/2511.15785}{{\ttfamily 2511.15785}}.

\bibitem{Zhou:2024aeu}
T.~Zhou, R.~Plestid, K.J.~Kelly, N.~Blinov and P.J.~Fox, \emph{{Long-lived
  vectors from electromagnetic cascades at SHiP}},
  \href{https://doi.org/10.1007/JHEP02(2025)107}{\emph{JHEP} {\bfseries 02}
  (2025) 107} [\href{https://arxiv.org/abs/2412.01880}{{\ttfamily
  2412.01880}}].

\bibitem{Han:2026ozp}
X.~Han and G.~Krnjaic, \emph{{New Thermal-Relic Targets for sub-GeV Dark Matter
  Direct Detection}},  \href{https://arxiv.org/abs/2603.03444}{{\ttfamily
  2603.03444}}.

\bibitem{Bauer:2020itv}
M.~Bauer, P.~Foldenauer and M.~Mosny, \emph{{Flavor structure of anomaly-free
  hidden photon models}},
  \href{https://doi.org/10.1103/PhysRevD.103.075024}{\emph{Phys. Rev. D}
  {\bfseries 103} (2021) 075024}
  [\href{https://arxiv.org/abs/2011.12973}{{\ttfamily 2011.12973}}].

\bibitem{Boehm:2013jpa}
C.~Boehm, M.J.~Dolan and C.~McCabe, \emph{{A Lower Bound on the Mass of Cold
  Thermal Dark Matter from Planck}},
  \href{https://doi.org/10.1088/1475-7516/2013/08/041}{\emph{JCAP} {\bfseries
  08} (2013) 041} [\href{https://arxiv.org/abs/1303.6270}{{\ttfamily
  1303.6270}}].

\bibitem{Nollett:2014lwa}
K.M.~Nollett and G.~Steigman, \emph{{BBN And The CMB Constrain Neutrino Coupled
  Light WIMPs}}, \href{https://doi.org/10.1103/PhysRevD.91.083505}{\emph{Phys.
  Rev. D} {\bfseries 91} (2015) 083505}
  [\href{https://arxiv.org/abs/1411.6005}{{\ttfamily 1411.6005}}].

\bibitem{NA64:2024klw}
{\scshape NA64} collaboration, \emph{{First Results in the Search for Dark
  Sectors at NA64 with the CERN SPS High Energy Muon Beam}},
  \href{https://doi.org/10.1103/PhysRevLett.132.211803}{\emph{Phys. Rev. Lett.}
  {\bfseries 132} (2024) 211803}
  [\href{https://arxiv.org/abs/2401.01708}{{\ttfamily 2401.01708}}].

\bibitem{Foldenauer:2024cdp}
P.~Foldenauer and J.~Hoefken~Zink, \emph{{How to rule out (g {\ensuremath{-}}
  2)$_{\mu}$ in $ \textrm{U}{(1)}_{L_{\mu }-{L}_{\tau }} $ with white dwarf
  cooling}}, \href{https://doi.org/10.1007/JHEP07(2024)096}{\emph{JHEP}
  {\bfseries 07} (2024) 096}
  [\href{https://arxiv.org/abs/2405.00094}{{\ttfamily 2405.00094}}].

\bibitem{Bellini:2011rx}
G.~Bellini et~al., \emph{{Precision measurement of the 7Be solar neutrino
  interaction rate in Borexino}},
  \href{https://doi.org/10.1103/PhysRevLett.107.141302}{\emph{Phys. Rev. Lett.}
  {\bfseries 107} (2011) 141302}
  [\href{https://arxiv.org/abs/1104.1816}{{\ttfamily 1104.1816}}].

\bibitem{Borexino:2017rsf}
{\scshape Borexino} collaboration, \emph{{First Simultaneous Precision
  Spectroscopy of $pp$, $^7$Be, and $pep$ Solar Neutrinos with Borexino
  Phase-II}}, \href{https://doi.org/10.1103/PhysRevD.100.082004}{\emph{Phys.
  Rev. D} {\bfseries 100} (2019) 082004}
  [\href{https://arxiv.org/abs/1707.09279}{{\ttfamily 1707.09279}}].

\bibitem{BaBar:2016sci}
{\scshape BaBar} collaboration, \emph{{Search for a muonic dark force at
  BABAR}}, \href{https://doi.org/10.1103/PhysRevD.94.011102}{\emph{Phys. Rev.
  D} {\bfseries 94} (2016) 011102}
  [\href{https://arxiv.org/abs/1606.03501}{{\ttfamily 1606.03501}}].

\bibitem{Belle-II:2024wtd}
{\scshape Belle-II} collaboration, \emph{{Search for a
  {\ensuremath{\mu}}+{\ensuremath{\mu}}- resonance in four-muon final states at
  Belle II}}, \href{https://doi.org/10.1103/PhysRevD.109.112015}{\emph{Phys.
  Rev. D} {\bfseries 109} (2024) 112015}
  [\href{https://arxiv.org/abs/2403.02841}{{\ttfamily 2403.02841}}].

\bibitem{ATLAS:2024uvu}
{\scshape ATLAS} collaboration, \emph{{Search for a new Z' gauge boson via the
  pp{\textrightarrow}W{\ensuremath{\pm}}(*){\textrightarrow}Z'{\ensuremath{\mu}}{\ensuremath{\pm}}{\ensuremath{\nu}}{\textrightarrow}{\ensuremath{\mu}}{\ensuremath{\pm}}{\ensuremath{\mu}}{\ensuremath{\mp}}{\ensuremath{\mu}}{\ensuremath{\pm}}{\ensuremath{\nu}}
  process in pp collisions at s=13{\,}{\,}TeV with the ATLAS detector}},
  \href{https://doi.org/10.1103/PhysRevD.110.072008}{\emph{Phys. Rev. D}
  {\bfseries 110} (2024) 072008}
  [\href{https://arxiv.org/abs/2402.15212}{{\ttfamily 2402.15212}}].

\bibitem{Gninenko:2014pea}
S.N.~Gninenko, N.V.~Krasnikov and V.A.~Matveev, \emph{{Muon g-2 and searches
  for a new leptophobic sub-GeV dark boson in a missing-energy experiment at
  CERN}}, \href{https://doi.org/10.1103/PhysRevD.91.095015}{\emph{Phys. Rev. D}
  {\bfseries 91} (2015) 095015}
  [\href{https://arxiv.org/abs/1412.1400}{{\ttfamily 1412.1400}}].

\bibitem{Gninenko:2018tlp}
S.N.~Gninenko and N.V.~Krasnikov, \emph{{Probing the muon $g_\mu$ - 2 anomaly,
  $L_\mu - L_\tau$ gauge boson and Dark Matter in dark photon experiments}},
  \href{https://doi.org/10.1016/j.physletb.2018.06.043}{\emph{Phys. Lett. B}
  {\bfseries 783} (2018) 24}
  [\href{https://arxiv.org/abs/1801.10448}{{\ttfamily 1801.10448}}].

\bibitem{NA64:2024nwj}
{\scshape NA64} collaboration, \emph{{Shedding light on dark sectors with
  high-energy muons at the NA64 experiment at the CERN SPS}},
  \href{https://doi.org/10.1103/PhysRevD.110.112015}{\emph{Phys. Rev. D}
  {\bfseries 110} (2024) 112015}
  [\href{https://arxiv.org/abs/2409.10128}{{\ttfamily 2409.10128}}].

\bibitem{NA64:2025ddk}
{\scshape NA64} collaboration, \emph{{Searching for Light Dark Matter and Dark
  Sectors with the NA64 experiment at the CERN SPS}},
  \href{https://arxiv.org/abs/2505.14291}{{\ttfamily 2505.14291}}.

\bibitem{Foldenauer:2018zrz}
P.~Foldenauer, \emph{{Light dark matter in a gauged $U(1)_{L_\mu-L_\tau}$
  model}}, \href{https://doi.org/10.1103/PhysRevD.99.035007}{\emph{Phys. Rev.
  D} {\bfseries 99} (2019) 035007}
  [\href{https://arxiv.org/abs/1808.03647}{{\ttfamily 1808.03647}}].

\bibitem{Albanese:2878604}
{\scshape SHiP} collaboration, \emph{{BDF/SHiP at the ECN3 high-intensity beam
  facility}},  Tech. Rep. \href{https://cds.cern.ch/record/2878604}{}, CERN,
  Geneva (2023).

\bibitem{Blinov:2025aha}
N.~Blinov, P.J.~Fox, K.J.~Kelly, R.~Plestid and T.~Zhou, \emph{{$L_\mu-L_\tau$
  gauge bosons in beam dumps and supernovae}},
  \href{https://arxiv.org/abs/2511.09619}{{\ttfamily 2511.09619}}.

\bibitem{Andreas:2012mt}
S.~Andreas, C.~Niebuhr and A.~Ringwald, \emph{{New Limits on Hidden Photons
  from Past Electron Beam Dumps}},
  \href{https://doi.org/10.1103/PhysRevD.86.095019}{\emph{Phys. Rev. D}
  {\bfseries 86} (2012) 095019}
  [\href{https://arxiv.org/abs/1209.6083}{{\ttfamily 1209.6083}}].

\bibitem{BaBar:2014zli}
{\scshape BaBar} collaboration, \emph{{Search for a Dark Photon in $e^+e^-$
  Collisions at BaBar}},
  \href{https://doi.org/10.1103/PhysRevLett.113.201801}{\emph{Phys. Rev. Lett.}
  {\bfseries 113} (2014) 201801}
  [\href{https://arxiv.org/abs/1406.2980}{{\ttfamily 1406.2980}}].

\bibitem{BaBar:2017tiz}
{\scshape BaBar} collaboration, \emph{{Search for Invisible Decays of a Dark
  Photon Produced in ${e}^{+}{e}^{-}$ Collisions at BaBar}},
  \href{https://doi.org/10.1103/PhysRevLett.119.131804}{\emph{Phys. Rev. Lett.}
  {\bfseries 119} (2017) 131804}
  [\href{https://arxiv.org/abs/1702.03327}{{\ttfamily 1702.03327}}].

\bibitem{Heeck:2018nzc}
J.~Heeck, M.~Lindner, W.~Rodejohann and S.~Vogl, \emph{{Non-Standard Neutrino
  Interactions and Neutral Gauge Bosons}},
  \href{https://doi.org/10.21468/SciPostPhys.6.3.038}{\emph{SciPost Phys.}
  {\bfseries 6} (2019) 038} [\href{https://arxiv.org/abs/1812.04067}{{\ttfamily
  1812.04067}}].

\bibitem{Coloma:2020gfv}
P.~Coloma, M.C.~Gonzalez-Garcia and M.~Maltoni, \emph{{Neutrino oscillation
  constraints on U(1)' models: from non-standard interactions to long-range
  forces}}, \href{https://doi.org/10.1007/JHEP01(2021)114}{\emph{JHEP}
  {\bfseries 01} (2021) 114}
  [\href{https://arxiv.org/abs/2009.14220}{{\ttfamily 2009.14220}}].

\bibitem{Denton:2018xmq}
P.B.~Denton, Y.~Farzan and I.M.~Shoemaker, \emph{{Testing large non-standard
  neutrino interactions with arbitrary mediator mass after COHERENT data}},
  \href{https://doi.org/10.1007/JHEP07(2018)037}{\emph{JHEP} {\bfseries 07}
  (2018) 037} [\href{https://arxiv.org/abs/1804.03660}{{\ttfamily
  1804.03660}}].

\bibitem{Wise:2018rnb}
M.B.~Wise and Y.~Zhang, \emph{{Lepton Flavorful Fifth Force and Depth-dependent
  Neutrino Matter Interactions}},
  \href{https://doi.org/10.1007/JHEP06(2018)053}{\emph{JHEP} {\bfseries 06}
  (2018) 053} [\href{https://arxiv.org/abs/1803.00591}{{\ttfamily
  1803.00591}}].

\bibitem{Coloma:2022umy}
P.~Coloma, M.C.~Gonzalez-Garcia, M.~Maltoni, J.P.~Pinheiro and S.~Urrea,
  \emph{{Constraining new physics with Borexino Phase-II spectral data}},
  \href{https://doi.org/10.1007/JHEP07(2022)138}{\emph{JHEP} {\bfseries 07}
  (2022) 138} [\href{https://arxiv.org/abs/2204.03011}{{\ttfamily
  2204.03011}}].

\bibitem{Ferber:2015jzj}
T.~Ferber, \emph{{Towards First Physics at Belle II}},
  \href{https://doi.org/10.5506/APhysPolB.46.2285}{\emph{Acta Phys. Polon. B}
  {\bfseries 46} (2015) 2285}.

\bibitem{Belle-II:2018jsg}
{\scshape Belle-II} collaboration, \emph{{The Belle II Physics Book}},
  \href{https://doi.org/10.1093/ptep/ptz106}{\emph{PTEP} {\bfseries 2019}
  (2019) 123C01} [\href{https://arxiv.org/abs/1808.10567}{{\ttfamily
  1808.10567}}].

\bibitem{Kling:2025udr}
F.~Kling, P.~Reimitz and A.~Ritz, \emph{{Dark Vector Boson Bremsstrahlung: New
  Form Factors for a Broader Class of Models}},
  \href{https://arxiv.org/abs/2509.09437}{{\ttfamily 2509.09437}}.

\bibitem{Harnik:2012ni}
R.~Harnik, J.~Kopp and P.A.N.~Machado, \emph{{Exploring nu Signals in Dark
  Matter Detectors}},
  \href{https://doi.org/10.1088/1475-7516/2012/07/026}{\emph{JCAP} {\bfseries
  07} (2012) 026} [\href{https://arxiv.org/abs/1202.6073}{{\ttfamily
  1202.6073}}].

\bibitem{Bilmis:2015lja}
S.~Bilmis, I.~Turan, T.M.~Aliev, M.~Deniz, L.~Singh and H.T.~Wong,
  \emph{{Constraints on Dark Photon from Neutrino-Electron Scattering
  Experiments}}, \href{https://doi.org/10.1103/PhysRevD.92.033009}{\emph{Phys.
  Rev. D} {\bfseries 92} (2015) 033009}
  [\href{https://arxiv.org/abs/1502.07763}{{\ttfamily 1502.07763}}].

\bibitem{Denton:2022nol}
P.B.~Denton and J.~Gehrlein, \emph{{New constraints on the dark side of
  non-standard interactions from reactor neutrino scattering data}},
  \href{https://doi.org/10.1103/PhysRevD.106.015022}{\emph{Phys. Rev. D}
  {\bfseries 106} (2022) 015022}
  [\href{https://arxiv.org/abs/2204.09060}{{\ttfamily 2204.09060}}].

\bibitem{Ilten:2015hya}
P.~Ilten, J.~Thaler, M.~Williams and W.~Xue, \emph{{Dark photons from charm
  mesons at LHCb}},
  \href{https://doi.org/10.1103/PhysRevD.92.115017}{\emph{Phys. Rev. D}
  {\bfseries 92} (2015) 115017}
  [\href{https://arxiv.org/abs/1509.06765}{{\ttfamily 1509.06765}}].

\bibitem{Eijima:2022dec}
S.~Eijima, O.~Seto and T.~Shimomura, \emph{{Revisiting sterile neutrino dark
  matter in gauged U(1)B-L model}},
  \href{https://doi.org/10.1103/PhysRevD.106.103513}{\emph{Phys. Rev. D}
  {\bfseries 106} (2022) 103513}
  [\href{https://arxiv.org/abs/2207.01775}{{\ttfamily 2207.01775}}].

\bibitem{Fernandez-Martinez:2025qsw}
E.~Fern{\'a}ndez-Mart{\'\i}nez, A.L.~Foguel, X.~Marcano, D.~Naredo-Tuero,
  V.~Syvolap and K.A.~Urqu{\'\i}a-Calder{\'o}n, \emph{{Leptogenesis and Dark
  Matter in an Inverse Seesaw from gauged B-L breaking}},
  \href{https://arxiv.org/abs/2512.17682}{{\ttfamily 2512.17682}}.

\end{thebibliography}\endgroup

\end{document}